\documentclass[longauth]{aa}
\usepackage[varg]{txfonts}

\usepackage{natbib}
\bibpunct{(}{)}{;}{a}{}{,} % to follow the A&A style
\usepackage{graphicx}   % Including figure files
\usepackage{amsmath}    % Advanced maths commands
\usepackage{placeins}
\usepackage[hyperfootnotes=false, linktocpage=true, breaklinks=true, colorlinks=true, linkcolor=blue, citecolor=blue, urlcolor=blue]{hyperref}

\makeatletter
\renewcommand*\aa@pageof{, page \thepage{} of \pageref*{LastPage}}

\usepackage{lipsum} % For generating dummy text
\usepackage{xcolor}

\usepackage{booktabs}
\usepackage{tablefootnote}
        
\usepackage{color, colortbl}
\usepackage{supertabular}
\usepackage{subcaption}
\usepackage{caption}

\newcommand{\funit}{~erg cm$^{-2}$ s$^{-1}$} 
\newcommand{\ergs}{~erg s$^{-1}$}

\newcommand{\cmsq}{~cm$^{-2}$ }   
\newcommand{\nh}{$\rm{N_{H}}$}
\newcommand{\chisq}{$\rm{\chi^{2}}$ }
\newcommand{\chisqr}{$\rm{\chi_{\nu}^{2}}$ }

\newcommand{\msun}{\rm M$_\odot$}

\usepackage{amssymb}% http://ctan.org/pkg/amssymb
\usepackage{pifont}% http://ctan.org/pkg/pifont
\newcommand{\cmark}{\ding{51}}%

\def\chandra{{\it Chandra}}

\def\gaia{{\it Gaia}}



\begin{document}

\title{EWOCS-II: X-ray properties of the Wolf-Rayet stars in the young Galactic super star cluster Westerlund 1}

\author{K. Anastasopoulou\inst{1,2}
\and M. G. Guarcello\inst{1}
\and E. Flaccomio\inst{1}
\and S. Sciortino\inst{1}
\and S. Benatti\inst{1}
\and M. De Becker\inst{3}
\and N. J. Wright\inst{4}
\and J. J. Drake\inst{5}
\and J. F. Albacete-Colombo\inst{6}  
\and M. Andersen\inst{7}
\and C. Argiroffi\inst{1,8} 
\and A. Bayo\inst{7}
\and R. Castellanos\inst{9}
\and M. Gennaro\inst{10,11}
\and E. K. Grebel\inst{12}
\and M. Miceli\inst{1,7}
\and F. Najarro\inst{8}
\and I. Negueruela\inst{13}
\and L. Prisinzano\inst{1}
\and B. Ritchie\inst{14}
\and M. Robberto\inst{10,15}
\and E. Sabbi\inst{10}
\and P. Zeidler\inst{16}}

   \institute{Istituto Nazionale di Astrofisica (INAF) – Osservatorio Astronomico di Palermo, Piazza del Parlamento 1, 90134 Palermo, Italy\\\email{konstantina.anastasopoulou@inaf.it}
   \and Center for Astrophysics $|$ Harvard \& Smithsonian, 60 Garden Street, Cambridge, MA 02138, USA 
     \and Space Sciences, Technologies and Astrophysics Research (STAR) Institute, University of Liège, Quartier Agora, 19c, Allée du 6 Aôut, B5c, B-4000 Sart Tilman, Belgium 
        \and
        Astrophysics Group, Keele University, Keele, Staffordshire ST5 5BG, United Kingdom
         \and     
        Lockheed Martin Solar and Astrophysics Laboratory, 3251 Hanover Street, Palo Alto, CA 94304, USA  
         \and Universidad de Rio Negro, Sede Atlántica - CONICET, Viedma CP8500, Río Negro, Argentina 
        \and European Southern Observatory, Karl-Schwarzschild-Strasse 2, D-85748 Garching bei M\"{u}nchen, Germany 
        \and Department of Physics and Chemistry, University of Palermo, Palermo, Italy
      \and
         Departamento de Astrofísica, Centro de Astrobiología, (CSIC-INTA), Ctra. Torrejón a Ajalvir, km 4, Torrejón de Ardoz, E-28850 Madrid, Spain
        \and
        Space Telescope Science Institute, 3700 San Martin Dr, Baltimore, MD, 21218, USA 
         \and
        The William H. Miller III Department of Physics \& Astronomy, Bloomberg Center for Physics and Astronomy, Johns Hopkins University, 3400 N. Charles Street, Baltimore, MD 21218, USA 
        \and
        Astronomisches Rechen-Institut, Zentrum f\"ur Astronomie der Universit\"at Heidelberg, M\"onchhofstr.\ 12--14, 69120 Heidelberg, Germany
        \and
        Departamento de Física Aplicada, Facultad de Ciencias, Universidad de Alicante, Carretera de San Vicente s/n, E-03690, San Vicente del Raspeig, Spain 
        \and
        School of Physical Sciences, The Open University, Walton Hall, Milton Keynes MK7 6AA, UK 
         \and
        Johns Hopkins University, 3400 N. Charles Street, Baltimore, MD 21218, USA   
        \and
        AURA for the European Space Agency (ESA), ESA Office, Space Telescope Science Institute, 3700 San Martin Drive, Baltimore, MD 21218, USA }

\date{Received xxxx / Accepted xxxx}

\abstract{}

 \abstract
   {Wolf-Rayet (WR) stars are massive evolved stars that exhibit particularly fast and dense stellar winds. Although they constitute a very short phase near the end of a massive star's life, they play a crucial role in the evolution of massive stars and have a substantial impact on their surrounding environment.}
   {We present the most comprehensive and deepest X-ray study to date of the properties of the richest Wolf-Rayet population observed in a single stellar cluster, Westerlund 1 (Wd1). By examining the X-ray signatures of WR stars, we aim to shed light on the hottest plasma in their stellar winds and gain insights into whether they exist as single stars or within binary systems.}
   {This work is based on 36 \chandra{} observations obtained from the "Extended Westerlund 1 and 2 Open Clusters Survey" (EWOCS) project, plus 8 archival \chandra{} observations. The overall exposure depth ($\sim$1.1\,Ms) and baseline of the EWOCS observations extending over more than one year enable us to perform a detailed photometric, colour, and spectral analysis, as well as to search for short- and long-term periodicity.}
   {In X-rays, we detect 20 out of the 24 known Wolf-Rayet stars in Wd1 down to an observed luminosity of $\sim$7$\times10^{29}$\ergs\  (assuming a distance of 4.23\,kpc to Wd1),
   with 8 WR stars being detected in X-rays for the first time. Nine stars show clear evidence of variability over the year-long baseline, with clear signs of periodicity. 
    The X-ray colours and spectral analysis reveal that the vast majority of the WR stars are hard X-ray sources (kT$\geq$2.0\,keV). The Fe~XXV emission line at $\sim$6.7\,keV, which commonly originates from the wind--wind collision zone in binary systems, is detected for the first time in the spectra of 17 WR stars in Wd1. 
    In addition the $\sim$6.4\,keV fluorescent line is observed in the spectra of three stars, which are among the very few massive stars exhibiting this line, indicating that dense cold material coexists with the hot gas in these systems.    
    Overall, our X-ray results alone suggest a very high binary fraction ($\geq$80\%) for the WR star population in Wd1. When combining our results with properties of the WR population from other wavelengths, we estimate a binary fraction of $\geq$92\%, which could even reach unity. This suggests that either all the most massive stars are found in binary systems within Wd1, or that binarity is essential for the formation of such a rich population of WR stars.}
   {}


\authorrunning{K. Anastasopoulou et al.}

\titlerunning{EWOCS Wolf-Rayet stars}
\maketitle

%%%%%%%%%%%%%%%%%%%%%%%%%%%%%%%%%%%%%%%%%%%%%%%%%%

%%%%%%%%%%%%%%%%% BODY OF PAPER %%%%%%%%%%%%%%%%%%

\section{Introduction}

Wolf-Rayet (WR) stars are evolved massive stars with typical current masses of 10-25 \msun. The minimum initial mass at solar metallicity for a single star to become a WR star is $\sim$25 \msun\ \citep[][and references therein]{crowther07}. WR stars usually fuse helium or heavier elements in their cores, having lost part or all of their outer hydrogen layer. However, the most massive among them, which can reach up to 80 \msun\ or even higher, can still be fusing hydrogen and exhibit substantial hydrogen content in their atmospheres \citep[for a review, see:][]{crowther07}.
WR stars are considered to mainly descend from O type stars, with the WR phase accounting for $\sim$5\% of their entire lifetime. They exhibit high mass-loss rates and dense winds that can reach terminal velocities of the order of $\sim1-2\times10^3$~km s$^{-1}$ \citep{crowther07}. Due to their high-velocity winds, they are characterised by broad emission line spectra, which, depending on their spectral type, show lines of N (WN spectral type), C (WC spectral type), or a combination of C and  O (rare WO spectral type) in addition to strong lines
of He. Studying the WR phase in the lifetime of massive stars is crucial for better understanding the final stages of stellar evolution. Moreover, through their high mass-loss rates and their ultimate death as core-collapse supernova, they dominate the feedback to the local interstellar medium (ISM), can trigger or pause the formation of new stars, and overall contribute to the chemical evolution of galaxies.

In WR stars, due to the high density and higher metal abundances of their winds, it is anticipated that X-rays would be either entirely or highly absorbed. This absorption is expected unless the X-rays are generated at locations farther out from the star, where the wind density is comparatively low. 
Indeed, theoretical simulations make evident the higher opacity of the WR winds ---especially for the WC type--- compared to O-type stars, with the winds of the WR stars remaining optically thick up to a few thousand stellar radii \citep[figure 2 in][]{oskinova16}. Consequently, the presence of X-rays in WR winds necessitates mechanisms capable of generating hot gas at significantly larger radii compared to the lower-density winds observed in OB stars.

X-rays from WR stars have been observed both for presumed single stars and colliding wind binary (CWB) systems, the latter being usually X-ray brighter. For single WR stars, the X-ray emission could be produced from a distribution of shocks embedded in their stellar winds, although alternative scenarios have been proposed \citep{oskinova16,rauw22}. 
The WN type is generally brighter than the WC type, whose wind opacity to soft X-rays is greater. In CWB systems (usually WR+OB), a part of the kinetic energy of the winds injected into the wind--wind interaction region is converted into heat and the shock-heated plasma emits additional X-rays \citep{stevens1992}. 

The X-ray spectra of WR stars are characteristic of optically thin thermal plasmas with emission lines of abundant elements (Mg, Si, S, Ar, Ca, Fe, etc.) and a temperature that ranges up to a few million degrees for single stars to tens of millions of degrees for CWBs. In particular, the Fe~XXV line (at $\sim$6.7\,keV) is observed for plasma temperatures of $>$10M degrees ($>$1\,keV) that cannot be generally reached in the winds of single non-magnetic massive stars and could be treated as a diagnostic of CWBs \citep{rauw22}.

In our Galaxy, 669 WR stars\footnote{\url{http://pacrowther.staff.shef.ac.uk/WRcat/index.php}} have been identified  so far (July 2023), with the vast majority located within the Galactic plane with a confirmed binary fraction of $\sim$30-40\% \citep[e.g.][]{vanderhucht01,crowther07,rosslowe15,dsilva20,dsilva22,dsilva23}. The young massive star cluster Westerlund\,1 (Wd1; l=339.55$^\circ$, b=-00.40$^\circ$) is an ideal laboratory with which to perform a comprehensive study of a large coeval population of WR stars. The Wd1 WR population is the richest known in a single star cluster, amounting to an impressive number of 24 WR stars thought to have been created in a single burst of star formation 3-5\,Myr ago \citep{clarknegueruela02,negueruelaclark05,groh06,crowther06,gennaro17,andersen17}\footnote{Recent papers mainly based on analysis of Gaia data have suggested older ages for the cluster, but these estimates are still in tension with the observed variegate population of massive stars of Wd1, and thus in this paper the 3-5\,Myr age is adopted.}. 
 For comparison, the Arches and Quintuplet clusters have 13 and 19 known WR stars, respectively \citep[e.g.][]{clark18a,clark18b,clarck19arches}.
Moreover, Wd1 is  considered to be the most massive young stellar cluster known within the Milky Way \citep[e.g.][]{clark05,brandner08,gennaro11,lim13,andersen17}, and it is  the closest super star cluster to the Sun  \citep[$\sim$4\,kpc][]{davies19,navarete22,negueruela22}. The Wd1 WR population comprises 16 WN type and 8 WC type stars, with binarity being firmly confirmed through optical and infrared studies for 4 out of the 24 WR stars \citep[e.g.][]{crowther06,bonanos07,clark11,koumpia12,clark20,ritchie22}. However, the true binary fraction is likely higher based on indirect signs typical for CWBs, such as infrared excess, hard X-ray colours, and flat radio spectra \citep{crowther06,skinner06,clark08,dougherty10,clark19}.

Twelve WR stars in Wd1 have already been detected in two pre-EWOCS \chandra{} observations \citep[$\sim$60\,ks;][]{skinner06,clark08}.
These WR stars exhibited hard  X-ray spectra and X-ray colours  overall, suggesting a binary nature.
While the early shallow X-ray observations provide very important insights into the WR population, only deeper X-ray observations can provide crucial information on the nature of all the known WR stars in Wd1, through detailed spectral and timing analysis.
The EWOCS observations (Extended Westerlund 1 and 2 Open Clusters survey), along with the pre-EWOCS \chandra{} archival observations, reach a total exposure time of $\sim$1.1\,Ms, and provide the unique opportunity to obtain a very deep X-ray look into the Wd1 WR population.

This is the second paper in the series of EWOCS papers, the aim of which is to study the WR population in detail. The paper is organised as follows: In Sect. \ref{observations} we provide a very short review of the data, in Sect. \ref{analysis} we present our analysis and results, including the WR source detection, variability (short- and long-term), X-ray colours, and spectral analysis. 
In Sect. \ref{discussion} we discuss our main findings regarding the spectral analysis and the implications for binarity.
In Sect. \ref{conclusions} we summarise our results. We assume a distance to Wd1 of 4.23\,kpc \citep{negueruela22}.

\section{Observations}\label{observations}

This work is based on 44 \chandra{} observations of Wd1, 36 of 
which have been taken as part of the EWOCS \chandra{} Large Program (Proposal number: 21200267, P.I. Guarcello; ACIS-I; 967.80\,ks), while 8 are pre-EWOCS archival observations (ACIS-S; 151.93\,ks). All EWOCS observations are pointed towards Wd1, while two of the pre-EWOCS observations are pointed at Wd1 and six at the magnetar CXO J164710.20-455217. The EWOCS observations were performed between June 2020 and August 2021, while the pre-EWOCs observations occurred between June 2005 and February 2018. A detailed description of all the observations, exposure times, technical characteristics, and data reduction is presented in \citet{guarcello24}. The list of \chandra{} observations used in this paper are contained in the Chandra Data Collection (CDC) 153\footnote{\href{https://doi.org/10.25574/cdc.153}{https://doi.org/10.25574/cdc.153}}.

\section{Analysis and results}\label{analysis}

\subsection{X-ray detection and photometry of Wolf-Rayet stars }\label{detection}

Due to source confusion, and the bright and irregular background in Wd1, a rather sophisticated strategy using four different methods was adopted for source detection in five energy bands. The source validation and photometry was performed using the ACIS-Extract software \citep{broos10}. A detailed description of the source detection and photometry of the 5963 validated sources in Wd1 can be found in \citet{guarcello24}. 

\begin{figure*}[!h]
        %\centering
        \includegraphics[width=2.0\columnwidth]{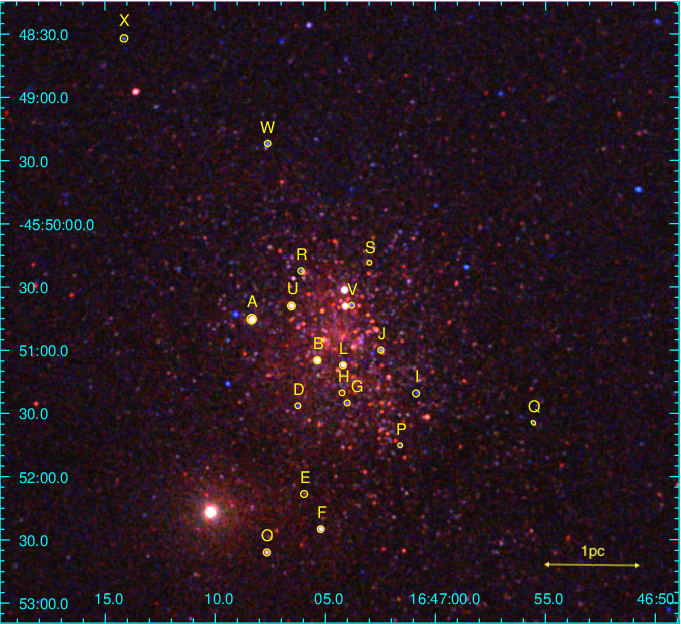}  
        \caption{Smoothed colour \chandra{} image of the central 4$\times$4 arcmin of Westerlund 1. Soft (0.5-2.0\,keV), medium (2.0-4.0\,keV), and hard (4-8.0\,keV) X-ray photons are shown with red, green, and blue colours, respectively. Yellow circles show 19 out of the 20 detected WR stars. The WR star N is not depicted, as it is located $\sim$ 4.5\,arcmin south of the centre of the star cluster.}
                \label{fig.wd1_color}
 \end{figure*}

\begin{figure*}[!h]
        \centering
        \includegraphics[width=0.30\columnwidth]{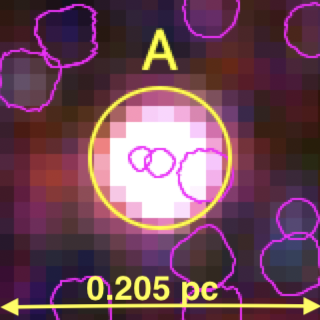} 
    \includegraphics[width=0.30\columnwidth]{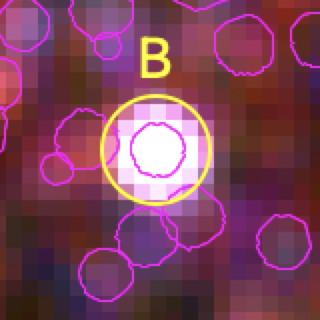} 
    \includegraphics[width=0.30\columnwidth]{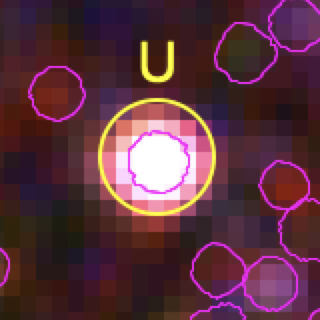} 
    \includegraphics[width=0.30\columnwidth]{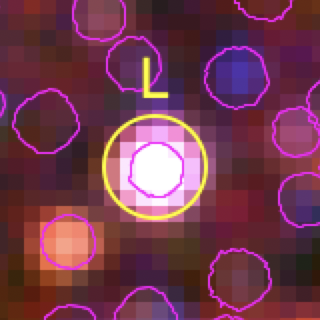} 
    \includegraphics[width=0.30\columnwidth]{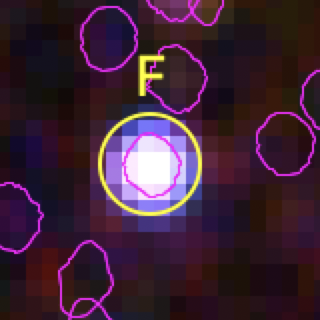} 
    \includegraphics[width=0.30\columnwidth]{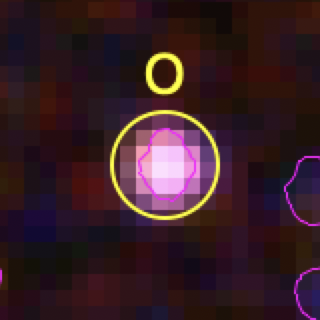} 
\includegraphics[width=0.30\columnwidth]{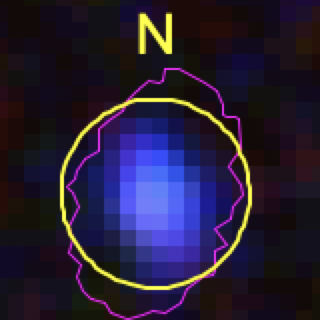} 
    \includegraphics[width=0.30\columnwidth]{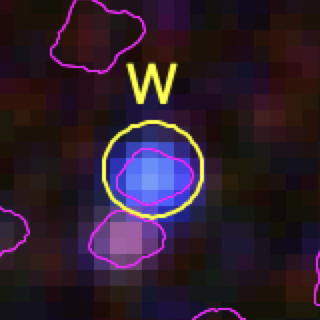} 
    \includegraphics[width=0.30\columnwidth]{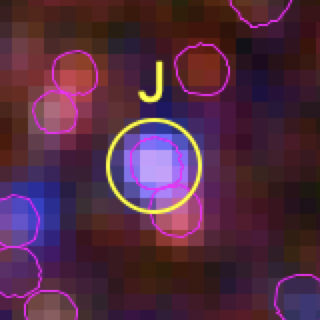} 
    \includegraphics[width=0.30\columnwidth]{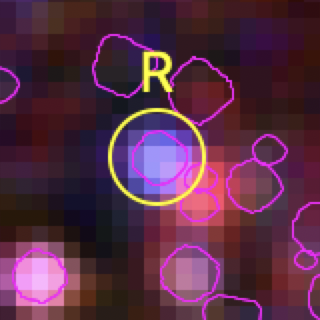} 
    \includegraphics[width=0.30\columnwidth]{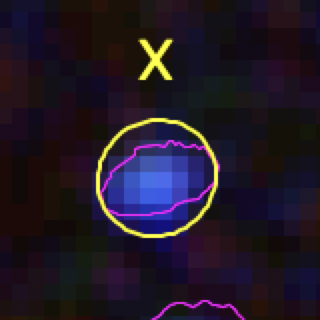} 
    \includegraphics[width=0.30\columnwidth]{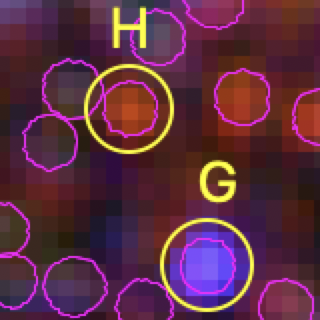} 
    \includegraphics[width=0.30\columnwidth]{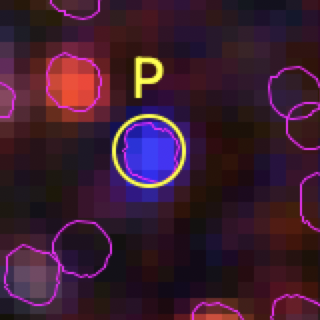} 
 \includegraphics[width=0.30\columnwidth]{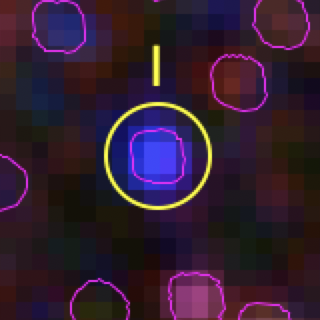}
    \includegraphics[width=0.30\columnwidth]{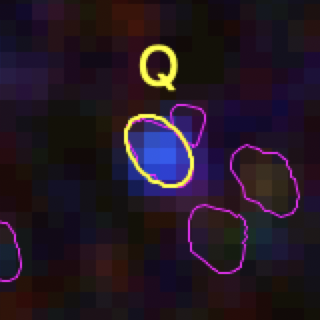} 
    \includegraphics[width=0.30\columnwidth]{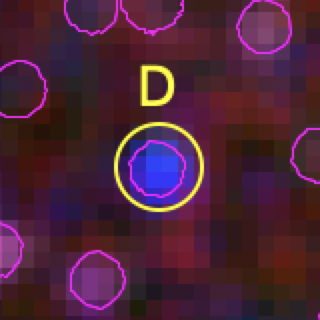} 
    \includegraphics[width=0.30\columnwidth]{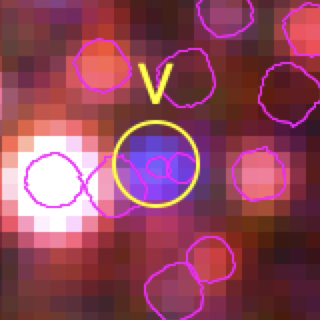} 
    \includegraphics[width=0.30\columnwidth]{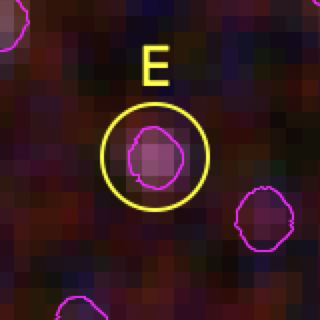} 
    \includegraphics[width=0.30\columnwidth]{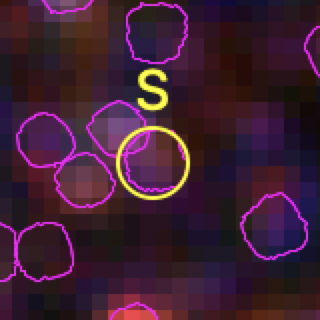} 
        \caption{ X-ray-detected WR population in Wd1. We present the X-ray-brightest WR star from the top left panel and moving rightwards the X-ray brightness decreases. Each panel has a size of 10\arcsec$\times$10\arcsec and the colour coding is the same as in Fig.\,\ref{fig.wd1_color}. The yellow regions are shown just for illustration purposes centred at the X-ray coordinates of the WR stars, while the magenta regions correspond to the extraction regions of detected sources used for photometry and spectral analysis, except for star A where a larger extraction region was used (Sect. \ref{detection}).}
                \label{fig.wr_color}
 \end{figure*}

We detected 20 out of the 24 WR stars known in Wd1. Among them, 8 stars (J, X, P, I, Q, V, H, and S) were detected in X-rays for the first time. The exact locations of the X-ray detected WR stars within the star cluster are shown in the \chandra{} colour image of Wd1 (Fig.\,\ref{fig.wd1_color}), while a close-up view of each X-ray detected WR star can be seen in Fig.\,\ref{fig.wr_color}. The extreme crowding characterising the core of the cluster is evident in both Fig.\,\ref{fig.wd1_color} and Fig.\,\ref{fig.wr_color}. Moreover, the close-up colour images of each WR star in Fig.\,\ref{fig.wr_color} reveal that  12 out of the 20 WR stars appear to be predominately hard (F, N, W, J, R, X, G, P, Q, D, and V) while 1 WR star (H) appears to be predominately soft.

The vast majority of the X-ray sources in Wd1 are very faint \citep{guarcello24}, and since WR stars are expected to be significantly brighter than X-ray-emitting low-mass stars, even in the case of contamination, the low-mass sources would contribute only a few photons to the signal we attribute to the WR stars. Thus, no impact on our study is expected, except for 
 the very faint WR stars.
We therefore inspected NIRCam JWST images in order to evaluate if the X-rays are generated indeed from the WR stars or from another nearby source. The WR stars are very bright and saturated and therefore it was impossible to detect very close sources. However, another much fainter infrared source is detected within the extraction region of  R and J, while within the extraction region of H we find another very bright IR source.

In Fig.\,\ref{fig.wr_color} we also show the extraction regions of the WR stars and nearby sources   with magenta colours. The sizes of the extraction regions were determined by the ACIS-Extract software, which takes into account the local point spread function (PSF) and reduces the region size where necessary in order to avoid overlap with nearby sources \citep{guarcello24}. For all WR stars, except for stars V and A, the extraction regions encircle 86-91\% of the PSF. Inside the yellow circle that illustrates the position of star V (Fig. \ref{fig.wr_color}), ACIS-Extract finds two sources of similar brightness. The source to the east corresponds to the optical position of star V. Due to overlap, the extraction region for star V is reduced to encircle 60\% of the integrated PSF. Similarly, for star A inside the yellow circle that illustrates its position (Fig.\ref{fig.wr_color}), ACIS-Extract finds two more sources, one to the east and one to the west of star A, which contain  130.7 and  233.7 net broad-band counts, respectively. Both  sources are much fainter than star A, but since the source in the east is very close to the WR star, ACIS-Extract reduces the size of the star A extraction region to 58\% of the integrated PSF (hereafter referred to as the small extraction region). Since star A is the brightest among the WR stars in the sample, and the nearby stars are comparatively very faint,  we decided to also use a larger extraction region corresponding to $\geq$90\% of the integrated PSF (hereafter referred to as the large extraction region) and contains 10295 net counts. The contribution of the neighbouring sources to the large extraction region is 3\% for the broad-band net counts (most of it originating from energies of higher than 2\,keV). In the following analysis (long- and short-term periodicity, colours, and spectra), we used both small and large extraction regions of star A. However, as no significant differences were observed, we present only the results from the large extraction region.

Basic properties of the detected WR stars are listed in Table\,\ref{tab.wr_stars}, where sources are sorted according to their X-ray brightness. 
In the broad band (0.5-8.0\,keV), four WR stars have more than 1000 net counts and are all of spectral type WN. Eight WR stars have between 100 and 1000 net counts (two WC and six WN), and another eight WR stars have between 20 and 100 net counts (two WC and six WN). The undetected sources are presented in Table\,\ref{tab.wr_stars_undetected} and are all of WC spectral type.

\begin{table*}[!h]
        \centering

                \caption{X-ray-detected Wolf-Rayet stars in Wd1}
        \setlength{\tabcolsep}{2pt} % Default value: 6pt
        %\renewcommand{\arraystretch}{1.5} % Default value: 1
                %\hspace*{-0.7cm}
  \begin{tabular}{@{}lccccrcccc@{}}
                        \toprule
                Name & ID & Other & RA& Dec&net counts &spectral type & info & P & P (this work) \\
  & & names &  h m s &\degr\ \arcmin\ \arcsec & && &days&days \\
  (1) & (2)&(3)  &(4) & (5)&(6)&(7) &(8)&(9) &(10)\\
                        \midrule
A & 3972 &    WR77sc, W72  &            16:47:08.35  &-45:50:45.4  &      10295$\pm$102 (l)   &  WN7b      &  earlyBIa  &  7.63\tablefootmark{a} & 81.75$\pm$4.63\\[2pt]
&   &&           & &     5412.8$\pm 74.3$ (s)    &     &    &   & 7.70$\pm$0.04\\[2pt]
B       &3299   &   WR77o       &       16:47:05.36  &-45:51:04.8  &    3046.2$\pm55.9$     &  WN7o      &  SB1  & 3.52\tablefootmark{b}  & 3.514$\pm$0.009\\[2pt]
&   &&           & &    &     &    &   & 2.222$\pm$0.004\\[2pt]
U       & 3577  &    WR77s, \#1\tablefootmark{c}         & 16:47:06.53  &-45:50:39.1  &        1996.9$\pm45.2$      &    WN6o     &    & & - \\[2pt]
L  & 3012       &   WR77k, W44       &  16:47:04.19  &-45:51:07.2  &    1221.9$\pm35.5$      &   WN9h      &  binary   &53.95\tablefootmark{d} & 46.41$\pm1.38$ \\[2pt]
F       & 3252  &   WR77n, W239         &       16:47:05.20  &-45:52:25.0  &      903.3$\pm30.3$       &  WC9d      &   SB1 &5.05\tablefootmark{e} & 2.263$\pm$0.004\\[2pt]
&    &  &        & &         &     &    &   & 5.15$\pm$0.02\\[2pt]
O       &  3830 &      WR77sb       &      16:47:07.65  &-45:52:36.0  &      387.7$\pm20.0$       &  WN6o           &    & & - \\[2pt]
N       & 2015  &    WR77b              &  16:46:59.91   &-45:55:25.6  &      375.2$\pm20.0$      &    WC9d        &  SB2?,  O?  & & 58.61$\pm$2.79 \\[2pt]
W       & 3815  &    WR77sa, \#3\tablefootmark{c}          &       16:47:07.61  &-45:49:22.1  &        243.1$\pm16.0$       &    WN6h              &    & &- \\[2pt]
J  &    2579    &     WR77e             &  16:47:02.47  &-45:50:59.9  &      233.5$\pm16.0$       &     WN5h      &    & & 5.15$\pm0.02 $ (u)\\[2pt]
R       &3488   &  WR77q, W14c               &     16:47:06.09  &-45:50:22.4  &      192.9$\pm14.8$       &  WN5o      &    & & - \\[2pt]
X       & 4798  &         WR77sd, \#2\tablefootmark{c}            &       16:47:14.13   &-45:48:32.0  &       154.2$\pm13.0$       &      WN5o           &    & &9.01$\pm$0.05 (u)\\[2pt]
G &     2967    &  WR77j, W39b               &  16:47:04.00  &-45:51:25.1  &      146.8$\pm13.0$       & WN7o       &    & & - \\[2pt]
 P  & 2381              &       WR77d, W57c          &     16:47:01.59  &-45:51:45.2  &        88.9$\pm10.0$        &   WN7o     &    & & 5.19$\pm$0.01  \\[2pt]
I &     2198    &     WR77c           &  16:47:00.87  &-45:51:20.6  &      86.6$\pm10.0$        &   WN8o        &    & &-\\[2pt]
Q &     1416    &     WR77a            &   16:46:55.54  &-45:51:34.5  &      86.5$\pm9.7$        &     WN6o      &    & & 7.4$\pm$0.04 (u) \\[2pt]
D       & 3523  &  WR77r                 & 16:47:06.25  &-45:51:26.4  &      69.3$\pm9.5$       &   WN7o     &    & & - \\[2pt]
V       &       2918 &       WR77h             &   16:47:03.79  &-45:50:38.7  &      66.8$\pm9.0$      &      WN8o       &    & & - \\[2pt]
  E  &  3445    &  WR77p, W241      &      16:47:05.96  &-45:52:08.3  &      56.3$\pm8.0$       & WC9 \tablefootmark{f}       &  RVbinary?  & & -\\[2pt]
H       & 3019  &      WR77l              &        16:47:04.23  &-45:51:20.2  &      26.3$\pm7.3$        &      WC9d    &  & & -\\[2pt]
    S   &2708   &    WR77f, W5            &        16:47:02.98  &-45:50:18.5  &      23.7$\pm5.8$        & WN10-11h/B0.5Ia+ &  & & -\\[2pt]
        \bottomrule
                        \end{tabular}   
   \tablefoot{Column 1 reports the most common names for the WR stars in Wd1. Column 2 reports the EWOCS catalogue source identification number 
 \citep{guarcello24}. Column 3 reports other WR names used in the literature. Columns 4 and 5 report the coordinates obtained from the X-ray source detection. Column 6 reports the net counts in the broad band (0.5-8.0\,keV) and the corresponding errors adopted from \citet{guarcello24}. For star A we also report the net counts corresponding to the large extraction region noted with (l). Column 7 presents the WR spectral type originally reported in \citet{crowther06} unless otherwise indicated. Column 8 presents some information on source binarity and companion properties adopted by \citet{crowther06}, \citet{clark20}, and \citet{ritchie22}. Columns 9 and 10 report the measured periods from other and this work, respectively. For the cases where we detect two periods for the same source, both are reported in order of highest significance. The letter (u) signifies that the measured period is uncertain (a detailed description is provided in Sect. \ref{longterm}).
   \tablefoottext{a}{\citet{bonanos07}} \tablefoottext{b}{\citet{koumpia12}} \tablefoottext{c}{as reported in \citet{groh06}}\tablefoottext{d}{\citet{ritchie22}} \tablefoottext{e}{\citet{clark11}} \tablefoottext{f}{\citet{clarknegueruela02}} }
                \label{tab.wr_stars}
                                \end{table*}

\begin{table}[!h]
        \centering
                \caption{X-ray-undetected Wolf-Rayet stars in Wd1}
  \setlength{\tabcolsep}{2pt}
  \begin{tabular}{@{}lcccc@{}}
                        \toprule
                Name  & Other &RA& Dec &spectral type  \\
  & names & h m s &\degr\ \arcmin\ \arcsec & \\
  (1) & (2)&(3) &(4) &(5) \\
                        \midrule

 C              & WR77m &  16:47:04.40  &-45:51:03.8        &  WC9d      \\
K  & WR77g  &      16:47:03.25  &-45:50:43.8        &  WC8      \\
M & WR77i, W66 &   16:47:03.96  &-45:51:37.5      &   WC9d      \\
T  & WR77aa      & 16:46:46.3   &-45:47:58.0        &    WC9d\tablefootmark{a}             \\ 
        \bottomrule
                        \end{tabular} \tablefoot{Same as in Table\,\ref{tab.wr_stars} apart from coordinate values that are obtained from \citet{clark20}. The luminosity upper limit  in the 0.5-8.0\,keV band for the undetected sources is 1.4$\times10^{29}$\ergs\ (we describe in detail its derivation in Sect. \ref{lxlbol}).\tablefoottext{a}{\citet{hopewell05}} }  \label{tab.wr_stars_undetected}
                                \end{table}

\subsection{Short-term variability}\label{shortterm}

In order to explore any possible short-term variability, which could indicate physical changes for the WR stars within a timescale of few hours, 
we extracted the probability-weighted light curves of the single observations using the CIAO tool \emph{glvary}. The tool utilizes the Gregory-Loredo algorithm variability test \citep{gl92}, through the \emph{glvary} implementation within the CIAO tool, and is optimal for detecting variability in a source since it corrects for instrumental variations that are not accounted for by the \emph{dmextract} tool\footnote{\url{https://cxc.cfa.harvard.edu/ciao/threads/variable/}}.
The \emph{glvary} tool splits the events of a single observation into multiple time bins, looks for significant deviations and then assigns a variability index (\textit{varindex}) to each observation based on the probability for a source to be variable. It also outputs the average standard deviation, and the average count-rate of the light curve. 
The variability index ranges from 0 to 10, and sources with \textit{varindex}$\geq$6 are categorised as definitely variable. 
We used a minimum time bin set at 1000 seconds (\textit{mintime}=1000sec), and explored the variability for the brightest WR stars ($>$200 net counts) in the broad (0.5-8\,keV), soft (0.5-2\,keV), and hard (2-8\,keV) energy bands.

We detected short-term variability for four WR stars, namely star A (OBSID 22988), star N (OBSID 23287), star U (OBSIDs 22983 and 22986) and star L (OBSID 22977). This is quite a rare event (2-5\% occurrence) taking into account that there are overall 41 observations for star A, 39 for stars U and L, and 38 for star N. 
The variability detected for stars A, U, and N is exclusively observed within the broad and/or hard X-ray bands, with no corresponding variation in the soft band. This suggests that the observed variability is more likely linked to temperature variations of the thermal plasma responsible for the X-ray emission. Temperature variations assuming a constant local absorption  would also result in variations of the soft flux, albeit at much lower levels not detected by the \emph{glvary} tool.  
On the contrary, star L shows short-term variability in the soft band alone, pointing to possible changes in the local absorption of the binary system within that observation. The short-term light curves of all observations where variability has been detected are presented in Fig.\,\ref{fig.lc_singleobs} in the Appendix.

\subsection{Long-term variability and periodicity search}\label{longterm}

Given their long baseline spanning more than one year, the EWOCS observations provide a unique opportunity to search for long-term variability of the WR stars. The identification of long-term periodic signals through X-ray data reveals modulations in the flux of the wind of the WR star. This modulation could be intrinsic to the wind, for example a corotating interaction region \citep{chene11,massa14} as suggested in the case of WR6 \citep{ignace13}, or could be due to orbital modulation revealing the presence of a companion.

In the search for variability and periodic signals we did not include the pre-EWOCS observations since they are separated by several years during which the sensitivity of the ACIS detector to soft X-rays has declined owing to contamination build up\footnote{The pre-EWOCS observations have systematically higher count-rate measurements than the EWOCS observations. This is the result of contamination accumulation and associated decline in the effective area over time \citep[i.e.][]{odell15} and of the large span in the observation dates (EWOCS: \chandra{} cycle 21,  pre-EWOCS: \chandra{} cycles 6 and 18), as well as the different chip used (EWOCS: ACIS-I, pre-EWOCS: ACIS-S).}, and we cannot be sure that any periodicity was stable for such a long time.
For the construction of the long-term light curve we used the probability-weighted light curve with the optimal time binning for each observation provided by the \textit{glvary} tool  (see Sect. \ref{shortterm}), as well as the raw light curve, with a single point per observation, provided by the \emph{dmextract} tool.
We initially performed a chi-square test for the probability-weighted light curves of all sources except for the five faintest, to access the significance of the long-term variability. We find that all sources show long-term variability at the 99\% confidence level except for star U which we do not include in the following analysis.
We then constructed the Lomb-Scargle (LS) periodogram which is designed to detect periodic signals in unevenly spaced observations using the \textit{astropy} class LombScargle\footnote{\url{https://docs.astropy.org/en/stable/api/astropy.timeseries.LombScargle.html\#id1}} \citep{zechmeister09,vanderplas12,vanderplas15}. For the construction of the LS periodogram we used the probability-weighted light curves provided by the \emph{glvary} tool.
The advantage of using the individual light curves instead of the single count-rate value for each observation, is that we can account for variability within the same observation. In addition, by utilising the \emph{glvary} light curves we can avoid instrumental variations affecting the light curve (i.e. bad pixels, chip gaps, etc.). One aspect that requires extra attention when using the \emph{glvary} light curves to construct periodograms, is the fact that the tool assigns multiple time bins even when the light curve is considered constant. This affects only the calculation of the false alarm probability (FAP) levels which result in lower power values. In order to address this issue, for observations with identical count-rate values in consecutive time bins, we retained only one value and the average time  stamp.

\begin{figure}[!h]
        %\centering     
        \includegraphics[width=1.0\columnwidth]{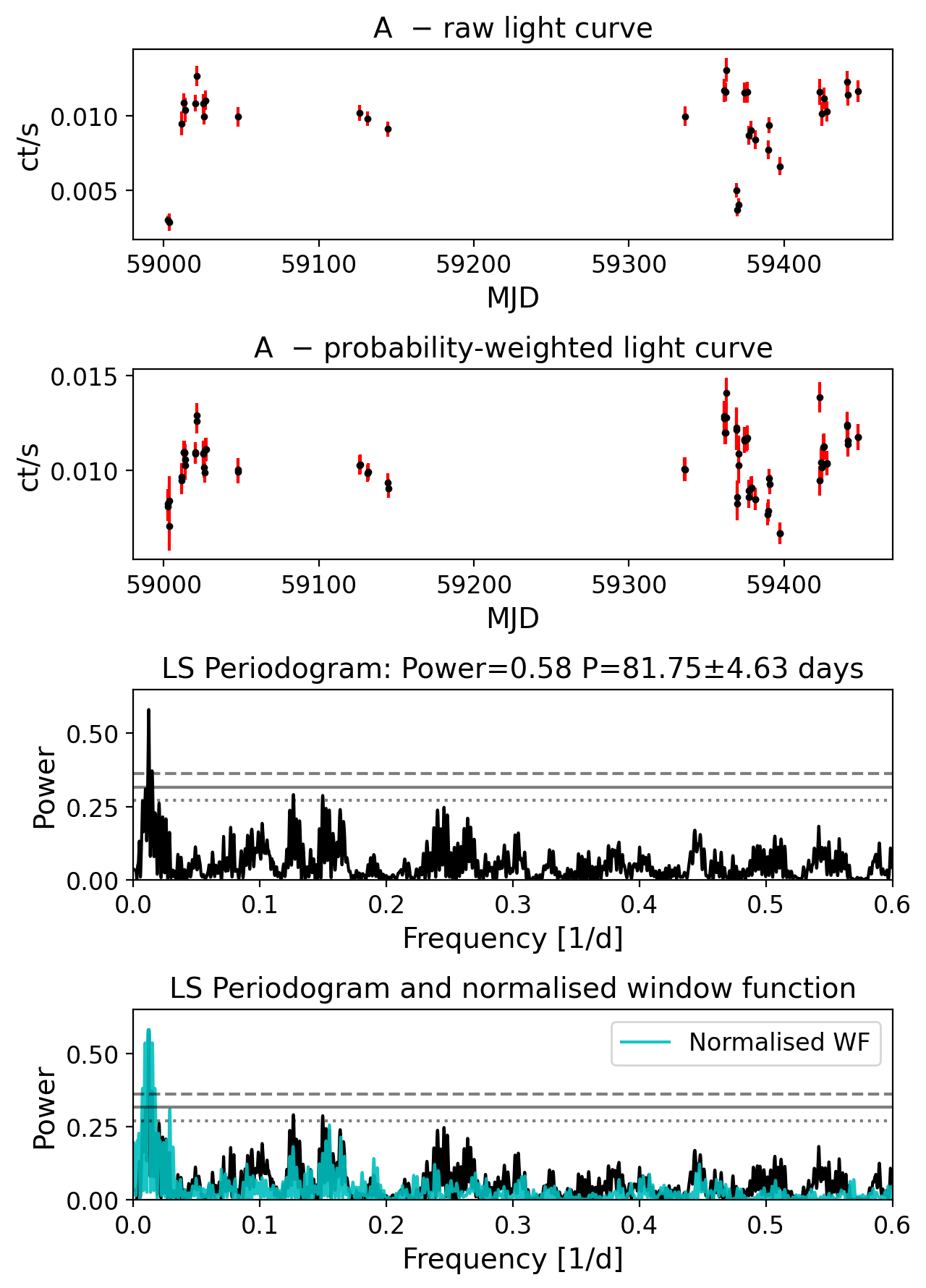}
        \caption{EWOCS light curves and periodogram of star A. Top panel: Broad-band raw light curve produced by the \emph{dmextract} tool. Second panel: Broad-band probability-weighted light curve produced by the  \emph{glvary} tool. Third panel: Resulting LS periodogram and the FAP levels of 0.01, 0.1, and 1 percent are marked with the dashed, solid, and dotted horizontal lines. A clear peak at about 81 days is detected. Bottom panel: Same as third panel, with the normalised window function overlaid on the periodogram.}
                \label{fig.lcbroadw72}
 \end{figure}

\begin{figure}[!h]
        %\centering     
  \includegraphics[width=0.9\columnwidth]{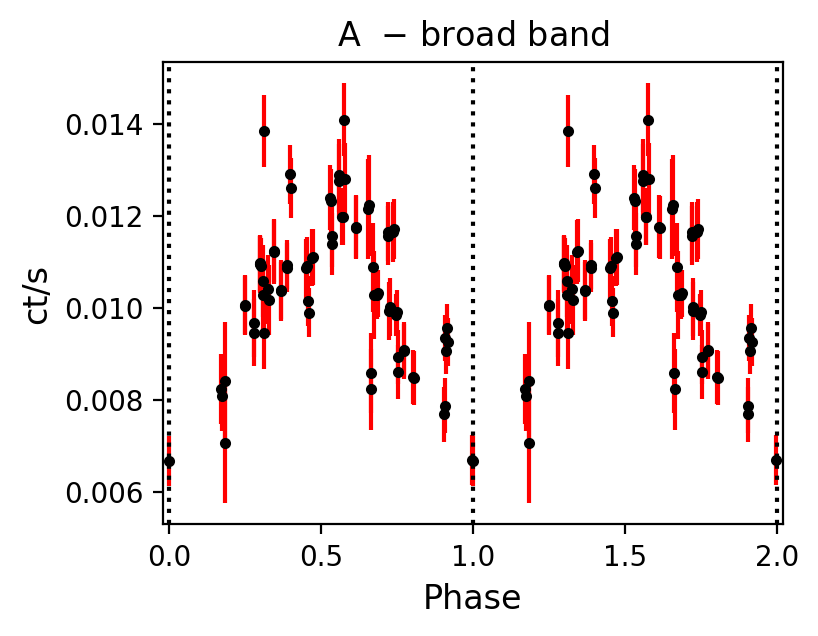}
 \caption{Phase-folded probability-weighted light curve of star A in the broad band.}
                \label{fig.phasedlcw72}
 \end{figure}

In the four panels of Fig.\,\ref{fig.lcbroadw72}, we display for the brightest WR star A, from top to bottom, the raw light curve produced by the \emph{dmextract} tool (one point per observation), the probability-weighted light curve produced by the \emph{glvary} tool (multiple points per observation), the LS periodogram, and the overlaid window function (WF) associated with the time series produced by the LombScargle astropy class. 
Inspecting the light curves in the first two panels, we notice that for a few observations, the raw light curve exhibits very low count-rates compared to that of the probability-weighted light curve. This happens as the source passes through the detector's chip gaps, and the corresponding decrease of the effective area is not accounted for by the \emph{dmextract} tool in the production of the raw light curve. In the third panel along with the LS periodogram, we indicate the highest significance period and its 1$\sigma$ error produced by the fit, as well as, with horizontal dashed, solid, and dotted lines, the 0.01\%, 0.1\%, and 1\% FAP levels respectively, which are calculated through the LombScargle class of astropy using the bootstrap method with 100000 iterations. In the bottom panel, we overplot on the LS periodogram the WF (cyan colour) which is centred and normalised at the frequency of the main periodicity. All the peaks of the  WF (apart from the highest)  correspond to aliases produced by the uneven sampling of the underlying "true" light curve. Therefore, the WF helps us understand which of the peaks present in the LS periodogram could correspond to real periodicities.
In Fig.\,\ref{fig.phasedlcw72} we show the phase-folded probability-weighted light curve for star A based on the highest significant period measured. For all the phase-folded light curves presented in this study, phase 0 is defined as the minimum flux in the broad band.

We performed the same analysis for all X-ray detected WR stars except for the five faintest ones (D, V, E, H, and S) since their signal-to-noise ratios are too low and no firm conclusions could be drawn. In order to be conservative, we consider significant periodicity signals those that exceed at least 0.1\% of the FAP, and whose periodograms do not exhibit many peaks with similar significance. These criteria, apart from star A, are fulfilled by another eight sources (B, L, F, N, J, X, P and Q), for which we present light curves and LS periodograms in Fig.\,\ref{fig.lcbroad_all} in the Appendix. The phase-folded probability-weighted light curves of the same eight WR stars are presented in the Appendix in Fig.\,\ref{fig.lcbroad_phase_all}. In Figs.\,\ref{fig.lcbroad_all_nondetected} and \ref{fig.lcbroad_phase_all_notdetected} in the Appendix, we retain, for future reference, the light curves, periodograms, and phase-folded probability-weighted light curves for the sources  that did not meet our criteria for significant periodicity detection (O, W, R, G, and I).  

For the nine sources with measured periods we also searched for possible secondary periodicity. To achieve this, we fitted a sinusoidal function to the light curve of the source. The period of the sinusoidal was set at the main period of the LS periodogram. We then computed the residual light curve which resulted after the subtraction of the best-fit sinusoidal from the initial light curve. As a final step, we constructed the LS periodogram of the residual light curve and searched for a secondary periodicity. We found secondary periodicities for three (A, B, and F) out of the nine sources with measured periods. The residual LS periodograms are presented in Fig.\,\ref{fig.lcsecondary} in the Appendix. For stars A and F, even though the significance of the measured period is not high (FAP<1\%) we present them since they coincide with optical periodicities.
 
In the following, we report and discuss the main results for the WR stars with measured main and secondary periodicities.
The light curve of star A shows significant long-term broad-band variability (first two panels of Fig.\,\ref{fig.lcbroadw72}), with a clear periodicity at 81.75$\pm 4.63$ days (power=0.58; third panel of Fig.\,\ref{fig.lcbroadw72}). We constructed the light curves and periodograms for the soft and hard bands and we find that they are similar and always agree within the errors. 
The residual time series after the removal of the main signal shows a shorter period of $7.70 \pm 0.04$ days (Fig.\,\ref{fig.lcsecondary}; FAP$<$1\%), which is compatible with the optical modulation of 7.63 days reported by \citet{bonanos07}.
These authors were not able to 
 detect the $\sim$81 day period since their observations roughly covered 20 days.
 
For star B, which is also a confirmed binary, we find a period of 3.514$\pm$0.009 days (power=0.47), which coincides with the period calculated from optical observations \citep[3.52 days;][]{koumpia12}. Even though we observe other highly significant peaks in the periodogram that may not be adequately described by the WF, we consider this result robust due to its agreement with the optical data.
The search of the residual time series resulted in a highly significant secondary period at $\sim$2.2 days.

For star L, which is a confirmed binary, we find a main periodicity at $46.41 \pm 1.38$ days (top right panel in Fig.\,\ref{fig.lcbroad_all}).
  If we zoom in on the highest significant peak in Fig.\,\ref{fig.lcw44} of the periodogram, we find that another significant peak (FAP $>$ 0.01\%) at approximately 53 days corresponds to the period of 53.95 days found by \citet{ritchie22} in the optical band. The peaks at approximately 46 days and 53 days are aliases of each other (Fig.\,\ref{fig.lcw44} in the Appendix).

Star F is a confirmed binary with a period of 5.05 days and a suspected triple system \citep{clark11}. 
We find a clear peak at $2.263 \pm 0.004$ days while the time series of the residuals reveals a second period at $5.15 \pm 0.02$ days (Fig.\,\ref{fig.lcsecondary}), which surpasses the FAP$=$1\%. We report it since it is compatible with the period found in the optical band by \citet{clark11}.

For star N, which is long suspected to be a binary due to its infrared excess and 
spectroscopy \citep{crowther06,ritchie22},  we recover a clear signal at FAP=0.01\% in the periodogram, with a relatively long-term periodicity of $58.61 \pm 2.79$ days. The inspection of the residual time series does not reveal any significant peak. 

Stars J and X exhibit their highest peaks at periods of $\sim$5.15 and $\sim$9.01 days, respectively. No secondary period is detected for these two stars. However, as the main periods barely pass the 0.1\% FAP threshold and a series of other peaks are located at relatively high power we characterise these periods as uncertain. 
Star P shows a peak slightly more significant than the 0.1\% FAP threshold at $\sim$5.19 days, and a clear periodicity in its phase-folded light curve (Fig.\,\ref{fig.lcbroad_phase_all}). Therefore, we mark also this period measurement as uncertain.
Finally, for star Q we measure a main period at 7.4 days. However, another significant peak that is not described by the WF is present at $\sim$130 days. The search of the residual time series for a secondary period did not result in any significant peaks. Due to the presence of these two significant peaks, we mark the main period of star Q as uncertain.

We present the highest significance period measurements of the nine WR stars in Column 10 of Table\,\ref{tab.wr_stars} and we mark as uncertain (with the letter u) the period of three stars according to the previous discussion.

\subsection{X-ray colours}\label{colors}

We computed the X-ray colours of the WR stars, 
both from the single and the combined observations, using only the EWOCS data. Our aim is to find possible variability of the spectral properties of the WR stars, even for those stars which are too faint for a detailed spectral analysis.

We define the X-ray colours as C$_1=\log_{10}$(S/M), C$_2 = \log_{10}$(M/H) and C$_3$ = $\log_{10}$(S/H) where S, M, and H are the net counts in the soft (0.5-2.0\,keV), medium (2.0-4.0\,keV), and hard (4.0-8.0\,keV) bands, respectively.  We calculate the X-ray colours and their corresponding uncertainties for the 20 detected WR stars using the \texttt{BEHR} tool \citep[Bayesian Estimation of Hardness Ratios;][]{park06}.
The tool provides reliable estimates and confidence limits even when one or both bands have very low counts, since it evaluates the posterior probability distribution of the X-ray colours taking into account the number of source and background counts.
Moreover, in order to estimate the spectral parameters of the WR stars (especially the faint ones) from their X-ray colours, we created grids by simulating absorbed thermal spectra.

\subsubsection{Combined observations}

\begin{figure*}[!h]
        \centering
        \includegraphics[width=0.95\linewidth]{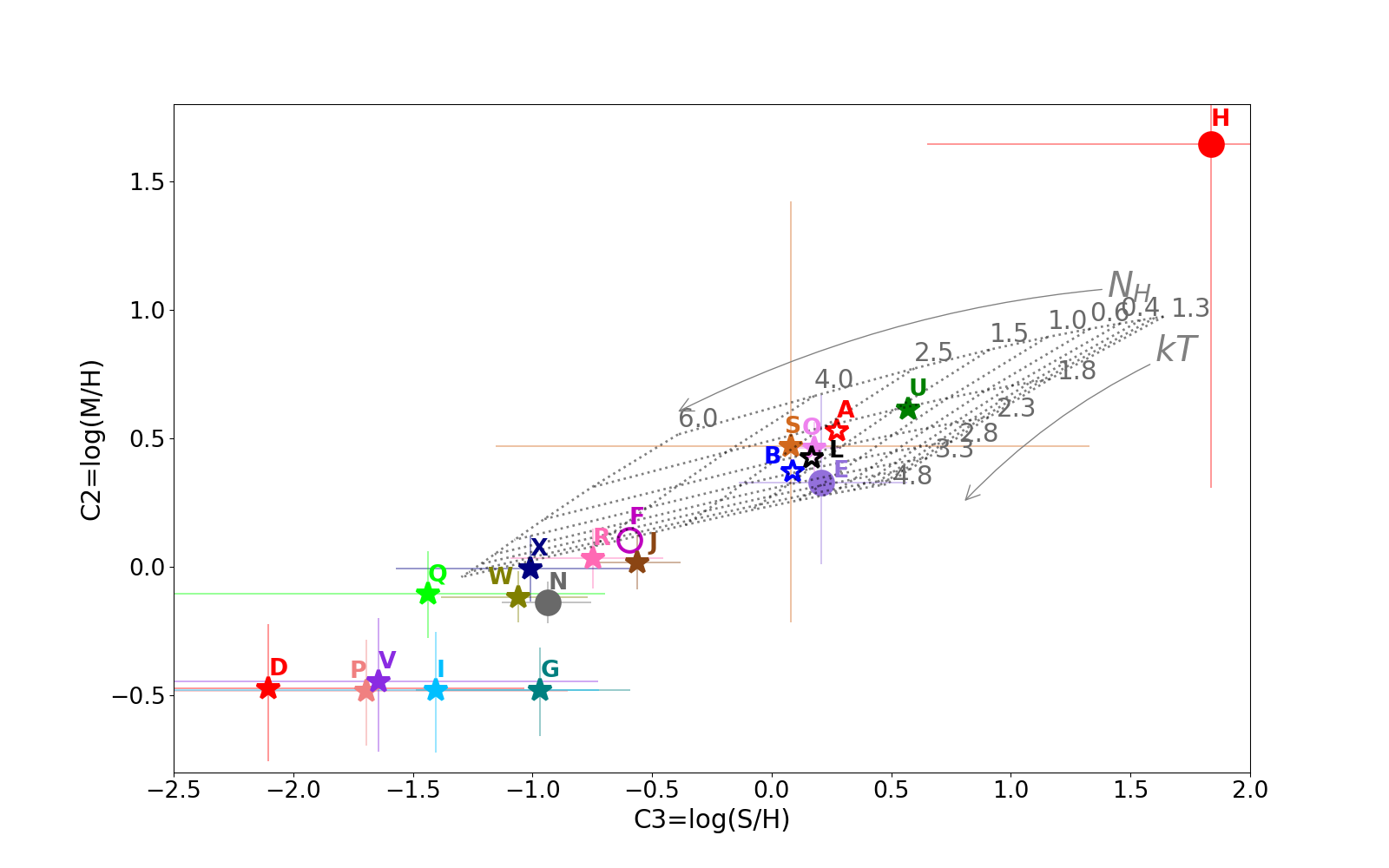}      
        \caption{X-ray colour--colour diagram of the WR stars. WN stars are marked with star symbols and WC stars with circles, while open symbols correspond to optically/infrared-confirmed binary systems. Error bars are reported at the 90\% confidence level. The grid shown with dotted grey lines represents the values  of \nh{} and kT obtained from simulated absorbed thermal spectra. The grey arrows  and the grey numbering indicate the direction in which the parameters of the two spectral components increase and their corresponding values, respectively.} 
                \label{fig.color23_all}
 \end{figure*}

We present the C$_2$ over C$_3$ diagram for the combined observations, along with the simulated absorbed thermal grid, in Fig.\,\ref{fig.color23_all}. 
Confirmed binary systems from optical studies are marked with open symbols, while WN and WC stars are marked with filled stars and circles, respectively. Errors are calculated with the BEHR tool at the 90\% confidence level.

In this colour--colour diagram the WR stars are concentrated mainly in two groups. The first group, hereafter referred to as ``medium colour group'', comprises seven WR stars (A, B, U, L, O, E, and S) located within the simulated grid of estimated spectral parameters at \nh=1-2.5$\times10^{22}$atoms \cmsq for the absorption and kT=1.8-5.0\,keV for the temperature of the emitting plasma.
These thermal plasma temperatures are typical for CWBs, and indeed three out of four confirmed optical binary systems belong to this group.

The second group, hereafter referred to as the ``hard colour group'', is located at the bottom-left of the colour--colour diagram and contains ten WN (J, R, W, G, X, P, I, Q, V, and D) and two WC (F and N) stars. This group contains the faintest non-confirmed binary WR stars of WN spectral type and the two brightest WC spectral type WR stars, all appearing hard in the colour \chandra{} image (see Fig.\,\ref{fig.wr_color}). 
 From their colours alone we cannot confirm whether their spectra are intrinsically hard, or whether they appear hard due to high absorption. The WC type stars N and F, are known to be dust-forming stars \citep{crowther06}, and therefore absorption could play a significant role. 
 
Only one WR star, the WC type star H, appears isolated at the softest side of the colour--colour diagram. Star H is among the two faintest X-ray detected WR stars and as a consequence its colours exhibit quite large errors. In fact, H is the only WR star appearing soft in the colour \chandra{} image (see Fig.\,\ref{fig.wr_color}). Taking the large errors into account, the source could fall on the medium colour group or remain at the soft side of the diagram.

\subsubsection{X-ray colour variations among the individual observations}

\begin{figure*}[!h]
        %\centering
\includegraphics[width=1.0\columnwidth]{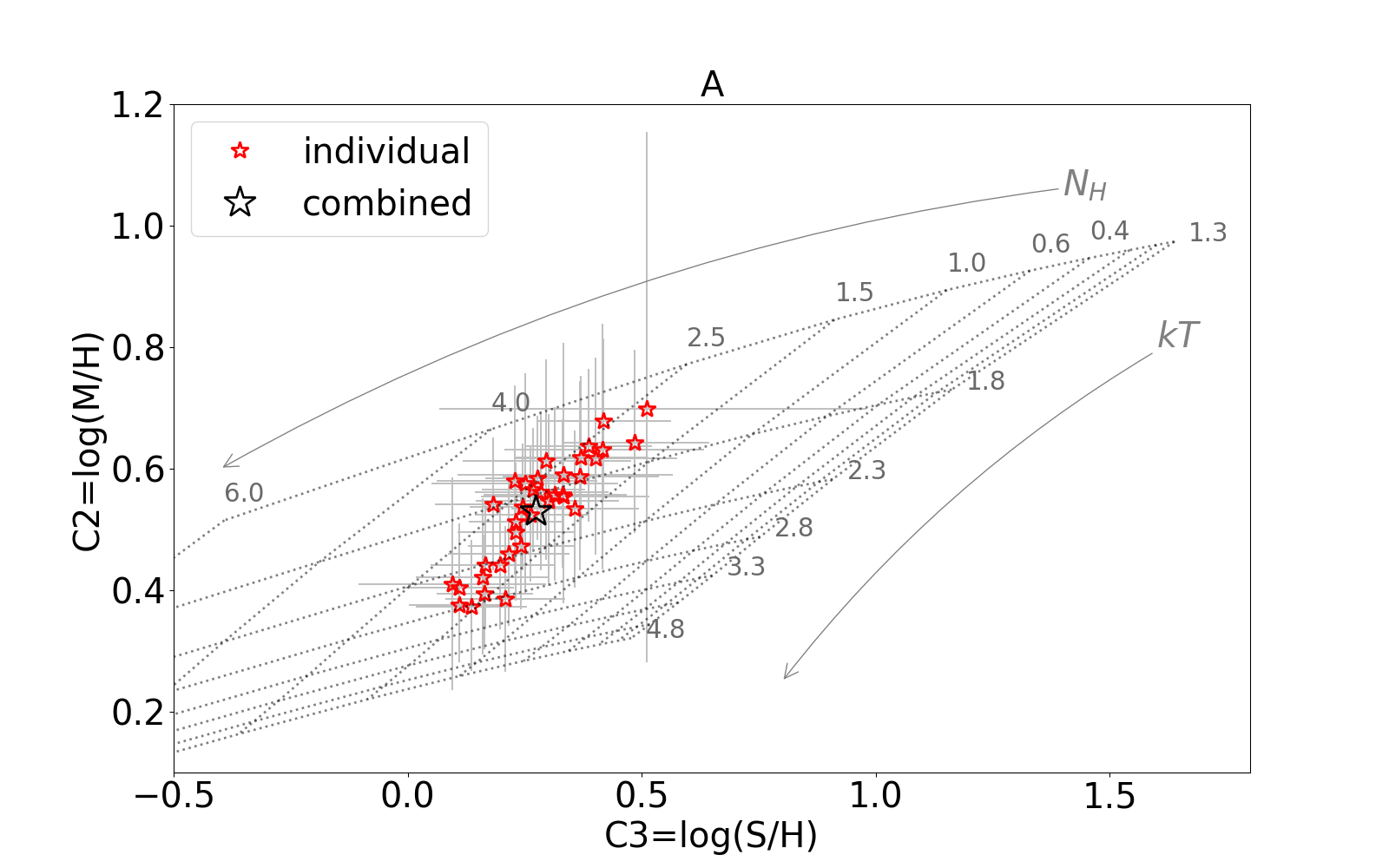}  
\includegraphics[width=1.0\columnwidth]{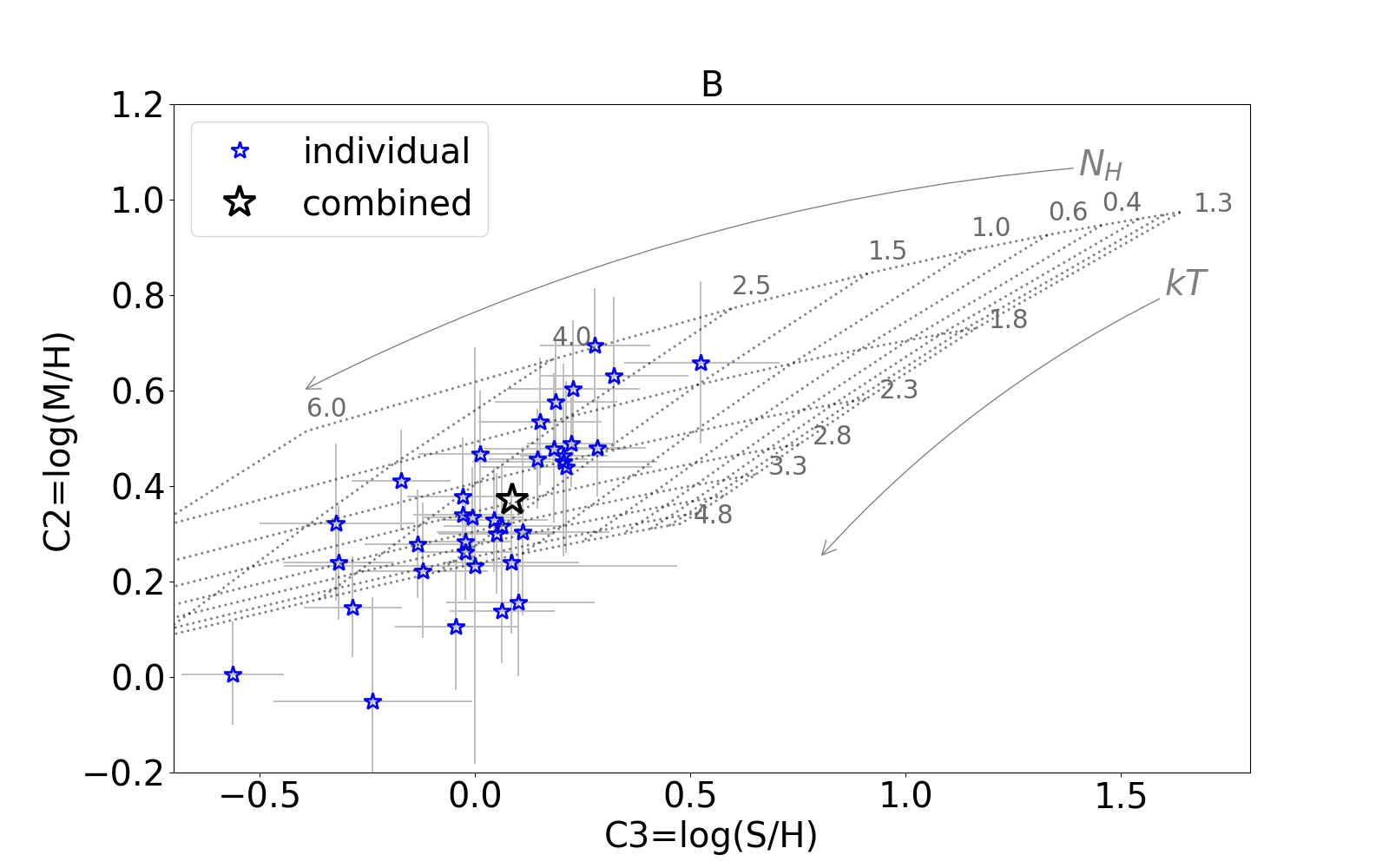}  
        \caption{Colour--colour diagrams for the two brightest WR stars in Wd1. Black large symbols correspond to the combined exposure while small symbols to the individual observations. Error bars are shown at the 90\% confidence level.}
                \label{fig.color23_individual}
 \end{figure*}

In order to explore possible spectral variation experienced by the WR stars over different timescales, we calculated their X-ray colours in each single observation.
We have used observations that have at least 10 net counts in the broad band in order to impose statistical confidence.
In Fig.\,\ref{fig.color23_individual} we show the colour--colour diagrams for the two brightest sources (A and B), while in Fig.\,\ref{fig.color23_individual_extra} we show the same diagram for other four bright WR stars. The remaining WR stars are too faint in the individual observations to extract meaningful results. We always present with small symbols the individual observations while the larger symbols refer to the combined one.

For star A, the colour--colour measurements overlaid on the simulated grid
are symmetrically distributed around the 2.1\,keV value obtained from the combined observation. They exhibit a rather constant absorption of \nh{}$\sim$2.0$\times10^{22}$atoms \cmsq, and a temperature varying between 1.5\,keV and 3.0\,keV. Individual observations of star B show variations in both \nh{} (1.0-3.5$\times10^{22}$atoms \cmsq) and kT (1.3-5.0\,keV) that cannot be attributed to colour uncertainties. Again, the symbol representing the combined observation falls amidst the individual observations. The same is true for the rest of the WR stars, although the error bars on the colours of the individual observations are quite large.

\subsection{Spectral analysis}\label{spectralanalysis}

To gain further insight into the X-ray properties of the WR population, the spectra of all available observations (EWOCS and pre-EWOCS) were fitted together.
Since the spectral properties of the WR stars can vary over time, particularly if they are in binaries, any fitting results obtained correspond to average measurements.
The details of the photon extraction and background region selection method are reported in \citet{guarcello24}.

To fit the spectra of the WR stars we used absorbed optically thin plasma models in thermal equilibrium, as have been typically used to describe the shock heated plasma observed in WR stars. We introduced a model of the form: \texttt{tbabs$\times$phabs$\times \sum$\texttt(vapec)}.
The model component \texttt{tbabs} is used to account for the ISM absorption fixed to the value of $2.05\times10^{22}$\,atoms\,cm$^{-2}$, which is the weighted average absorption towards Westerlund 1 using the \nh{} tool of HEASARC \citep{nhtool}. When we convert this value to optical extinction using the relation \nh{}=1.9$\times10^{21}A_{V}$ \citep{zombeck06}, we find an optical extinction of $A_{V}\sim$10.8, which is the exact same value as the average extinction towards Wd1 calculated by \citet{negueruela10}. This value is also in agreement with the value $A_{V}$=11.40$\pm$2.40 reported by \citet[][]{damineli16}. The second absorption parameter, \texttt{phabs}, was let free to vary in order to account for local variations of the absorption of mainly the stellar winds and/or of the ISM. The model \texttt{vapec}  corresponds to collisionally ionised diffuse gas with the possibility to vary the abundances. The plasma temperature is reported in keV, and its normalisation is given in units of $\frac{10^{-14}}{4\pi D^2}\int n_e n_H dV$, where $n_e$ and $n_H$ are the electron and hydrogen densities integrated over the volume V of the emitting region, and D is the distance to the source in cm \citep{smith01}. In all cases we used the abundance table provided by \citet{wilms00} as it is the most updated table to be used with the \texttt{tbabs} model for ISM absorption.

There has been evidence for non-equilibrium plasma in wide WR+O systems, while in shorter period WR binaries electron and ion temperatures were found to be equal \citep[see review by][]{rauw22}.
For this reason, we have also tested the \texttt{vpshock} model which allows one to account for non-thermal equilibrium \citep{borkowski01}. The model resulted in good fits and in many cases fewer spectral components were necessary in comparison to the \texttt{vapec} model. However, the ionisation parameter was in most cases unconstrained, and the best-fit plasma temperature was either very high and/or only constrained by a lower limit particularly for the low-luminosity sources. In addition, the plasma temperature was almost in all cases degenerate with the abundances parameter. Despite these problems, for two cases (stars L and O) the non-equilibrium model worked slightly better by fitting the strengths of the emission lines with one thermal component instead of two.
Regarding particular cases (stars A and F) where three thermal components were required in order to fit the equilibrium case (described later on in this section), the non-equilibrium case required two thermal plasma components with the second thermal component constrained only by a lower limit for star A, while for star F three plasma components were necessary with the ionisation parameter unconstrained and the softer thermal component frozen. Therefore, the need for up to three thermal components cannot  necessarily be attributed  to the use of a specific model but rather to the spectral complexity of these bright sources.
Overall, a non-equilibrium model could work well in some cases but the degeneracies and/or low counts of most of the spectra  do not allow to distinguish between the two scenarios (equilibrium and non-equilibrium). Therefore, for clarity and uniformity, we only present the thermal equilibrium emission model \texttt{vapec}.

Abundances in WR stars are expected to deviate strongly from solar, and therefore we also performed the fitting using typical WN and WC abundances reported in \citet{vanderhucht86}\footnote{A number of more recent studies have been dedicated to determining WR abundances \citep[i.e.][]{dessart00,herald01,crowther2006c,morris04,crowther24abund}. We tested our spectral analysis using these abundance measurements and found that our results remain unchanged.} for the local absorption component and the thermal components (model: \textit{tbabs$\times$ vphabs$\times \sum$vapec}).  In the case of a binary system, the two sets of abundances (solar and WR) are expected to represent the  two extreme situations: one in which the entire X-ray emission is dominated by the wind of the WR star and the second by a possible OB stellar companion. The reality is much more complicated, and it is likely a mixture of both, and for that reason we present both cases in Tables \ref{tab.spectralfit_solar} and \ref{tab.spectralfit_WR}.  

In XSPEC abundances are reported with respect to hydrogen and solar values, and in addition the hydrogen abundance cannot be set for the local absorption. Therefore, the hydrogen abundance cannot be set to zero. For this reason, and following similar procedure reported in \citet{naze21}, we set the helium abundance in XSPEC to an arbitrary value of 100 and then scale all the other element abundances relative to helium.  This results in XSPEC WN abundances of  5.2, 809.7, 5.8, 73.7, 85.4, 113.5, 40.6, and 46.5 for elements C, N, O, Ne, Mg, Si, S, and Fe respectively, and WC abundances of  16336, 0.0 , 3871.6, 208.3, 1058, 359.5, 120.8, and 138.7 for elements C, N, O, Ne, Mg, Si, S, and Fe respectively. For Al, Ar, and Ca, for which we generally have only weak or no spectral information, we set them at values of 40 and 100 for the WN and WC abundances, respectively. In one or two cases of the brightest stars, lines from these elements are more prominent; we address these in Sect.~\ref{discussion}. Choosing another arbitrary value for the helium abundance affected only the absorbing column and the normalisation of the thermal components without affecting the other parameters, the fluxes, or the fit statistic.

For the six brightest stars, we grouped the spectra using a minimum number of 15 total counts per bin, and we perform \chisq{} spectral fitting using the XSPEC v.12.13.0c software \citep{arnaud96}. For the faintest WR stars (net counts$<$370), we used the wstat statistics of the CIAO \textit{sherpa} tool \citep[][]{freeman01}. Using the wstat statistics, we grouped the spectra in bins of 5 total counts and calculated the null hypothesis probability in order to estimate the quality of the fit\footnote{The wstat statistic is the analogue of the Cash statistic used in XSPEC, but with the difference that the background data do not need to be modelled.}.
In Tables \ref{tab.spectralfit_solar} and \ref{tab.spectralfit_WR}, we present the best-fit spectral parameters and their corresponding 1$\sigma$ errors, along with the fluxes and luminosities of the best-fit models for solar and WR abundances respectively. We find that the best-fit values of the parameters as well as the fluxes agree between the two abundances.
Fluxes were calculated using the \textit{flux} command of XSPEC, and the \textit{sample\_flux} of Sherpa, and the flux errors are reported at the 68\% confidence level. 
Throughout this work, whenever we refer to corrected fluxes or luminosities, we are always referring to the corrected values solely for the ISM absorption. We are not presenting the corrected values for the entire column density (i.e., for the ISM and local absorption) primarily because, in WR stars the wind itself is the source of the X-ray emission (through shocks embedded in the wind). Therefore, only the energy that escapes the wind and is released into the ISM corresponds to the intrinsic luminosity  of the star \citep[for more details: Sect. 2.2 and references therein of][]{rauw22}.
Moreover, due to the moderate  spectral resolution of the ACIS detector, we cannot constrain well the abundances. Therefore, these were generally kept frozen, except for specific cases and for elements where a significant improvement in the fit was noted when allowed to vary. These cases are reported for each WR star in the corresponding Tables.
The spectra along with their best-fit models for the brightest WR star, A, are presented in Fig.\,\ref{fig.spectraw72}, while for the remaining sources they are presented in Fig.\,\ref{fig.spectra}.

\begin{table*}[htbp]
        \centering
                \caption{Solar abundances model: \textit{tbabs$\times$ phabs$\times \sum$vapec}}
        \setlength{\tabcolsep}{2pt} % Default value: 6pt  
        %\renewcommand{\arraystretch}{1.5} % Default value: 1
                %\hspace*{-0.7cm}
            \begin{tabular}{@{}lllllcc@{}}
                        \toprule                
                Name &  $\rm{N_{H}^{local}}$ & kT & norm   &$\chi^2_{\nu}$ or Prob ($\chi^2$/dof)  & F$_X$[F$\mathrm{_X^{ISMcor}}$] &  L$_X$[L$\mathrm{_X^{ISMcor}}$] \\[2pt]
  &  $10^{22}$  \cmsq  &  keV  & &&  \funit & \ergs\\
\midrule
        
      %\multicolumn{7}{c}{Solar abundances \textit{tbabs$\times$ phabs$\times \sum$vapec} } \\[3pt]   
A & 4.18$_{-0.49}^{+0.51}$ & 0.11$_{-0.01}^{+0.02}$ &(0.08$_{-0.07}^{+0.25}$)e2 &1.05(296.39/283)&(1.68$_{-0.1}^{+0.01}$[2.51])e$-$13 &(3.59$_{-0.22}^{+0.02}$[5.37])e32 \\ [3pt]
&&0.76$_{-0.09}^{+0.06}$ &(1.94$_{-0.42}^{+0.69}$)e$-$3&&&\\[3pt]
&&3.8$_{-1.0}^{+1.1}$ &(1.57$_{-0.36}^{+0.84}$)e$-$4&&&\\[3pt]
B & 3.14$_{-0.38}^{+0.49}$ &0.77$_{-0.12}^{+0.11}$ &(3.4$_{-0.92}^{+1.89}$)e$-$4  & 1.11(162.61/146)&(5.81$_{-0.44}^{+0.03}$[7.92])e$-$14 &(1.24$_{-0.09}^{+0.01}$[1.7])e32 \\ [3pt]
&&5.3$_{-1.2}^{+1.7}$ &(5.7$_{-1.0}^{+1.5}$)e$-$5&&&\\[3pt]
U\tablefootmark{a} & 1.88$_{-0.36}^{+0.51}$ &0.66$_{-0.16}^{+0.3}$ &(2.33$_{-0.82}^{+2.32}$)e$-$4  & 1.05(102.76/98)&(3.07$_{-0.42}^{+0.01}$[5.65])e$-$14 &(6.58$_{-0.9}^{+0.02}$[12.09])e31 \\ [3pt]
&&2.53$_{-0.53}^{+2.53}$ &(0.4$_{-0.24}^{+0.21}$)e$-$4 &&&\\[3pt]
L & 2.96$_{-0.95}^{+1.62}$ &0.53$_{-0.25}^{+0.47}$ &(18$_{-13}^{+235}$)e$-$5  &  0.90(59.99/67)&(2.18$_{-0.56}^{+0.01}$[3.1])e$-$14 &(4.66$_{-1.2}^{+0.02}$[6.63])e31 \\ [3pt]
&&2.49$_{-0.54}^{+1.61}$ &(0.47$_{-0.24}^{+0.16}$)e$-$4&&&\\[3pt]
F\tablefootmark{b} & 6.7$_{-2.3}^{+3.7}$ &  0.08$_{-0.03}^{+0.07}$ &(0.92$_{-0.92}^{+592.67}$)e2 & 1.19(54.73/46)&(3.23$_{-0.75}^{+0.01}$[3.77])e$-$14 &(6.91$_{-1.6}^{+0.02}$[8.08])e31 \\ [3pt]
&&0.62$_{-0.4}^{+0.82}$ &(15$_{-13}^{+376}$)e$-$5 &&&\\[3pt]
&&5.7$_{-2.2}^{+3.2}$ &(4.26$_{-0.99}^{+2.48}$)e$-$5&&&\\[3pt]
O \tablefootmark{c}& 1.58$_{-0.98}^{+1.37}$ &0.53$_{-0.15}^{+0.2}$ &(4.5$_{-3.3}^{+16.7}$)e$-$5 &0.94(18.78/20)&(1.03$_{-0.89}^{+0.25}$[1.6])e$-$14 &(2.21$_{-1.9}^{+0.53}$[3.43])e31 \\ [3pt]
&&$>9.2$ &(8.0$_{-2.1}^{+1.5}$)e$-$6 &&&\\[3pt]
*N & 20.3$_{-5.6}^{+4.7}$ &0.1$_{-0.03}^{+0.04}$ &(27$_{-27}^{+1040}$)e2  & 0.1(79.85/65)&(1.4$_{-0.71}^{+0.01}$[1.53])e$-$14 &(2.99$_{-1.52}^{+0.02}$[3.29])e31 \\ [3pt]
&&2.17$_{-0.58}^{+1.08}$ &(1.06$_{-0.56}^{+0.86}$)e$-$4&&&\\[3pt]
*W & 21.5$_{-7.8}^{+17.9}$ &0.25$\pm0.1$ &(8.7$_{-8.5}^{+219.4}$)e$-$2  &  0.25(48.78/43)&(0.97$_{-0.58}^{+0.01}$[1.06])e$-$14 &(2.08$_{-1.24}^{+0.02}$[2.28])e31 \\ [3pt]
&&4.6$_{-2.6}^{+3.2}$ &(0.27$_{-0.1}^{+0.65}$)e$-$4&&&\\[3pt]
*J & 6.8$_{-3.9}^{+10.7}$ &0.18$_{-0.16}^{+0.31}$ &(7.5$_{-7.5}^{+396.9}$)e$-$3  &  0.06(58.54/43)&(6.68$_{-3.54}^{+0.10}$[7.67])e$-$15 &(1.43$_{-0.76}^{+0.02}$[1.64])e31 \\ [3pt]
&&5.4$_{-3.4}^{+6.2}$ &(1.1$_{-0.35}^{+3.69}$)e$-$5&&&\\[3pt]
*R\tablefootmark{d} & 15.7$_{-7.5}^{+10.3}$ &0.28$_{-0.13}^{+0.23}$ &(1.2$_{-1.2}^{+21.8}$)e$-$2 &  0.8(27.89/35)&(6.14$_{-5.31}^{+0.10}$[6.9])e$-$15 &(1.31$_{-1.14}^{+0.02}$[1.48])e31 \\ [3pt]
&&3.3$_{-1.5}^{+3.9}$ &(1.22$_{-0.70}^{+2.43}$)e$-$5&&&\\[3pt]
*X & 6.0$_{-2.3}^{+3.2}$ &$>3.4$ &(8.1$_{-2.1}^{+6.0}$)e$-$6 & 0.77(23.96/30)&(5.52$_{-2.22}^{+0.16}$[6.11])e$-$15 &(11.82$_{-4.75}^{+0.34}$[13.09])e30 \\ [3pt]
*G\tablefootmark{e}& 44 & 3.96$_{-0.93}^{+1.92}$ &  (3.6$\pm1.0$)e$-$5 &   0.02(47.71/30)& (6.31$_{-1.19}^{+0.28}$[8.0])e-15 & (1.35$_{-0.26}^{+0.06}$[1.71])e31  \\[3pt]
*P & 32$_{-11}^{+24}$ &4.1$_{-2.5}^{+5.0}$ &(2.0$_{-1.0}^{+9.4}$)e$-$5   & 0.12(23.8/17)&(4.66$_{-4.66}^{+0.01}$[4.88])e$-$15 &(9.98$_{-9.98}^{+0.02}$[10.44])e30 \\ [3pt]
*I & 35$_{-11}^{+12}$ &0.1$_{-0.04}^{+0.03}$ &(3.4$_{-3.4}^{+117.6}$)e4  &  0.15(19.45/14)&(3.53$_{-3.39}^{+0.01}$[3.78])e$-$15 &(7.56$_{-7.26}^{+0.02}$[8.09])e30 \\ [3pt]
&&5.1$_{-2.7}^{+6.4}$ &(0.13$_{-0.05}^{+0.22}$)e$-$4&&&\\[3pt]
*Q & 8.4$_{-3.6}^{+5.3}$ &$>2.7$ &(6.9$_{-2.7}^{+8.0}$)e$-$6   & 0.28(17.69/15)&(3.8$_{-3.25}^{+0.02}$[4.16])e$-$15 &(8.14$_{-6.97}^{+0.04}$[8.9])e30 \\ [3pt]
*D & 62$_{-41}^{+29}$ &$>1.0$ &(2.2$_{-2.2}^{+10}$)e$-$4  & 0.73(10.47/14)&(2.75$_{-2.75}^{+0.01}$[2.88])e$-$15 &(5.89$_{-5.89}^{+0.02}$[6.16])e30 \\ [3pt]
*V & 99$_{-27}^{+43}$ &0.1 &(16$_{-16}^{+718}$)e8  & 0.54(10.82/12)&(5.01$_{-4.95}^{+0.01}$[5.23])e$-$15 &(10.73$_{-10.59}^{+0.02}$[11.2])e30 \\ [3pt]
&&1.66$_{-0.54}^{+4.78}$ &(0.35$_{-0.32}^{+2.45}$)e$-$3&&&\\[3pt]
*E & 0.06$_{-0.06}^{+1.84}$ &3.0$_{-1.4}^{+2.4}$ &(2.43$_{-0.76}^{+2.76}$)e$-$6   & 0.66(7.67/10)&(1.53$_{-0.8}^{+0.05}$[2.99])e$-$15 &(3.28$_{-1.71}^{+0.12}$[6.4])e30 \\ [3pt]
*H & 4.5$_{-3.6}^{+9.7}$ &0.28$_{-0.17}^{+0.48}$ &(2.8$_{-2.8}^{+81.1}$)e$-$4   & 0.01(19.47/7)&(2.97$_{-2.97}^{+0.01}$[7.21])e$-$16 &(6.35$_{-6.35}^{+0.02}$[15.44])e29 \\ [3pt]
*S & $2.4_{-2.4}^{+6.4}$ & $1.7_{-1.1}^{+13}$ & $(2.0_{-2.0}^{+3.2})\text{e-6}$ & $0.61(2.72/4)$ & $(5.1_{-5.1}^{+0.08}[7.1])\text{e-16}$ & $(1.1_{-1.1}^{+0.02}[1.5])\text{e}30$ \\ [3pt]
\bottomrule
    \end{tabular}%}
    \tablefoot{Sources that are fitted with the wstat statistic are marked with an asterisk; the null hypothesis probability is reported instead of the \chisqr. We show the absorbed 0.5-8.0\,keV flux, the corresponding luminosity, and in brackets the ISM-corrected values (not corrected for local absorption). Errors are not reported for parameters that were kept frozen (Sect.\,\ref{descriptionspectra}) . Abundances of specific elements that were varied: \tablefoottext{a}{S=1.39$\pm0.50$} \tablefoottext{b}{Ar=8.7$_{-7.7}^{+11.3}$, Ca=21$_{-14}^{+26}$}
    \tablefoottext{c}{S:9.77 Ar:28.08} \tablefoottext{d}{A Gaussian line was added in the model (lineE=5.21$_{-0.11}^{+0.15}$\,keV, norm=(6.7$_{-4.4}^{+5.6})\times10^{-8}$)} \tablefoottext{e}{A model of the form \textit{tbabs$\times$ pcfabs$\times \sum$vapec} was used in this particular case with the parameter for the covering fraction frozen at 0.95.} 
    }
    \label{tab.spectralfit_solar}
\end{table*}

\subsubsection{Description of spectral fitting results}\label{descriptionspectra}

\begin{figure*}[!h]
        \centering
        \includegraphics[width=1\columnwidth]{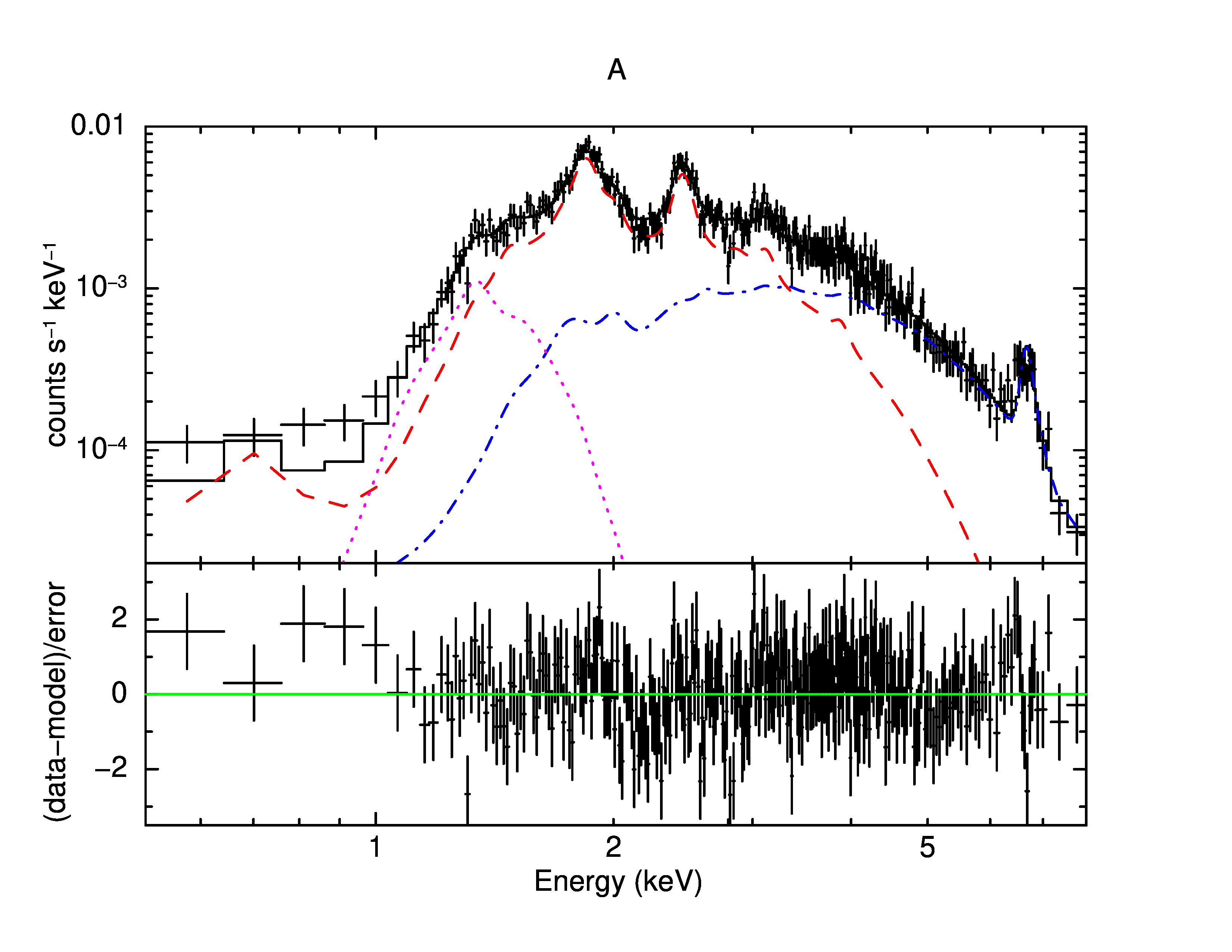}  
          \includegraphics[width=1\columnwidth]{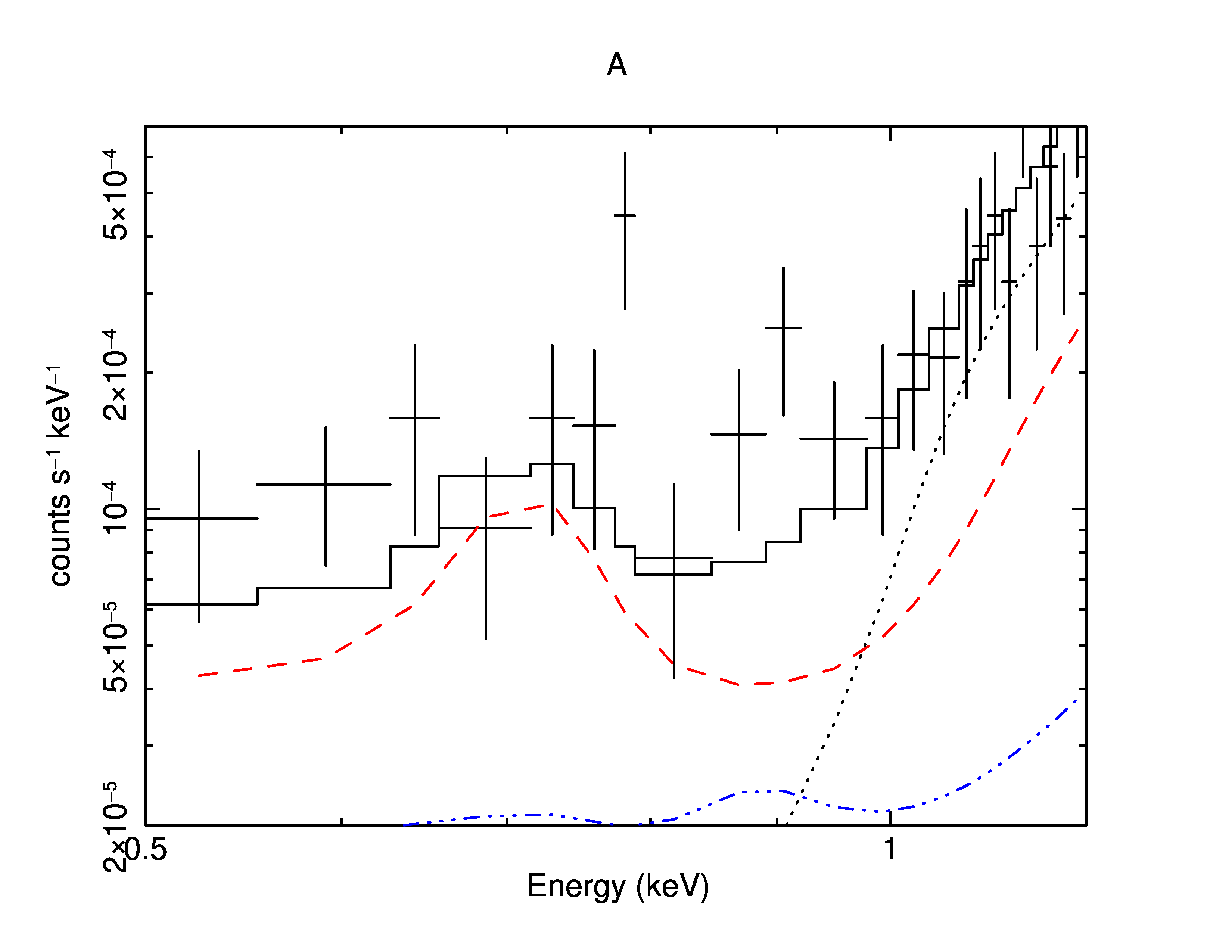}
           \caption{Spectral modelling of star A. Left: Combined spectrum of star A along with the best-fit model consisting of three vapec components (for solar composition). The fit residuals are shown in terms of sigma with error bars of size 1$\sigma$.
           Right: Same spectrum as in the left panel but at low energies ($<$1.5\,keV), using a grouping of 5 counts per bin.}
                \label{fig.spectraw72}
 \end{figure*}

%W72  
In Table\,\ref{tab.lines} we present a summary of the emission lines visible in the X-ray spectra of 17 WR stars in order of X-ray brightness. The number of lines we observe is correlated with the luminosity of the WR star and all 17 sources exhibit the iron line at 6.7\,keV. In more detail, regarding the individual spectra, for star A, three thermal plasma components were necessary in order to obtain a good fit (\chisqr=1.05).
We present the spectrum along with its best-fit model in the left panel of Fig.\,\ref{fig.spectraw72}. 
The spectrum shows characteristic lines of a $>$10MK thermal plasma, with the lines of Si, S, and Fe being quite prominent. 
%In particular the strong iron line at 6.7\,keV, which is typical of colliding wind binaries, is observed for the first time in the spectrum of WRA. 
The iron line shows significant residuals at lower energies ($\sim$6.4\,keV), which we investigate further in Sect. \ref{fluorescent}.
Moreover, we observe a soft flux excess with respect to the best-fit model below 1\,keV. We had difficulty fitting this excess even after introducing an additional very soft ($<$0.5\,keV) thermal component.
Soft excess of this kind has been seen, for example, during the low state of the wind colliding system gamma-2 Vel = WR\,11 (WC+O) where a radiative recombination continuum could be present \citep{schild04}. This system has a similar orbital period (78.5d) to that of star A, and shows a strong iron line.
However, we noticed that most of the soft excess disappears if we adopt a smaller spectral grouping (5 counts per bin) in an attempt to highlight more details and see the emission lines in the soft part of the spectrum (right panel in Fig.\,\ref{fig.spectraw72}). The observed bump at $\sim$0.7\,keV roughly corresponds to the O~VIII emission line, while a small remaining excess at $\sim$0.9\,keV could correspond to the Fe-L line. We conclude that part of the observed soft excess may be due to emission lines that cannot be fitted well when the spectra are binned to provide higher counts (i.e. 15 counts) per spectral bin. 

\begin{table*}[!h]
        \centering
                \caption{Wolf-Rayet abundances model: \textit{tbabs$\times$ vphabs$\times \sum$vapec}}
        \setlength{\tabcolsep}{2pt} % Default value: 6pt  
        %\renewcommand{\arraystretch}{1.5} % Default value: 1
                %\hspace*{-0.7cm}
            \begin{tabular}{@{}lllllcc@{}}
                        \toprule                                
        Name &  $\rm{N_{H}}$ & kT & norm   &$\chi^2_{\nu}$ or Prob ($\chi^2$/dof)  & F$_X$[F$\mathrm{_X^{ISMcor}}$] &  L$_X$[L$\mathrm{_X^{ISMcor}}$] \\
  &  $10^{22}$  \cmsq  &  keV  & &&  \funit & \ergs\\ 
        \hline
\multicolumn{7}{c}{WN abundances with He/H=100
solar} \\[2pt]       
A & (9.71$_{-0.64}^{+0.71}$)e$-$2 & 0.11$\pm0.01$ &(0.64$_{-0.42}^{+1.96}$)&1.04(294.3/282)&(1.66$_{-0.06}^{+0.01}$[2.43])e$-$13 &(3.56$_{-0.14}^{+0.02}$[5.2])e32 \\ [3pt]
&&0.83$_{-0.05}^{+0.04}$ &(5.35$_{-0.82}^{+1.17}$)e$-$5 &&&\\[3pt]
&&6.0$_{-1.4}^{+1.5}$ &(2.54$_{-0.31}^{+0.52}$)e$-$6 &&&\\[3pt]
B\tablefootmark{a} & (3.52$_{-0.55}^{+0.61}$)e$-$2 &0.84$\pm0.15$ &(2.78$_{-0.67}^{+1.13}$)e$-$6  &1.13(164.38/145)&(5.93$_{-0.24}^{+0.07}$[8.1])e$-$14 &(1.27$_{-0.05}^{+0.02}$[1.73])e32 \\ [3pt]
&& 4.55$_{-0.82}^{+1.16}$ &(1.8$_{-0.27}^{+0.31}$)e$-$6  &&&\\[3pt]
U\tablefootmark{b} & (2.29$_{-0.58}^{+0.78}$)e$-$2 &0.6$_{-0.14}^{+0.17}$ &(2.7$_{-1.0}^{+2.9}$)e$-$6  &1.13(110.71/98)&(3.06$_{-0.24}^{+0.02}$[5.66])e$-$14 &(6.55$_{-0.52}^{+0.05}$[12.12])e31 \\ [3pt]
&& 2.16$_{-0.32}^{+0.45}$ &(1.36$_{-0.34}^{+0.41}$)e$-$6  &&&\\[3pt]
L\tablefootmark{c} & (6.2$_{-3.2}^{+3.1}$)e$-$2  & 0.26$_{-0.08}^{+0.39}$ &(0.43$_{-0.42}^{+8.7}$)e$-$4  &0.89(58.86/66)&(2.16$_{-0.36}^{+0.01}$[3.04])e$-$14 &(4.63$_{-0.77}^{+0.02}$[6.51])e31 \\ [3pt]
&& 2.06$_{-0.41}^{+0.8}$ &(1.66$_{-0.65}^{+0.87}$)e$-$6  &&&\\[3pt]
O\tablefootmark{d} & (1.05$_{-0.92}^{+1.6}$)e$-$2 &0.63$_{-0.2}^{+0.28}$ &(2.7$_{-1.7}^{+10.0}$)e$-$7 & 1.1(22.05/20)&(1.03$_{-0.9}^{+0.04}$[1.71])e$-$14 &(2.2$_{-1.92}^{+0.08}$[3.66])e31\\[3pt]
&& $>9.9$ &(2.09$_{-0.37}^{+0.44}$)e$-$7  &&&\\[3pt]
*W & (3.4$_{-1.2}^{+2.5}$)e$-$1 &0.25$_{-0.09}^{+0.11}$ &(2.2$_{-2.1}^{+47.7}$)e$-$3 &  0.31(46.93/43)&(1.0$_{-0.66}^{+0.01}$[1.09])e$-$14 &(2.14$_{-1.41}^{+0.02}$[2.34])e31 \\ [3pt]
&& 4.2$_{-2.4}^{+3.6}$ &(7.9$_{-2.9}^{+17.3}$)e$-$7 &&&\\[3pt]
*J & (9.3$_{-5.2}^{+14.7}$)e$-$2  & 0.18$_{-0.16}^{+0.53}$ &(8.8$_{-8.8}^{+370.6}$)e$-$5 & 0.07(57.48/43)&(6.78$_{-3.47}^{+0.10}$[7.79])e$-$15 &(1.45$_{-0.74}^{+0.02}$[1.67])e31 \\ [3pt]
&&  5.13$_{-3.12}^{+7.01}$ &(0.03$_{-0.01}^{+0.1}$)e$-$5 &&&\\[3pt]
*R\tablefootmark{e} & (3.1$_{-1.1}^{+1.7}$)e$-$1 &0.24$_{-0.1}^{+0.12}$ &(2.6$_{-2.6}^{+43.5}$)e$-$3 & 0.61(33.01/36)&(5.91$_{-4.56}^{+0.10}$[6.66])e$-$15 &(1.27$_{-0.98}^{+0.02}$[1.43])e31 \\ [3pt]
&&  3.9$_{-1.8}^{+2.9}$ &(4.7$_{-1.7}^{+8.6}$)e$-$7 &&&\\[3pt]
*X & (8.59$_{-3.32}^{+4.48}$)e$-$2 &$>3.38$ &(2.35$_{-0.57}^{+1.49}$)e$-$7  & 0.79(23.5/30)&(5.68$_{-2.25}^{+0.16}$[6.28])e$-$15 &(12.16$_{-4.83}^{+0.35}$[13.44])e30 \\ [3pt]
*G\tablefootmark{f}& 44 & 3.96$_{-0.93}^{+1.79}$ &  (9.4$\pm2.5$)e-7 &   0.02(48.63/30)& (6.52$_{-1.13}^{+0.41}$[8.16])e-15 & (1.39$_{-0.24}^{+0.09}$[1.74])e31  \\[2pt]
*P & (4.5$_{-1.6}^{+3.2}$)e$-$1 &4.6$_{-2.8}^{+4.6}$ &(4.5$_{-1.9}^{+19.8}$)e$-$7   & 0.16(22.8/17)&(4.79$_{-3.98}^{+0.05}$[5.01])e$-$15 &(10.25$_{-8.53}^{+0.11}$[10.72])e30 \\ [3pt]
*I & (5.1$_{-1.7}^{+2.0}$)e$-$1 &0.11$_{-0.04}^{+0.05}$ &(3.0$_{-3.0}^{+55}$)e2 &  0.16(19.25/14)&(3.64$_{-3.43}^{+0.01}$[3.88])e$-$15 &(7.79$_{-7.35}^{+0.02}$[8.32])e30 \\ [3pt]
&& 4.9$_{-2.6}^{+7.0}$ &(3.5$_{-1.5}^{+3.3}$)e$-$7 &&&\\[3pt]
*Q & (1.12$_{-0.43}^{+0.75}$)e$-$1 &$>3.1$ &(1.75$_{-0.49}^{+1.79}$)e$-$7  & 0.3(17.29/15)&(3.93$_{-2.17}^{+0.15}$[4.29])e$-$15 &(8.41$_{-4.64}^{+0.33}$[9.18])e30 \\ [3pt]
*D & (9.9$_{-6.6}^{+4.4}$)e$-$1 &$>1.0$ &(5.3$_{-5.3}^{+14.5}$)e$-$6  & 0.71(10.74/14)&(2.76$_{-2.76}^{+0.01}$[2.89])e$-$15 &(5.91$_{-5.91}^{+0.02}$[6.18])e30 \\ [3pt]
*V & (1.39$_{-0.41}^{+0.67}$) &0.1 &(3.8$_{-3.8}^{+148.5}$)e6  & 0.58(10.39/12)&(5.12$_{-5.12}^{+0.01}$[5.35])e$-$15 &(10.96$_{-10.96}^{+0.02}$[11.45])e30 \\ [3pt]
&&  1.78$_{-0.63}^{+5.35}$ &(5.8$_{-5.1}^{+39.3}$)e$-$6 &&&\\[3pt]
*S & (6.1$_{-6.1}^{+10.1}$)e$-$2 &1.35$_{-0.74}^{+3.61}$ &(8.1$_{-8.1}^{+93.9}$)e$-$8  & 0.83(1.47/4)&(4.84$_{-4.84}^{+0.05}$[6.74])e$-$16 &(1.04$_{-1.04}^{+0.01}$[1.44])e30 \\ [3pt]
 \multicolumn{7}{c}{WC abundances with He/H=100
solar} \\[2pt]
F\tablefootmark{g} & (3.7$_{-1.1}^{+1.2}$)e$-$3  &  0.06$_{-0.02}^{+0.03}$ &2.5$_{-2.5}^{+4563.2}$& 1.03(49.66/48)&(3.28$_{-0.67}^{+0.01}$[3.82])e$-$14 &(7.03$_{-1.43}^{+0.02}$[8.19])e31 \\ [3pt]
&& 0.7$_{-0.25}^{+0.34}$ &(2.9$_{-2.1}^{+13.1}$)e$-$7  &&&\\[3pt]
&& 5.6$_{-1.4}^{+2.6}$ &(1.17$_{-0.17}^{+0.23}$)e$-$7&&&\\[3pt]
*N & (8.3$\pm1.9$)e$-$3 &0.08$_{-0.02}^{+0.08}$ &(4.9$_{-4.8}^{+433.4}$) &  0.24(72.87/65)&(1.42$_{-0.24}^{+0.03}$[1.55])e$-$14 &(3.03$_{-0.5}^{+0.06}$[3.32])e31 \\ [3pt]
&& 1.61$_{-0.23}^{+0.21}$ &(2.47$_{-0.93}^{+1.38}$)e$-$7 &&&\\[3pt]
 *E & (0.29$_{-0.29}^{+70868.82}$)e$-$8 &2.9$_{-1.4}^{+3.2}$ &(6.4$_{-1.4}^{+4.5}$)e$-$9 & 0.62(8.14/10)&(1.49$_{-0.56}^{+0.14}$[3.14])e$-$15 &(3.2$_{-1.19}^{+0.31}$[6.73])e30 \\ [3pt]
*H & (1.77$_{-0.81}^{+1.07}$)e$-$3 &0.28 &(12.8$_{-8.5}^{+22.9}$)e$-$8 &  0.08(12.88/7)&(3.13$_{-1.22}^{+0.59}$[7.2])e$-$16 &(6.7$_{-2.6}^{+1.2}$[15.4])e29 \\ [3pt]
\bottomrule
    \end{tabular}%}
    \tablefoot{ Same as in Table \ref{tab.spectralfit_solar}. Abundances of specific elements that were varied relative to that of helium (for reference, the generic values are reported in Sect. \ref{spectralanalysis}) \tablefoottext{a}{ S:126$_{-36}^{+46}$} \tablefoottext{b}{S=96.1$\pm 31.0$}\tablefoottext{c}{S=85.09$\pm 38.0$} \tablefoottext{d}{S=829, Ar=1000} 
    \tablefoottext{e}{A Gaussian line was added in the model at 5.21\,keV (see Notes in Table\,\ref{tab.spectralfit_solar}).} \tablefoottext{f}{A model of the form \textit{tbabs$\times$ pcfabs$\times \sum$vapec} was used in this particular case with the parameter for the covering fraction frozen at 0.95.} \tablefoottext{g}{Si=2930, Ar=3100, Ca=8173, Fe=460}}
    \label{tab.spectralfit_WR}
\end{table*}

\begin{table*}[htbp]
        \centering
                \caption{Lines visible in the X-ray spectra of the WR stars listed in decreasing order of X-ray brightness}
        \setlength{\tabcolsep}{2pt} % Default value: 6pt
        %\renewcommand{\arraystretch}{1.5} % Default value: 1
                %\hspace*{-0.7cm}
  \begin{tabular}{@{}lccccrcccc@{}}
                        \toprule
Name & Mg~XI-XII & Si~XIII  &S~XV  & Ar~XVII& S~XVI & Ca~XIX&  Ca~XIX-XX &  Ca~XX   & Fe~XXV\\
& $\sim$1.3-1.5\,keV & $\sim$1.8\,keV  & $\sim$2.5\,keV& $\sim$3.1\,keV & $\sim$3.5\,keV &$\sim$3.9\,keV &  $\sim$4.9\,keV & $\sim$5.4\,keV   & $\sim$6.7\,keV \\
                        \midrule
A       & \cmark & \cmark& \cmark & \cmark&&\cmark&&&\cmark \\[2pt]
B  &&\cmark&\cmark&\cmark&&\cmark&&&\cmark \\[2pt]
U  &\cmark&\cmark&\cmark&\cmark&&\cmark&&&\cmark \\[2pt]
L       &&\cmark&\cmark&\cmark&\cmark&&&\cmark& \cmark\\[2pt]
F &&\cmark&\cmark&\cmark&&\cmark&\cmark&&\cmark \\[2pt]
O       &&\cmark&\cmark&\cmark&&\cmark&&&\cmark \\[2pt]
N       &&\cmark&&&&&&& \cmark\\[2pt]
W       &&\cmark&\cmark&&&&&& \cmark\\[2pt]
J       &&\cmark&&&&&&& \cmark\\[2pt]
R       &&\cmark&\cmark&&& \cmark&&\cmark&\cmark \\[2pt]
X       &&&\cmark&\cmark&&\cmark&&& \cmark \\[2pt]
G  &&&&&&&&& \cmark \\[2pt]
P  &&&&&&\cmark&&&\cmark \\[2pt]
I  &&&&\cmark&&&&& \cmark\\[2pt]
Q  &&&\cmark&&&&&&\cmark \\[2pt]
D       &&&&&&&&& \cmark \\[2pt]
V  &&&&&&&&& \cmark\\[2pt]
        \bottomrule
                        \end{tabular}   
   %\tablefoot{}
                \label{tab.lines}
                                \end{table*}

WR star B exhibits a very similar spectrum to that of star A (panel \subref{subfig:panel3} in Fig.\,\ref{fig.spectra}). 
 Its spectrum was fitted well with two thermal components. However, some residuals remain close to the iron line, which may indicate the existence of a lower ionisation Fe line, which is discussed in more detail in Sect. \ref{fluorescent}. 

Similar to star A is the combined spectrum of star U (panel \subref{subfig:panel4} in Fig.\,\ref{fig.spectra}). The spectrum is fitted well by two thermal plasma components. The WR star U has not been confirmed as a binary by optical/infrared studies, but it exhibits a hard spectrum (kT$\sim$2\,keV), and a hint of the Fe line above 6\,keV. When we group the spectrum using 5 total counts per bin (panel \subref{subfig:panel5} in Fig.\,\ref{fig.spectra}) we clearly see the line at $\sim 6.7$\,keV. 

Star L is a known long period binary, and apart from the emission lines seen in other WN bright stars, shows a S~XVI at $\sim$3.5\,keV and a hint of the Ca~XX line at $\sim$5.4\,keV (see panel \subref{subfig:panel6} in Fig.\,\ref{fig.spectra}). The best-fit model consists of two thermal components. However, residuals remain around the Ar~XVII (3.1\,keV), S~XVI (3.5\,keV) and Ca~XX (5.4\,keV) lines.

Star F is the brightest WC star and it is a confirmed binary. Its X-ray spectrum is very different from those of the bright WN stars. It is harder with more prominent lines of Ar and Ca, compared to those of Si and S (see panel \subref{subfig:panel8} in Fig.\,\ref{fig.spectra}).
This could be the result of high local absorption, or an actual hardening of the spectrum due to intrinsic properties of the system, since the S line is much weaker than the Si line although less affected by absorption.
The best-fit model is obtained using a three-temperature plasma model, where the elements of Ar and Ca for solar composition are free to vary, while for WR composition in addition to Ar and Ca also the abundances of Fe and Si are free to vary. However, residuals remain around the emission line of Ar ($\sim$3.1\,keV; panel \subref{subfig:panel8} in Fig.\,\ref{fig.spectra}). 

The spectrum of star O (panel \subref{subfig:panel9} in Fig.\,\ref{fig.spectra} is fitted well by two thermal plasma components, with the strong lines requiring the abundances of the elements S, and Ar to be variable. We can only obtain a lower limit on the temperature of the hotter component.
The star does not show an Fe line; however, when we use a spectral grouping of 5 counts per bin, we see a hint of a line at 6.7\,keV (panel \subref{subfig:panel11} in Fig.\,\ref{fig.spectra}).

WR star N, the second brightest source in the hard colour group, is of WC spectral type. It is highly suspected to be a binary  \citep{crowther06,ritchie22} and  its X-ray spectrum shows a strong emission line of Fe (panel \subref{subfig:panel13} in Fig.\,\ref{fig.spectra}).
Stars W and J show similar spectra to that of N, (panel \subref{subfig:panel14} and \subref{subfig:panel15} in Fig.\,\ref{fig.spectra}).
All three sources require two thermal components in order to fit well the soft part of the spectrum, but stars N and W also require a high local absorption.

In the spectrum of star R we see for the second time a line at $\sim$5.3\,keV (as in L). The best-fit model consists of two thermal components plus a Gaussian line at 5.3\,keV to fit the Ca~XX line which is very strong, while the S~XVI line, which is weaker, is not fitted well by the model (panel \subref{subfig:panel18} in Fig.\,\ref{fig.spectra}).
The spectrum of star X is hard, and is fitted by a single thermal model (panel \subref{subfig:panel20} in Fig.\,\ref{fig.spectra}).

WR star G shows a very absorbed spectrum with a strong iron line (panel \subref{subfig:panel22} in Fig.\,\ref{fig.spectra}). All attempts to fit this spectrum with our default model have failed. Only a single thermal component, seen through a partial fraction absorber \texttt{pcfabs} can adequately fit its spectrum. However, the partial absorber components were kept fixed at the best-fit values of an initial attempt to fit the spectrum. Due to poor statistics, letting these parameters free in order to calculate the errors resulted in none of the spectral parameters to be constrained. Since even in this case, the null hypothesis probability is very low,  these spectral fitting results, except for the best-fit absorbed flux, should be treated with caution. 

For star P we obtain the best fit when using a single thermal model with a high temperature (kT$\sim$4-5\,keV) and high local absorption (panel \subref{subfig:panel24} in Fig.\,\ref{fig.spectra}). 
The spectrum of star I is well fitted adopting two thermal plasma model (panel \subref{subfig:panel26} in Fig.\,\ref{fig.spectra}), with a low and high temperature component (kT$\sim$0.1 and $\sim$5.0\,keV) and high local absorption. 
For star Q, a single thermal model is required to provide good fitting results. However, the temperature is not constrained, giving only a lower limit of $\sim$3\,keV (panel \subref{subfig:panel30} in Fig.\,\ref{fig.spectra}).

The spectrum of the WR star D is fitted by a single thermal model that yielded only a lower temperature limit of $\sim$1\,keV (panel \subref{subfig:panel32} in Fig.\,\ref{fig.spectra}).
The spectrum of star V appears highly absorbed, flat, and with only the Fe line visible (panel \subref{subfig:panel34}). One thermal component leaves significant residuals at soft energies, and therefore, we introduced a second thermal soft component. The soft component after a initial fit converged to the value of 0.1\,keV. However, when we calculated the uncertainties, and due to low statistics, the overall fit was unstable, and for that reason the temperature of the soft component was frozen at the value of 0.1\,keV.

We fitted the spectra of the faintest sources E, H, and, S, with a single thermal component. Except for the spectrum of star H, which resulted in a very low null hypothesis probability, for the other stars we obtain reasonable statistics. The best-fit model of star E (panel \subref{subfig:panel36} in Fig.\,\ref{fig.spectra}),  indicates low local absorption and a high temperature (kT$\sim$3.0\,keV).
The spectrum of star H is soft and moderately absorbed (kT$\sim$0.28\,keV), with no emission above $\sim$3\,keV (panel \subref{subfig:panel37} in Fig.\,\ref{fig.spectra}). 
The spectrum of star S is flatter (i.e harder) than the other two faint sources. It requires a moderate local absorption, and a relatively high temperature (kT$\sim$1.3-1.7\,keV; panel \subref{subfig:panel38} in Fig.\,\ref{fig.spectra}).

\subsubsection{Fluorescent Fe~K$\alpha$ line }\label{fluorescent}

\begin{figure*}[htbp]
        \centering
        \includegraphics[width=0.66\columnwidth]{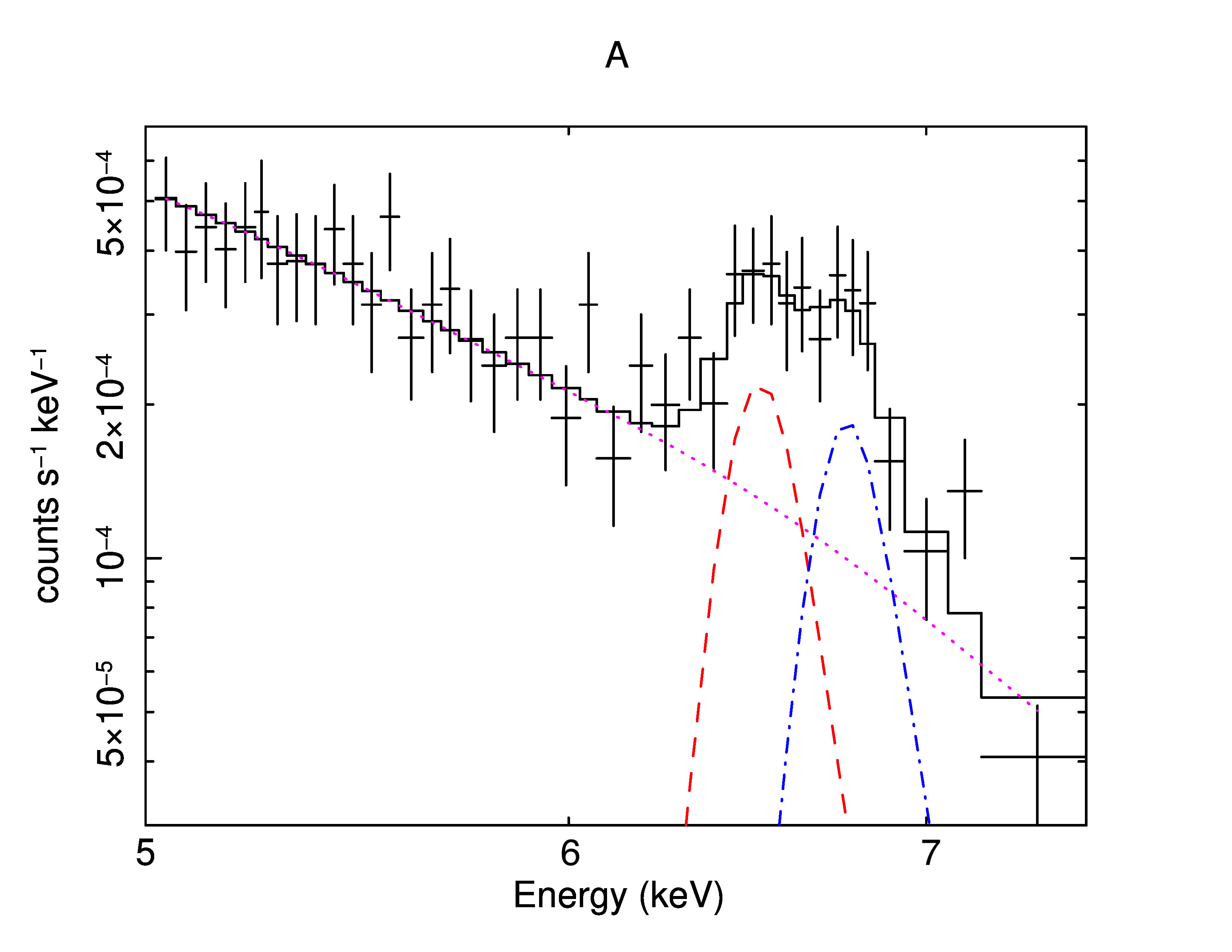}  
\includegraphics[width=0.66\columnwidth]{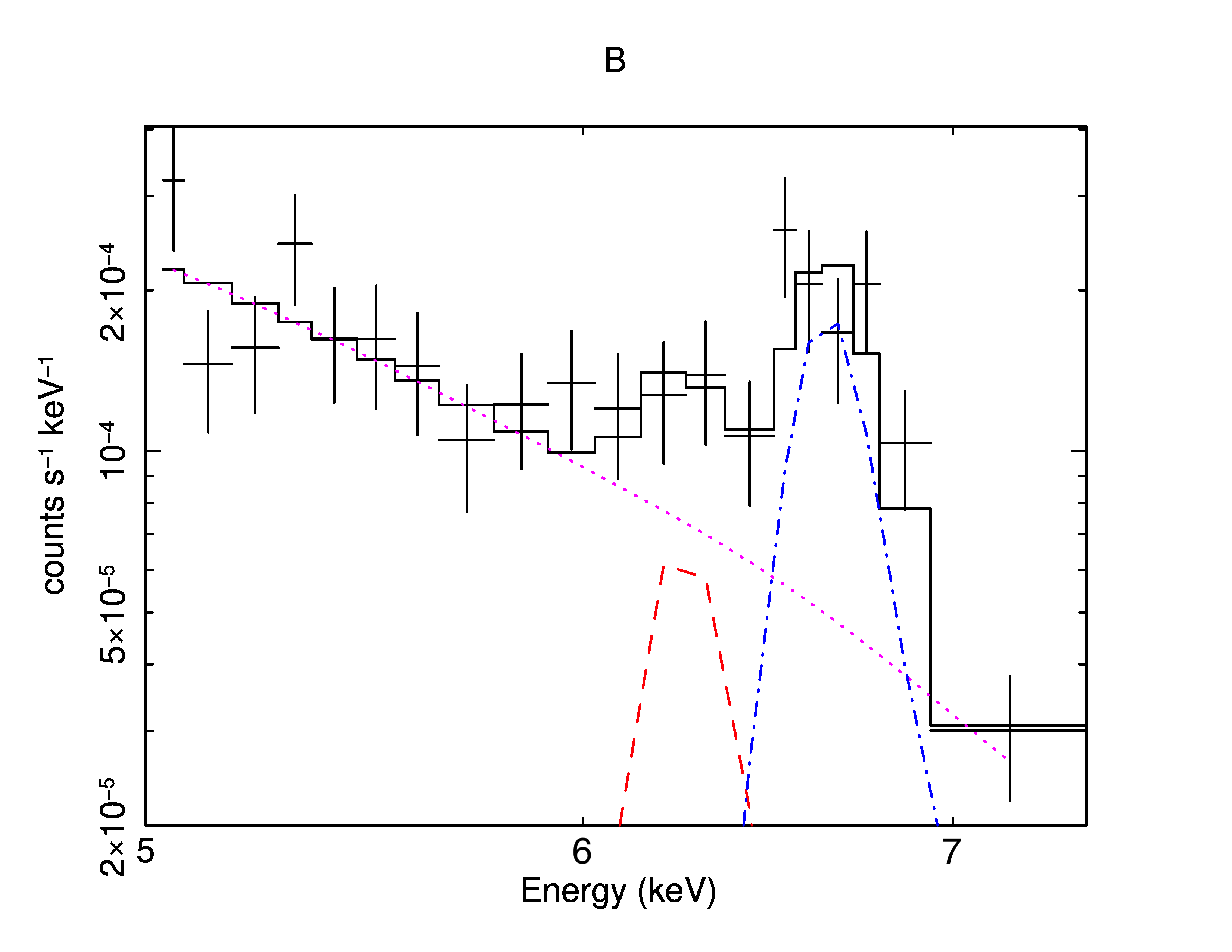}   \includegraphics[width=0.66\columnwidth]{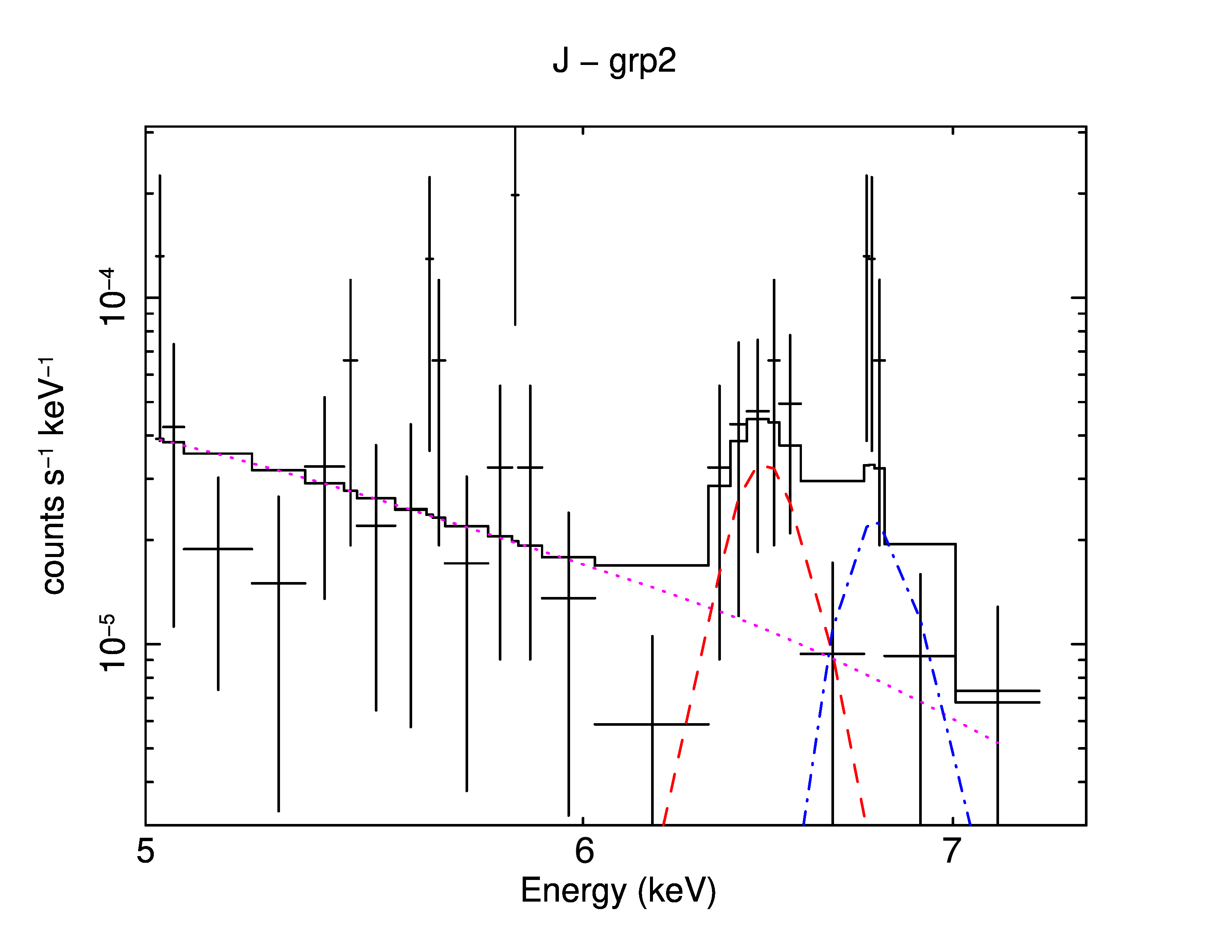} 
           \caption{Candidate fluorescent Fe line in the spectra of stars A, B, and J. We show the spectra between 5 and 7.5\,keV, fitting the continuum with a bremsstrahlung model (magenta dotted line) and two Gaussian line models to represent the fluorescent (red dashed line) and hot gas emission (blue dashed dotted line).
           }
                \label{fig.spectra_fe6.4}
 \end{figure*}

In the spectrum of stars A, B, and J, together with the Fe line at $\sim$6.7\,keV, we observe a line at $\sim$6.4\,keV. The wide separation is not compatible with a redshifted $\sim$6.7\,keV line, but is consistent with a fluorescent Fe line.
In Fig.\,\ref{fig.spectra_fe6.4} we present the spectra of these stars between 5 and 7.5\,keV, using spectral groupings where the lower energy Fe line is clearly visible. The continuum is fitted with a bremsstrahlung model (model in XSPEC: \texttt{bremss}; magenta dotted line), and the Fe lines with two Gaussian line models (model in XSPEC: \texttt{gauss}; red dashed and blue dash dotted lines).
The spectral fitting results for the two lines are presented in Table\,\ref{tab.fe_6.4}.

\begin{table}[!h]
        \centering
                \caption{Double Fe line fitting}
  \begin{tabular}{@{}lccc@{}}
                        \toprule
Star              & A & B & J \\
                        \midrule
 LineE$_1$ (keV)& 6.50$\pm0.06$&  6.25$_{-0.23}^{+0.15}$& 6.46$\pm0.08$ \\ [2pt]   
 norm$_1$ ($\times$10$^{-7}$)& 4.24$\pm 1.33$& 1.26$\pm 0.7$& 0.65$\pm 0.4$ \\ [2pt]        
 EW$_1$ (eV) & 288$_{-104}^{+123}$ & 270$\pm 160$ &  386$_{-284}^{+475}$ \\ [2pt]        
 LineE$_2$ (keV)        &6.75$\pm0.06$ &  6.64 $_{-0.06}^{+0.07}$  & 6.77$\pm0.10$ \\ [2pt]        
 norm$_2$ ($\times$10$^{-7}$)& 4.1$\pm 1.5$& 4.2$\pm 1.2$ & 0.55$\pm 0.41$\\ [2pt]   
 EW$_2$ (eV) &296$_{-108}^{+140}$&  1008$_{-218}^{+369}$& 328$_{-286}^{+429}$\\ [2pt]   
 Null hypothesis Prob&  0.99& 0.69& 0.20 \\ [2pt]       
        \bottomrule
                        \end{tabular}   
                \label{tab.fe_6.4}
                                \end{table}

The fitting results confirm that the Gaussian line at lower energies is consistent with a fluorescent $\sim$6.4\,keV line in all cases, with higher signal to noise for A. The line energies indicate that the line could arise from any ionisation state of Fe below Fe~XVII \citep[for an example on the expected centroid of FeK$\alpha$ line on the ionisation level see:][]{yamaguchi14}. The presence of a fluorescent line indicates that cool and dense gas coexists with the hot gas responsible for the $\sim$6.7\,keV.
The line might arise from photoionised cool material located within the wind of the system, possibly close to the wind--wind collision zone, when it is illuminated by the energetic photons produced in the shock-heated plasma. This is the first clear evidence of fluorescent Fe K$\alpha$ emission for WR binary systems. Fluorescent lines have already been observed for the putative single WR star WR6 \citep{oskinova12}, while both thermal and fluorescent iron lines have been observed for colliding wind binaries such as $\eta$ Carina with similar equivalent width measurements, and also for other massive stars such as $\gamma$ Cas stars \citep[e.g.][]{corcoran98,corcoran01,panagiotou18,rauw22}.

\section{Discussion}\label{discussion}

\subsection{Overall X-ray properties}

We observe that overall the spectral fitting results do not show any significant differences, regardless of whether we use solar or WR abundances (Tables\,\ref{tab.spectralfit_solar} and \ref{tab.spectralfit_WR}). Moreover, all sources (except for H) exhibit hard spectra. Our fitting results, similarly to a number of other works on WR stars \citep[e.g.][]{skinner10, zhekov12, naze21},  show for the majority of the sources that two temperature models can provide acceptable fits requiring a cool and a hot plasma component (kT$<$0.8\,keV and kT$>$1.5\,keV). However, for the brightest WN and WC sources an additional thermal component at kT$<$0.2\,keV is necessary, while for the fainter sources just one thermal component, usually hard (kT$>$1.0\,keV) is adequate.

The stars in the hard colour group (F, N, J, R, X, G, P, I, Q, D, and V) show systematically higher local absorption than the stars in the medium colour group (Tables\,\ref{tab.spectralfit_solar}, and \ref{tab.spectralfit_WR}). However, we should stress that for fainter sources obtaining physically relevant results was challenging due to the quality of the spectra.
For example, degeneracies were often present among the temperature of the hotter thermal plasma component, the local absorption, and the abundance parameters leading in some cases to poorly constrained plasma temperatures (low null hypothesis probability as shown in Tables\,\ref{tab.spectralfit_solar} and \ref{tab.spectralfit_WR}). For that reason these results should be treated with caution.
 
All optically known WN stars, but only half of the WC stars, are detected in X-rays. This is expected, due to the different chemical composition and  higher opacity of WC stellar winds \citep[e.g.][]{oskinova16}. When we fit the spectra using the generic WR abundances reported in \citet{vanderhucht86}, we find that in some cases a higher S abundance of a factor of 2 to 3 is required (Table\,\ref{tab.spectralfit_WR}). This is in agreement with previous analyses, indicating that the S abundance might need to be higher \citep[e.g.][]{skinner10,zhekov12}. In addition, the spectra of stars O and F presented very strong emission lines that could only be fitted with very high abundances of specific elements for both solar and WR abundances (notes in Tables\,\ref{tab.spectralfit_solar} and \ref{tab.spectralfit_WR}).

Non-thermal emission for colliding wind binaries has been so far detected only above 20\,keV \citep[e.g. $\eta$ Car and Apep][]{leyder10,hamaguchi18,NTApep2023}. We explored for possible signs of a non-thermal component in the spectrum of star A in the \chandra{} energy range using the EWOCS data, both because it is the brightest WR star in Wd1 ($\sim$10000 net counts) and radio observations have shown signs of non-thermal emission \citep{dougherty10}. However, we found no signs of non-thermal emission. This indicates that any non-thermal X-ray emission, if present, must be significantly weaker in this energy range than the thermal emission observed, and difficult to detect with the current instrumentation \citep{debecker07}.

\begin{figure}[!h]
        \centering
    \includegraphics[width=0.95\columnwidth]{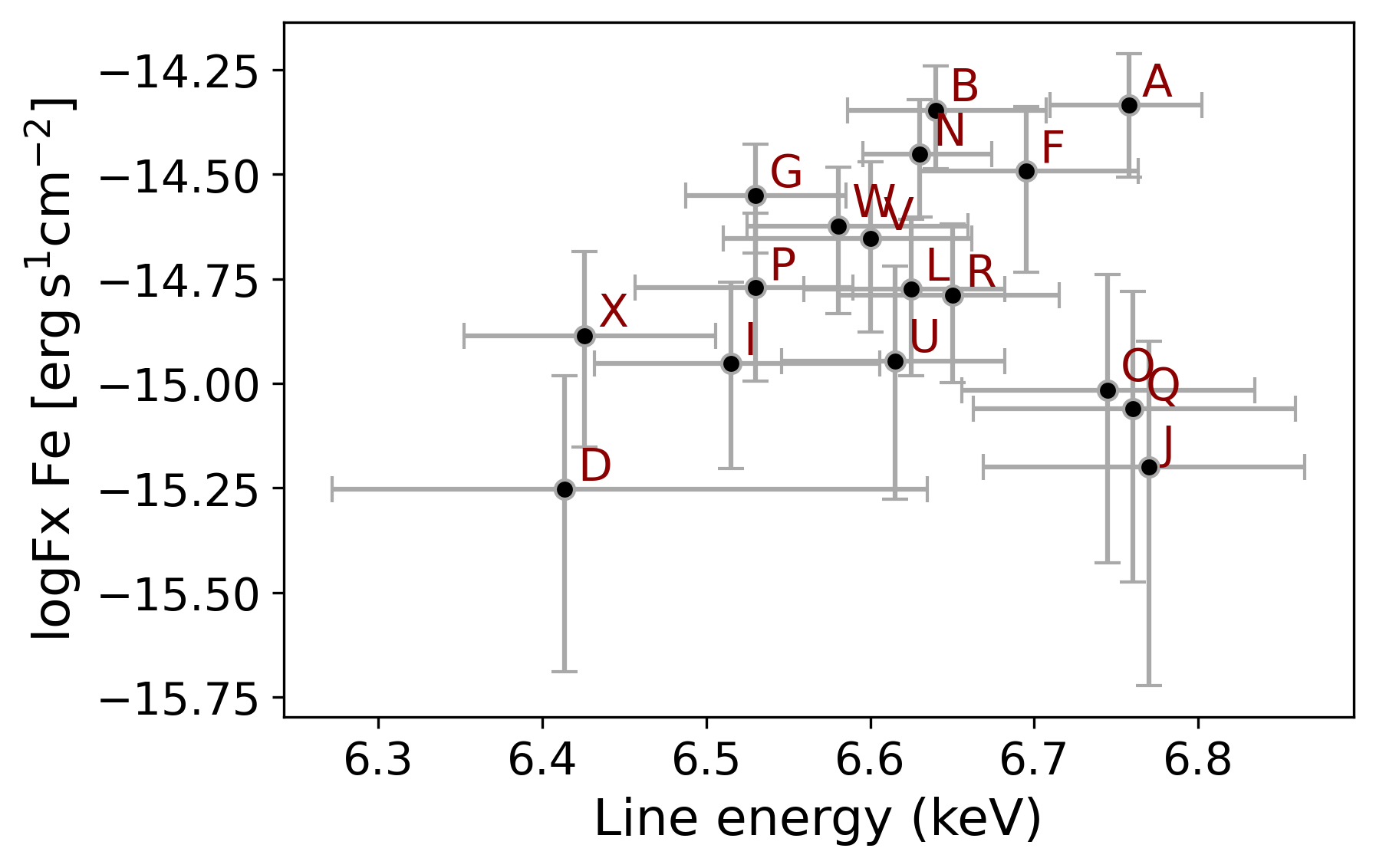}  
        \caption{Fe~XXV line properties for WR population. Fe~XXV line flux
        over the central line energy.}          
        \label{fig.fe}
 \end{figure}

The 1\,Ms deep Chandra EWOCS observations reveal for the first time that the vast majority of the WR stars in Wd1 exhibit  an iron line at $\sim$6.7\,keV. In order to present the properties of the iron line for the entire WR star population, we fit all the spectra between 5.0 and 8.0\,keV with a simple bremsstrahlung spectrum and a Gaussian line model (\texttt{bremms+gauss}). When another line is present near the 6.7 keV line, as discussed in Sect. \ref{fluorescent}, we include an additional \texttt{gauss} component. 
We calculated the flux of the 6.7\,keV line by including the \texttt{cflux} component in the model. The line width values were always set to zero. We present the results for the flux and the energy of the line in Fig.\,\ref{fig.fe}. The brightest lines are observed in  stars A, B, F, and N, which are the two brightest WN and WC stars respectively. The line energies for the four brightest stars are consistent with Fe~XXV emission ($\sim$6.67\,keV). For almost all the remaining WR stars, the line energies agree within the errors with being an Fe~XXV line, except for X, whose line centroid is found at a lower ($\sim$6.4\,keV) energy.

Overall, we measure integrated absorbed fluxes (0.5-8.0\,keV) that span from $\sim$3$\times10^{-16}$\funit{} (for star H) to 1.7$\times10^{-13}$\funit (for star A). The corresponding  observed luminosities are 6$\times10^{29}$\ergs and 3.6$\times10^{32}$\ergs.
For all WR stars, the mean values of the observed  and corrected luminosity distribution are $\log$\,$L_{\mathrm{X}}$=31.19, and $\log$\,$L_{\mathrm{X}}$=31.32 respectively (with $L_{\mathrm{X}}$ in units of\ergs)
The WC stars, as expected due to the higher opacity of their dense metal rich winds, are generally fainter than WN stars, with 50\% detection compared to 100\%, and mean absorbed luminosities of $\log$\,$L_{\mathrm{X}}$=31.25 (WN) and $\log$\,$L_{\mathrm{X}}$=30.92 (WC) and mean corrected luminosities of $\log$\,$L_{\mathrm{X}}$=31.36 (WN) and $\log$\,$L_{\mathrm{X}}$=31.16 (WC).
For different WN spectral types we see no particular differences between the observed mean luminosity and the corrected mean luminosity.
However, the ISM-corrected luminosity distribution of the WNE WR stars is shifted to higher values with median corrected luminosity $\log$\,$L_{\mathrm{X}}$(WNL)=31.11 and $\log$\,$L_{\mathrm{X}}$(WNE)=31.35. This difference is small but nevertheless agrees with previous results where WNE are found to be more luminous than WNL \citep[e.g.][]{oskinova05,oskinova16b}.

Additionally, the EWOCS observations provide a unique opportunity to investigate potential variations in colour and spectra parameters along the orbits of known binaries. 
This analysis, combining X-ray and optical/infrared spectroscopy, will be presented in future papers of the EWOCS series.
Here we briefly mention some results on star A. The colours of the individual observations show that although \nh{} seems rather constant, kT values vary (see Fig.\,\ref{fig.color23_individual}). However, no specific correlation was found between orbital phase and colours/spectral parameters. We also performed Lomb-Scargle analysis to investigate whether the colours and median-photon energy vary over time, but still we did not obtain conclusive results.
Moreover, since we detected a $\sim$6.4\,keV fluorescent line in the spectrum of star A (see Sect.\,\ref{fluorescent}), we investigate how the Fe lines change at four different orbital phases using a period of 81.75 days.
We fitted the combined spectra corresponding to the four different phases between 5 and 8.5\,keV using a \texttt{bremss+gaussian+gaussian} model in XSPEC and a spectral grouping of five counts per bin.
We have always detected two Fe lines, and found that they are centred at 6.40-6.42\,keV and at 6.61-6.74\,keV for the low flux observations, while for the highest flux observations the lines are centred at slightly higher energies 6.50-6.57\,keV and at 6.75-6.80\,keV.

\subsection{X-ray luminosity versus bolometric luminosity}\label{lxlbol}

For O type stars (both single and binaries) a linear relationship between the X-ray luminosity ($L_{\mathrm{X}}$) and the bolometric luminosity ({$L_{\mathrm{bol}}$}) has been observationally established, and it is of the form $L_{\mathrm{X}} \sim 10^{-7} L_{\mathrm{bol}}$ \citep[e.g.][]{long80,pallavicini81,chlebowski89,sana06,naze11,rauw15ob}. 
For WR stars a number of studies analysing data mostly for WN stars have not found a clear connection between X-ray and bolometric luminosities, with the relation exhibiting a large scatter of more than an order of magnitude around the $L_{\mathrm{X}}-L_{\mathrm{bol}}$  of O type stars \citep[e.g.][]{wessolowski96,oskinova05,skinner10,skinner12,naze21,crowther22}. In some cases, studies have found that  there is a connection between the kinetic properties of the wind and X-ray luminosity for single and binary stars \citep[e.g.][]{skinner10,skinner12,zhekov12}, but in some other cases no clear correlation between these properties is evident \citep[e.g.][]{wessolowski96,naze21}.
The situation is also not very clear in terms of binarity, since binary WR stars are generally found to exhibit values of  $L_{\mathrm{X}}/L_{\mathrm{bol}}\geq 10^{-7}$ \citep[e.g.][]{oskinova05,crowther22}, while a recent study of 18 confirmed binary systems revealed that not all binary systems are necessarily X-ray bright, and binaries exhibiting the same $L_{\mathrm{bol}}$ have X-ray luminosities that can differ by an order of magnitude or more \citep{naze21}.

\begin{figure}[htbp]
        \centering
     \includegraphics[width=0.95\columnwidth]{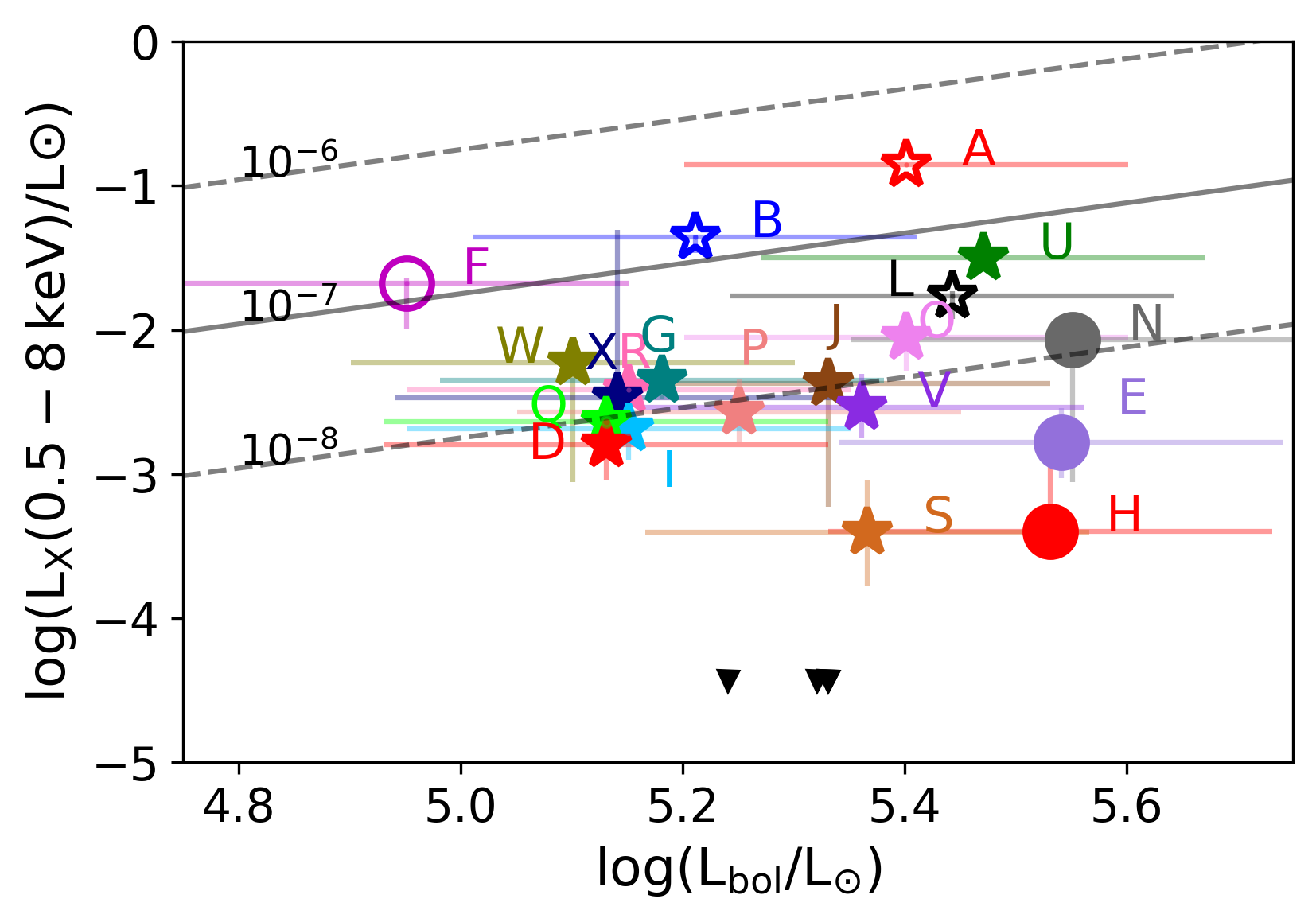}  
      \includegraphics[width=0.95\columnwidth]{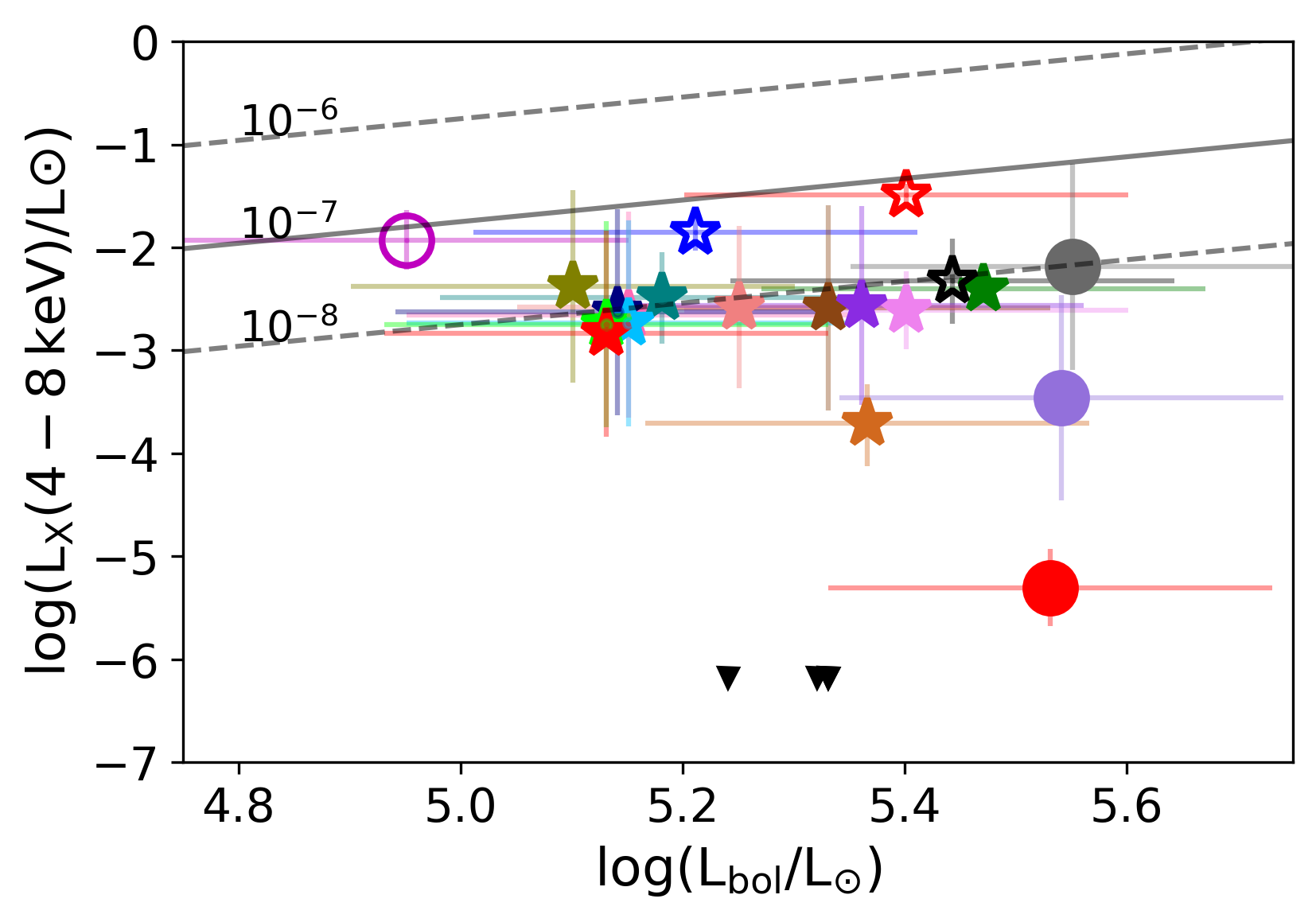}  
        \caption{X-ray versus bolometric luminosity for the WR stars in Wd1. The top panel depicts the "classical" correlation using the 0.5-8.0\,keV band in the X-rays, while the bottom panel shows the same correlation using the 4-8.0\,keV X-ray band. Upper limits on X-ray-undetected WR stars are shown with black triangles. X-ray luminosities are obtained using solar abundances.}           
        \label{fig.lxlbol}
 \end{figure}

In this section we examine the $L_{\mathrm{X}}-L_{\mathrm{bol}}$ for the WR stars in Wd1 at an initial level and in light of binarity content. More thorough analysis on the  connection of X-ray and wind properties will follow in a subsequent publication including all massive stars in Wd1.  
We use the bolometric luminosities provided in \citet{rosslowethesis} scaled to the adopted distance for Wd1. For the X-ray luminosities we use the ISM corrected $L_{\mathrm{X}}$ in the energy band 0.5-8.0\,keV provided in Table \ref{tab.spectralfit_solar} for the detected stars, while for the undetected stars we have  calculated upper limits\footnote{We calculated the upper limits using the source detection limit of the EWOCS observations which is five net counts \citep{guarcello24}. Then we converted the count-rate to flux using PIMMS (\url{https://heasarc.gsfc.nasa.gov/cgi-bin/Tools/w3pimms/w3pimms.pl}) assuming an absorbed thermal spectrum with a temperature of kT=1\,keV and \nh{} equal to the Galactic value towards Wd1. In the vicinity of the sources in question, the typical background value is $\sim$3 counts. Therefore, the upper limits are reported at the 3.0$\sigma$ level following the method described in \citet{vinay10}.}. In the top panel of Fig.\,\ref{fig.lxlbol} we show the $L_{\mathrm{X}}$ versus $L_{\mathrm{bol}}$ plot for all the WR stars in Wd1. The $L_{\mathrm{X}}$$\sim$10$^{-7} L_{\mathrm{bol}}$ is shown with a solid line while the dashed lines represent the $L_{\mathrm{X}}$$\sim$10$^{-8} L_{\mathrm{bol}}$ and $L_{\mathrm{X}}$$\sim$10$^{-6} L_{\mathrm{bol}}$ relations. The nomenclature is the same as in Fig.\,\ref{fig.color23_all}, while the upper limits are shown with black triangles.  We observe that the entire detected population with the exception of the very faint stars E, S, and H, lies within the  $L_{\mathrm{X}}\sim 10^{-8} L_{\mathrm{bol}}$ and $L_{\mathrm{X}}\sim 10^{-6} L_{\mathrm{bol}}$ lines, in agreement with what has been observed in previous works \citep[e.g.][]{wessolowski96,skinner10,crowther22}.
Three of the brightest WR stars (A, B, and F) and firmly confirmed binaries show values higher than $10^{-7}$, with star A being the only one clearly surpassing $L_{\mathrm{X}}\sim 10^{-7} L_{\mathrm{bol}}$.
We also present in the bottom panel of Fig.\,\ref{fig.lxlbol} the same relation for the 4-8\,keV X-ray luminosity range. This provides us with the $L_{\mathrm{X}}/L_{\mathrm{bol}}$ relation free of any absorption effects (foreground or local). In this energy range there is a clearer separation between the WR stars with A, B, and F all being a bit below the $10^{-7}$ line, and the rest of the WR stars (apart from the faintest  E, S, and H) nicely following the $10^{-8}$ line. 

Overall, apart from stars A, B, and F, all other WR stars show suppressed values of the $L_{\mathrm{X}}/L_{\mathrm{bol}}$ relation compared to what has been found for O type stars.
According to \citet{owocki13} lower $L_{\mathrm{X}}/L_{\mathrm{bol}}$ is expected for optically thick winds and for increasing mass-loss rate.
\citet{owocki13} also discuss how orbital separation affects their results, suggesting that close, short (day to week) period binaries should exhibit lower $L_{\mathrm{X}}/L_{\mathrm{bol}}$ values. Our results are consistent with these findings since for star A, which is the only WR star clearly surpassing $L_{\mathrm{X}}$$\sim$10$^{-7} L_{\mathrm{bol}}$, we measure a long period ($\sim$81 days). This finding aligns with
results presented for a number of Galactic long period WR binaries \citep[i.e. WR25: P$\sim$208 days; WR138: P$\sim$1521 days; WR133: P$\sim$112 days;][]{oskinova05}, as well as in the case of Melnick 34, an X-ray bright WR system (WN5h + WN5h, P$\sim$155 days) in the 30 Dor region of the Large Magellanic Cloud \citep{tehrani19}.

\subsection{Comparison to previous X-ray studies}

Using the first two \chandra{} observations of Wd1 ($\sim$60\,ks), \citet{skinner06} detected 12 of the 22 known WR stars at the time. One of them, the WC star K, remains undetected through the EWOCS data. 
However, the detection of stars K and E was considered marginal with 9$\pm$3 and 6$\pm$3 counts in the 0.5-7.0\,keV band, respectively. Moreover, star K was only detected in the first \chandra{} observation (18.8\,ks). \citet{skinner06} also do not report one of the brightest sources, star U, which was then not identified as WR star.  For the stars present also in our catalogue, the unabsorbed luminosities they list in their table 1 are higher by up to an order of magnitude compared to the ones we report. This can be explained by the fact that for 10 out of the 12 detected sources \citet{skinner06} use PIMMs conversion assuming a spectrum softer (kT=1.0\,keV) than the actual spectrum (see Tables\,\ref{tab.spectralfit_solar} to \ref{tab.spectralfit_WR}). Moreover, they report the spectral properties of star A detecting only the strong lines of S and Si. Their results agree within the errors with the spectral properties we obtain. They conclude that probably all their detections are binaries, particularly the WC stars.

\citet{clark08} used the same \chandra{} observations as \citet{skinner06}, and examined the global properties of the X-ray sources along with photometric and spectroscopic information from optical and radio datasets. They consider the marginal detection of star K by \citet{skinner06} to be in error, and they present the results of the then newly discovered WR star U. They present the spectral fit results for the five brightest sources (A, B, U, L, and F) where in none of those the iron line at 6.7\,keV was observed. 
The observed fluxes they present agree well with the ones we obtain.

Our deeper EWOCS observation produces spectra with much more detail including many more lines, the Fe line at 6.4\,keV and 6.7\,keV for the first time (see Sect. \ref{spectralanalysis}), and a better characterisation of the softer part of the spectrum ($<$1\,keV). For this reason in most cases more than one thermal plasma model is required to fit adequately the spectra. Moreover, we constrain better and at lower values the temperatures for stars L and F using thermal emission components in thermal equilibrium, while for star B we always obtain higher temperatures for the thermal plasma. 

In addition \citet{clark08} present the intensity colour diagram for all the detected sources, and find that all WR stars exhibit hard colours especially in comparison to other massive stars. Only E has a soft colour (kT$\sim$0.6\,keV) and it is the faintest source in their sample. Our results agree  for all the sources except for source E. We find instead that star E has hard colours similar to that of the brightest sources of the `medium colour group', which is also confirmed by its spectrum (kT$\sim$3.0\,keV) .

\subsection{The nature of stars N and T}

Stars N and T are the most distant WR stars from the centre of the cluster, at a projected distance of about 4.5 arcminutes. Both are WC9 dusty stars, with star N also being an SB2 candidate, and are therefore, considered to be binaries with O-type companions \citep[][]{crowther06,ritchie22}. X-ray emission is detected only from star N, and the EWOCS data confirm its binary nature (hard spectrum, strong Fe 6.7\,keV emission line, P=58~days). \citet{crowther06} discussed the location of the two stars outside of the cluster and suggested that they were ejected due to either dynamical interactions within the cluster or following a supernova kick. Moreover they commented that neither mechanism can explain the observed binarity since dynamical interactions favour single stars while the kick would require that the companion has already undergone a supernova explosion.

\begin{figure}[htbp]
        \centering
     \includegraphics[width=0.95\columnwidth]{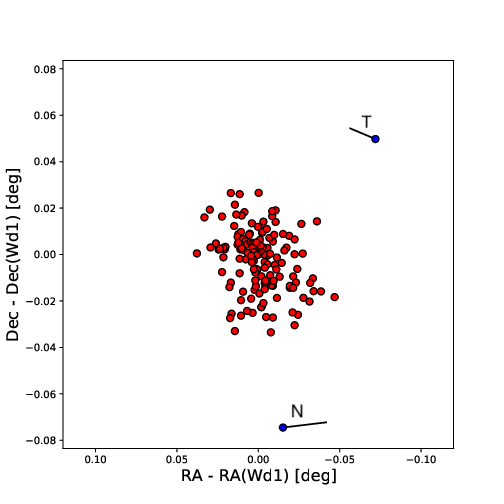} 
        \caption{Spatial distribution of Wd1 cluster members (red) and WR stars N and T (blue) with proper motion vectors shown for the latter (black lines) over a 0.2\,Myr baseline. The proper motions shown are those relative to that of the cluster, which was calculated from the membership of \citet{negueruela22}.
        }               
        \label{fig.gaia}
 \end{figure}

We plot the proper motions of the two stars relative to the cluster members using  \gaia{} EDR3 astrometry (Fig.\,\ref{fig.gaia}), where positions and proper motions have been plotted relative to the cluster members in Wd1. The red colour points show the cluster members from \citet{negueruela22}, while blue indicates the WR stars T and N along with their proper motion vectors over a 0.2\,Myr baseline. Neither proper motion vector points back towards the cluster, suggesting that neither WR star has been recently ejected. 
In the reference frame of the cluster, the proper motions of these stars are (PM$_\alpha$, PM$_\delta$) = ($0.5 \pm 0.1$, $-0.04 \pm 0.08$) and ($-0.28 \pm 0.07$, $-0.08 \pm 0.06$) mas yr$^{-1}$ for N and T respectively. Relative to the inherent velocity dispersion of the cluster (with the effect of uncertainties removed) of $\sigma_\alpha = 0.252$ and $\sigma_\delta = 0.280$ mas yr$^{-1}$, these proper motions are not very large and are both within 2$\sigma$ of the central velocity of the cluster.

For these two binary systems to lie outside of the cluster core and to have 
proper motions that do not point back towards the cluster makes it very unlikely that they were formed at the center of the cluster and then ejected.
It seems more likely that these stars formed outside the cluster core in a more dispersed population surrounding Wd1. Many star clusters are known to be surrounded by dispersed populations \citep[][]{wright20}. Detecting this dispersed population in the optical and infrared, particularly the more-numerous low-mass component, is challenging due to the typical large contamination of memberships in the outskirts of the clusters. However, in the X-rays with the EWOCS data, a dispersed population around the cluster core is observed \citep{guarcello24}.
In fact, out of the 24 WR stars only 4 (N, T, X, and W) are located farther than 2 arcminutes from the cluster core, while the number of X-ray validated sources more distant than 2 arcmin from the core is 2559 out of the 5963. If we restrict the sample to an energy between 2 and 3.5\,keV (where most of the members are) there are 1678 sources more distant than 2 arcminutes out of the 3253. 
While 83\% of the WR stars are located within 2 arcmin, this percentage decreases to approximately 49\%-57\% for the general population of the X-ray sources. This reveals that a significant population of dispersed, low to intermediate-mass young stars exists in the surroundings of Wd1.

\subsection{Binarity}\label{binarity}

A large percentage of WR stars are members of binary systems with massive stellar companions (typically OB stars).
The presence of a companion can be directly revealed through optical photometry, when eclipses or periodic signals are present in the light curve of the star, or from large periodic variations of radial velocity.  In spectroscopic data, the companion's existence can also be detected through specific absorption features \citep[e.g.][]{bonanos07,clark11,ritchie22}. 
While these methods offer direct and unambiguous evidence of a companion's existence, they also pose various challenges in detecting complete samples. These challenges encompass factors such as the orientation of the orbit, the duration of the data collection period, and, particularly for WR stars, the hindering of the companion's spectral information by their broad line spectra.

When optical spectroscopy and photometry are insufficient to reveal the existence of a companion, indirect methods for detecting the companion come into play.
For example in X-rays, since isolated massive stars have typically soft X-ray spectra coming from self-shocks developed in the stellar wind, it is generally assumed that hard colours or spectra ($\geq$2\,keV) are indication of high temperatures developing in the wind--wind collision zone of binary systems. Moreover, the presence of the Fe line at 6.7\,keV in the X-ray spectrum is considered a diagnostic for the identification of CWBs, since it is formed only at temperatures higher than about 10 million degrees. Such high temperatures cannot be developed in the winds of single WR stars unless they are highly magnetised, which is quite rare since very few cases have been discovered \citep[i.e.][]{magnetic1,magnetic2,magnetic3}.
At radio frequencies, binary systems can exhibit significant deviations from the expected thermal emission from single winds due to the presence of additional non-thermal emission, leading to flat or even negative spectral indexes \citep{williams97,dougherty00,dougherty05,pacwbcata}. 
Excess near to mid-IR emission is also considered as an indication of binarity in persistent \citep[i.e. WR104;][]{tuthill99} and episodic \citep[i.e. WR140;][]{han22} dusty WC systems. The formation of dust is favoured in WR+O binary systems since they provide the necessary ingredients (elements C and H) and the required high density in the wind--wind collision zone. In addition the wind--wind collision zone may provide protection over ionising photons that tend to destroy the dust \citep[e.g.][]{williams05,crowther06a,crowther07}.
However, the absence of dust emission in WR stars does not exclude binarity, since an episodic dust production could have prevented us from observing it.

Based on the above criteria, previous studies of the WR population in Wd1 have estimated a high binary fraction (62-70\%) that could even reach unity \citep{crowther06,clark08,dougherty10}. Along the same lines, an updated estimate on the binary WR fraction in Wd1, based on the new \chandra{} results from the EWOCS observations, is presented in this work. We are currently analysing the dust production from Wd1 JWST data which will be presented in a dedicated study as part of the EWOCS series.

Below we summarise all signs of binarity for the WR stars in Wd1 with a stress on the new indications provided by the EWOCS data.
Star A is unambiguously a binary system, through optical photometric, infrared spectroscopic, radio and previous X-ray analysis \citep{bonanos07,crowther06,dougherty10,skinner06,clark08}.
With the present work we determine a main period of 81.75 days and a secondary of 7.7 days, with a hard X-ray spectrum and the identification, for the first time, for this source of a strong Fe line at 6.7\,keV.
The identification of two periodic signals could indicate either the presence of a third body or the existence of corotating interaction regions within the stellar wind \citep[e.g.][]{chene11,massa14}. Given the specific characteristics of this system, and assuming a Keplerian orbit, the period of 7.63 days would imply that, the WR wind directly hits the surface of the companion star and therefore, no wind--wind interaction could develop, which increases the likelihood of the longer period being the orbital period. However, a colliding wind scenario could still develop in the case of the shorter period, if mutual radiative effects slow down the WR wind and prevent it from hitting the stars surface. Therefore,, more detailed modelling is necessary to understand the nature of this system.  

WR star B is a confirmed binary through photometric and previous X-ray data \citep{bonanos07,koumpia12,skinner06,clark08}. We detect the iron line at 6.7\,keV in its spectrum, and two periodicities, the main at 3.51 days (compatible with the optical), and the secondary at $\sim$2.2 days.
Star L is a confirmed binary through X-ray and radial velocity measurements \citep{clark08,ritchie22}. The EWOCS data reveal its hard X-ray spectrum with an Fe line at 6.7\,keV, with a long period of $\sim$46 days, although a significant peak at $\sim$53 days is also detected (Sect. \ref{longterm}).
Star F, the X-ray brightest WC star, is a confirmed binary through optical, infrared, radio and X-ray data \citep{clark11,crowther06,dougherty10,clark08}, while long-term IR variability suggests the presence of a third body \citep[][]{clark11}. We find a very hard spectrum with a strong Fe line at 6.7\,keV and two measured periods of $\sim$2 and $\sim$5 days, which either strengthen the third body scenario or support a corotating interaction region origin in either star of the binary.

WR star N shows signs of binarity through its hard X-ray colours, mid to NIR excess, and infrared spectroscopy \citep{skinner06,clark08,crowther06,ritchie22}.  
With the EWOCS data we confirm this finding, since the X-ray spectrum of star N is hard and shows a particularly strong Fe line at $\sim$6.7\,keV. Moreover, we find that star N exhibits a long period of $\sim$58 days. 

WR stars U, O, W, R, G, and D  were considered most probably binaries due to their hard  X-ray colours \citep{clark08}. We confirm these identifications through our colour and spectral analysis which reveal their hard spectra typical of CWBs. Moreover, all of them show the iron line at 6.7\,keV.

We find that star E has colours and spectra (see  Sects. \ref{colors} and \ref{spectralanalysis}) compatible with a hard X-ray source (kT$\sim$3.0\,keV). No iron line is visible but this could be due to the low count-rate of the source ($\sim$56 net counts) in the EWOCS observations. Its nature was debated since the source appeared soft (kT$\sim$0.6\,keV) in \citet{clark08,clark19}. However, we deem that result highly uncertain because of the very low number of net counts ($\sim$8) based only on two \chandra{} observations ($\sim$60\,ks). Due to its hard spectrum our data indicate that E is most likely a binary.

The only indication of binarity for WR star V comes from radio observations \citep{dougherty10}. The source is detected for the first time in the X-rays through the EWOCS data that reveal the very hard X-ray nature of the source (kT$>$1.5\,keV; hard colour group), with a strong iron line at 6.7 keV, making it a strong CWB candidate. 
The WR stars J, X, P, I, and Q are detected for the first time in X-rays through EWOCS. We consider all of them to be binaries due to their very hard spectra and iron lines (kT$>$3.0\,keV; hard colour group).

Star S has an intriguing nature. It has been suggested that it is the remnant of a short period binary system disrupted in a supernova explosion, and now appears to be single \citep{clark14}. With EWOCS it is detected for the first time in X-rays, and it is the faintest WR star (23 net counts) in the sample. Its colour and spectrum reveal its hard X-ray nature although the number of counts do not allow for further characterisation.

Star H is considered a possible CWB through its infrared excess \citep{crowther06}. However, in the X-rays it presents very soft (kT$\sim$0.5\,keV) colours and spectra, so its nature is quite puzzling. Within the \chandra{} extraction region of H, two bright infrared sources are observed through NIRCam JWST (Sect. \ref{detection}). Therefore,, if the X-ray detection truly corresponds to emission from the WR star and not its nearby source, H could be a single star that is an unusual dust emitter or it could be a binary system with a very particular projected geometry that at the epoch of observation allows us to see only the soft wind of one of the two stars. 

Three out of four X-ray non-detected WR stars, C, M, and T, are all dusty WC stars \citep{crowther06}. In addition, large radial velocities have been detected for star M \citep{ritchie22} further confirming its binary nature.  Therefore, we consider all of them to be binaries. They are probably not detected in X-rays because their very opaque and dense winds do not allow for any of the X-ray emission to escape (at least down to our detection limit).
Star K is also not detected in X-rays, and among the undetected WR stars is the only non-dusty WC star. 
Based on the available information, we cannot reach a conclusion regarding its single or binary nature.

To conclude, in our sample, given the hard X-ray spectra ($\geq$1.5\,keV) observed for 19 out of the 20 detected WR stars, in addition to the iron line at 6.7\,keV for 17 of them, we infer a binary fraction of 80\% only on the basis of the X-ray data. Combining our results with infrared observations \citep{crowther06}, we conclude that all these WR stars, except probably H, and K, are binaries, resulting in a binary fraction in the WR star population of $\sim$92\% (100\% for WN and 75\% for WC). This fraction could indeed approach unity as also predicted by previous works \citep{crowther06,clark08,clark19}.

A WR binary fraction potentially reaching unity in Wd1, is higher than current estimates for our Galaxy \citep[about 30-40\%;][]{vanderhucht01, dsilva20, dsilva22, dsilva23}. However, when corrected for biases, the estimated binary fraction is significantly higher, nearly reaching unity for WC stars (0.96+0.04\%) and being somewhat lower for WN stars (0.52+0.14\%); the higher frequency of short-period WN stars suggests different formation mechanisms \citep{dsilva22, dsilva23}.
The discrepancy between the overall Galactic binary fraction and that measured for Wd1 may be attributed to the unique nature of Wd1, recognised as the most massive young star cluster in the Milky Way \citep{clark05, brandner08, gennaro11, lim13, andersen17}. On the other hand, the high binary WR fraction in Wd1 would require that either all O-type stars, considered to be the progenitors of WR stars, are in binary systems or that this fraction is reached only for the O stars that happen to evolve into WR stars. 

In any case, a very high fraction of binaries aligns with the widely accepted notion that a significant proportion of massive stars exist in binary or multiple systems, with evidence suggesting that the binary fraction increases with the primary star's mass. O-type stars, for instance, exhibit a binary fraction exceeding 70\% \citep[see the review by][]{marchantrev}.
Confirming a high binary fraction for massive stars carries significant implications for interpreting their stellar evolution and the nature of young star clusters. For instance, 
different dynamical evolution scenarios may have different impacts on stellar multiplicity, and these should be considered in any modelling study. Furthermore, since many physical properties of young massive star clusters are currently calculated mainly assuming single-star evolution, factors such as age, mass, and consequently, the shape of the high-end of the initial mass function could be miscalculated.
This highlights the significance of incorporating binarity as a parameter not only in studies calculating these quantities but also in extragalactic studies utilising population synthesis models.

\section{Summary}\label{conclusions}

In this paper, we present the X-ray analysis of the WR population in the Galactic young and very massive star cluster Westerlund\,1 using 36 new \chandra{} observations from the EWOCS project (Extended Westerlund 1 and 2 Open Clusters Survey),  in addition to 8 archival shallow observations, reaching a total depth of $\sim$1.1\,Ms. The WR population in Westerlund 1 is the richest in our Galaxy, comprising 24 WR stars, 8 carbon type, and 16 nitrogen type. Our main findings, made possible by the exceptional depth and long time baseline of the EWOCS dataset, are summarised below.

\begin{itemize}
    \item We detect 20 out of the 24 WR stars in X-rays, 8 of which are detected for the first time in  X-rays, reaching down to an observed luminosity of $\sim$7$\times10^{29}$\ergs and a maximum observed luminosity of 3.6$\times10^{32}$\ergs for an adopted distance to Wd1 of 4.23\,kpc.
    
    \item The 4 undetected WR stars are all of carbon spectral type and are probably not detected due to their very dense, opaque stellar winds.
    
     \item We find evidence for short-term X-ray variability in 4 WR stars, and find clues in one case (star L) that suggest that the variability is connected to variations in local absorption.
     
     \item Most WR stars show clear signs of long-term variability, with defined periodicities measured for nine stars. For three systems, we measure two periods, while for five we measure a periodical signal for the fist time through the X-ray data. We retrieve periods consistent with the optical period for three stars.
     
    \item The majority of the WR stars have hard X-ray colours. Only one star (the carbon type star H) appears to be soft, although contamination from a nearby infrared source is possible. 
    
    \item Our spectral analysis reveals details seen for the first time in the spectra of WR stars in Westerlund 1, such as emission lines of  Mg, S, Si, Ca, and Fe, and it confirms the hard-X-ray nature (kT$>$2\,keV) for all but one WR star).
    
    \item The iron emission line at $\sim$6.7\,keV (Fe~XXV) is detected for the first time in the spectra of 17 out of the 20 X-ray-detected WR stars in Wd1. The presence of the line is typically associated with the very hot plasma located in the wind--wind collision zone of colliding wind binaries.  

    \item We observe the fluorescent Fe line at $\sim$6.4\,keV in the spectra of three WR stars. This is the first clear detection of this line in WR binary systems. The presence of the fluorescent line points to the existence of cold gas that could be photoionised by the hot continuum produced in the wind--wind collision zone.

\end{itemize}

Overall, the hard X-ray spectra of the WR population and the presence of the iron line in the majority of them, along with other binarity diagnostics from optical, infrared, and radio observations, allow us to estimate a binarity fraction of 92\%, which could also reach unity with more detailed and continuous observations in other wavelengths. This result has important implications for advancing our understanding of the evolution of massive stars and the formation of massive star clusters.

\begin{acknowledgements}
K. A. acknowledges support of INAF-OAPa Director Grant n. 72/2022. K. A. thanks Vinay L. Kashyap, Anthony Moffat, and Sean Gunderson for the helpful recommendations/discussions.
M. G. G., C. A., E. F., L. P., and S. S. acknowledge the INAF grant 1.05.12.05.03. J.
F.A.C. is a researcher of CONICET and acknowledges their support. Support for this work was also provided by the National Aeronautics and Space Administration through \emph{Chandra} Proposal 21200267 issued by the \emph{Chandra} X-ray Center, which is operated by the Smithsonian Astrophysical Observatory for and on behalf of the National Aeronautics Space Administration under contract NAS8-03060. 
I.N. is partially supported by the Spanish Government Ministerio de Ciencia e Innovaci\'on (MCIN) and Agencia Estatal de Investigaci\'on (MCIN/AEI/10.130~39/501~100~011~033/FEDER, UE) under grant PID2021-122397NB-C22, and also by MCIN with funding from the European Union NextGenerationEU and Generalitat Valenciana in the call Programa de Planes Complementarios de I+D+i (PRTR 2022), project HIAMAS, reference ASFAE/2022/017.
M. G. G. also acknowledge partial support from the Grant INAF 2022 YODA. The scientific results reported in this article are based on observations made by the \emph{Chandra} X-ray Observatory. We also made use of NASA’s Astrophysics Data System Bibliographic Services. 
\end{acknowledgements}


\bibliographystyle{aa} % style aa.bst
\bibliography{references} % if your bibtex file is called example.bib

%%%%%%%%%%%%%%%%%%%%%%%%%%%%%%%%%%%%%%%%%%%%%%%%%%
%%%%%%%%%%%%%%%%%%%%%%%%%%%%%%%%%%%%%%%%%%%%%%%%%%

%%%%%%%%%%%%%%%%%%%%%%%%%%%%%%%%%%%%%%%%%%%%%%%%%%
%%%%%%%%%%%%%%%%% APPENDICES %%%%%%%%%%%%%%%%%%%%%

\begin{appendix}

\onecolumn

\section{Short-term light curves of the WR stars that show variability}

\begin{figure*}[!h]
        \centering      
                \includegraphics[scale=0.33]{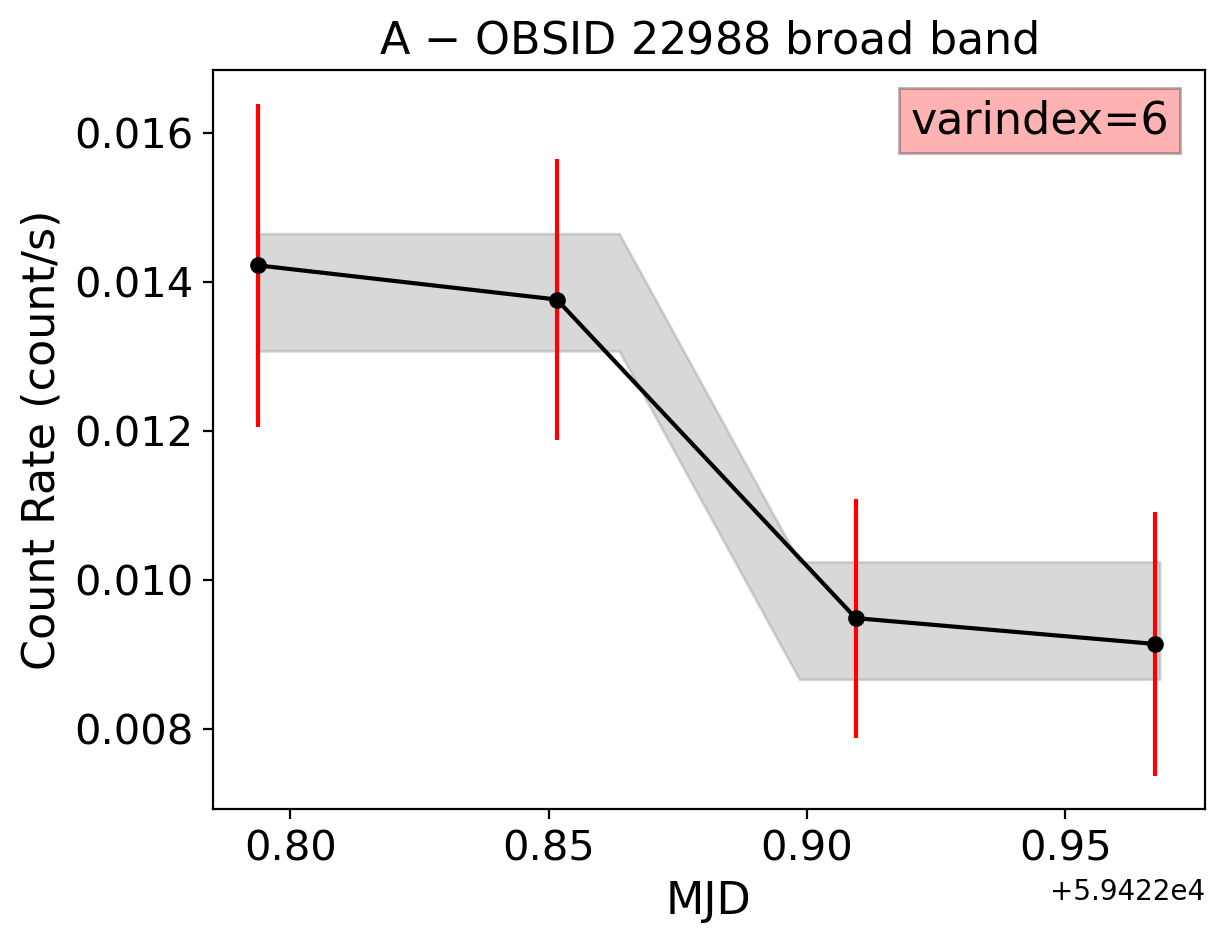}
   \includegraphics[scale=0.33]{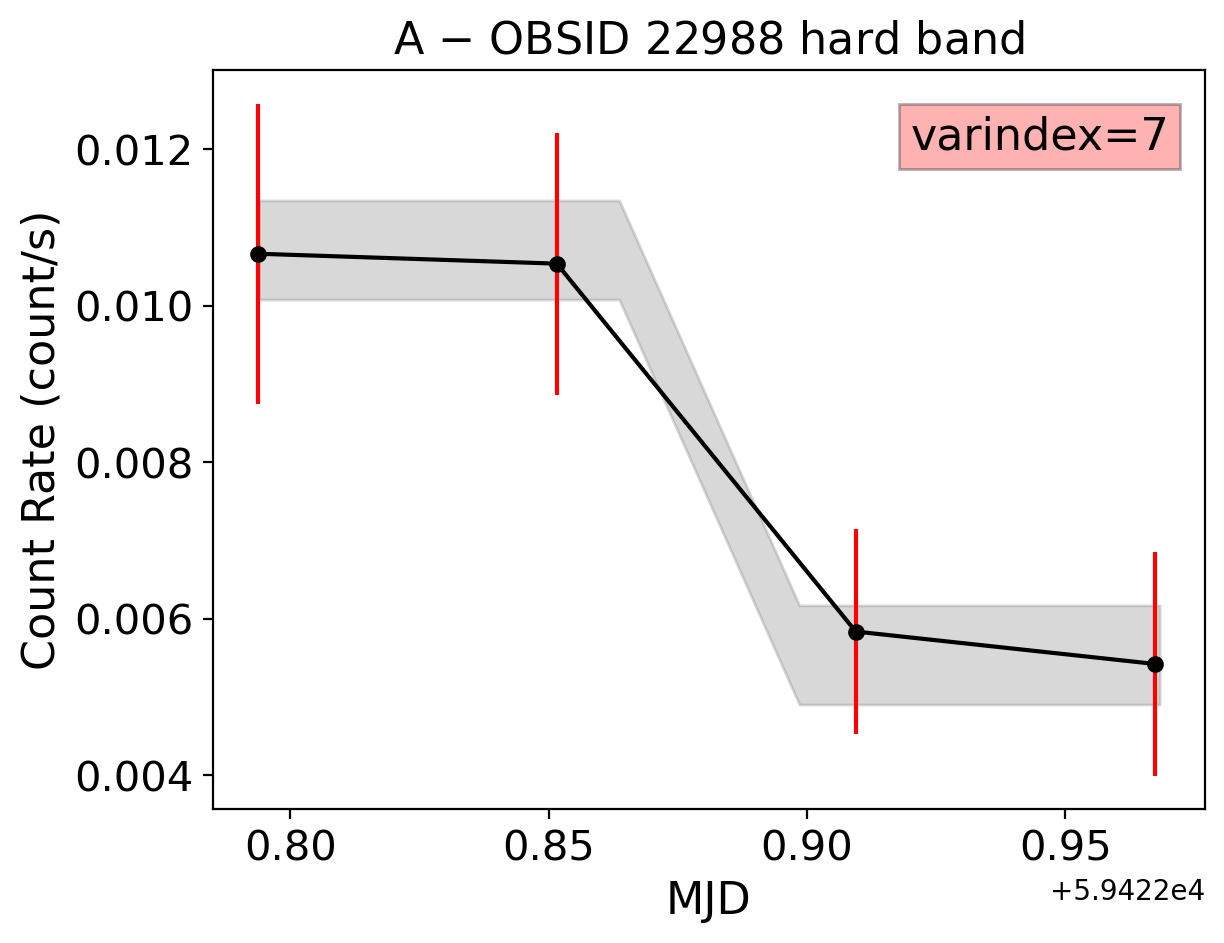}
\includegraphics[scale=0.33]{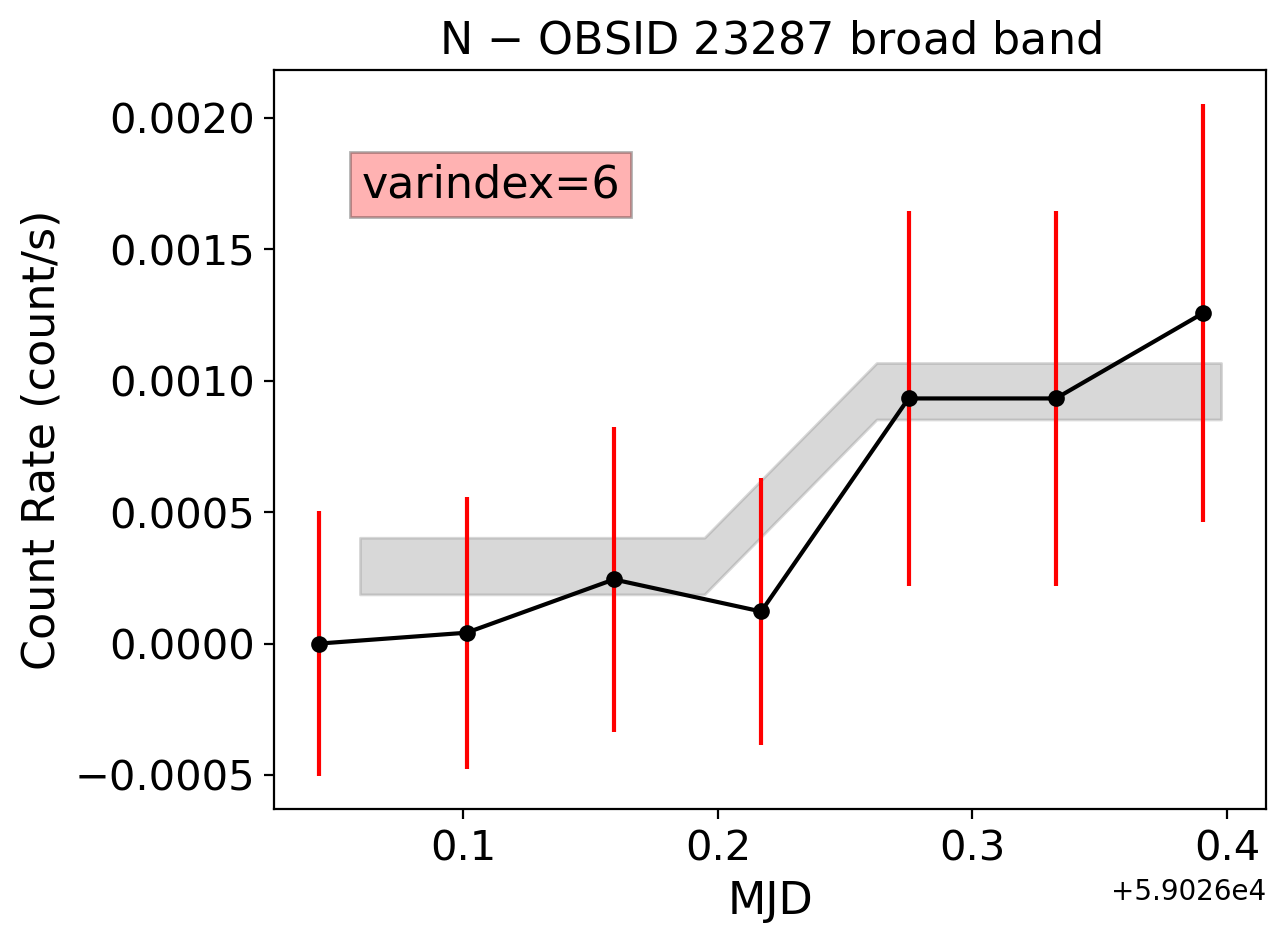}
\includegraphics[scale=0.33]{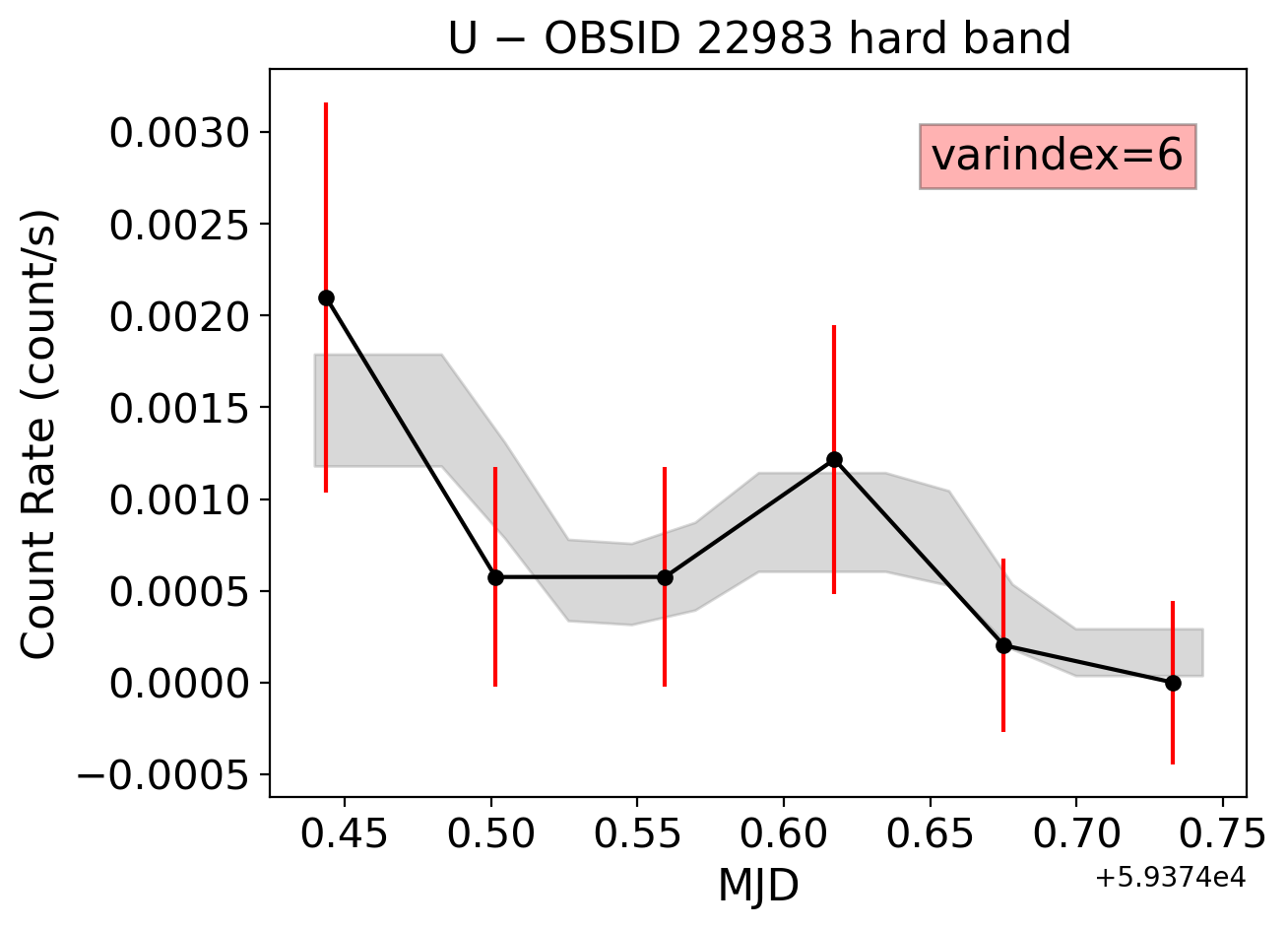}
\includegraphics[scale=0.33]{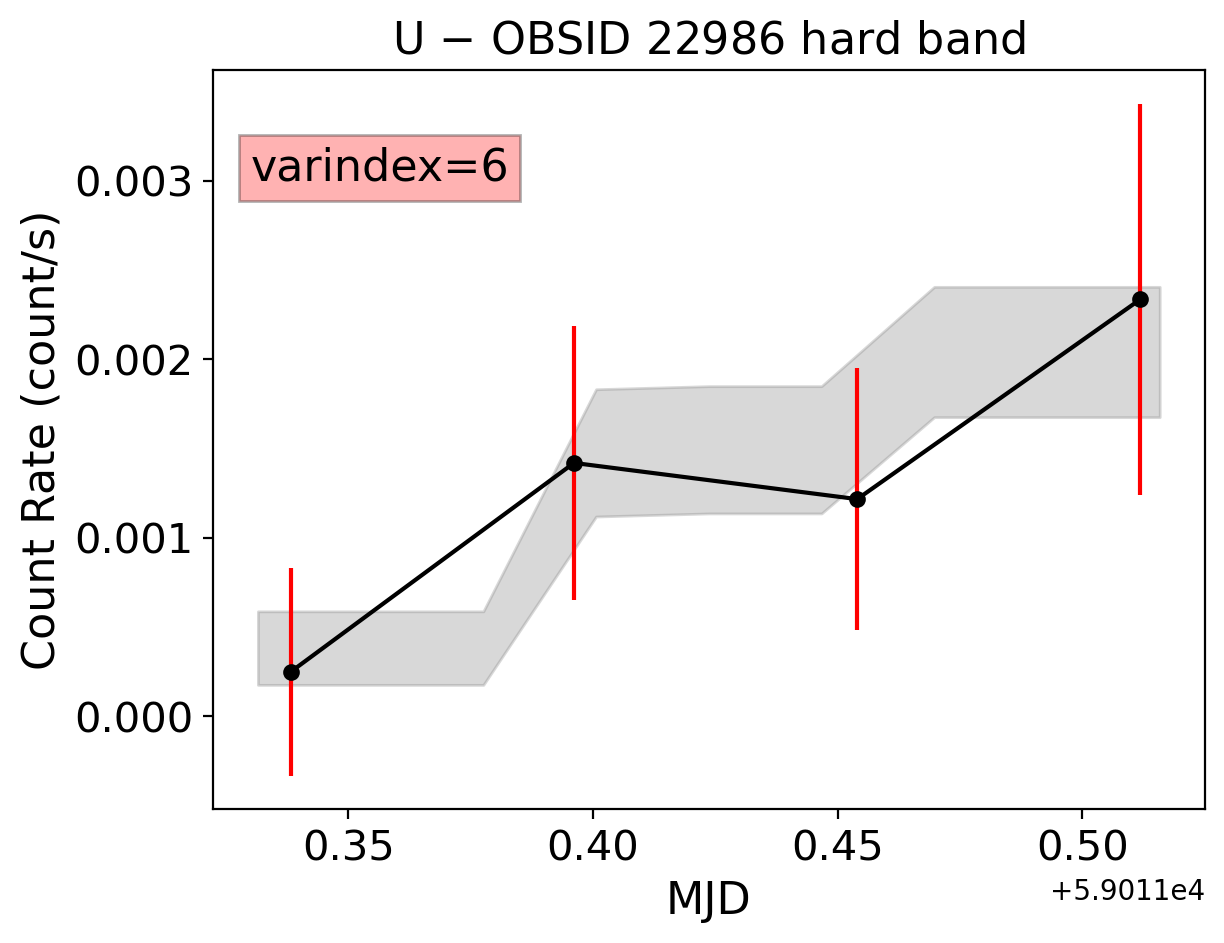}
\includegraphics[scale=0.33]{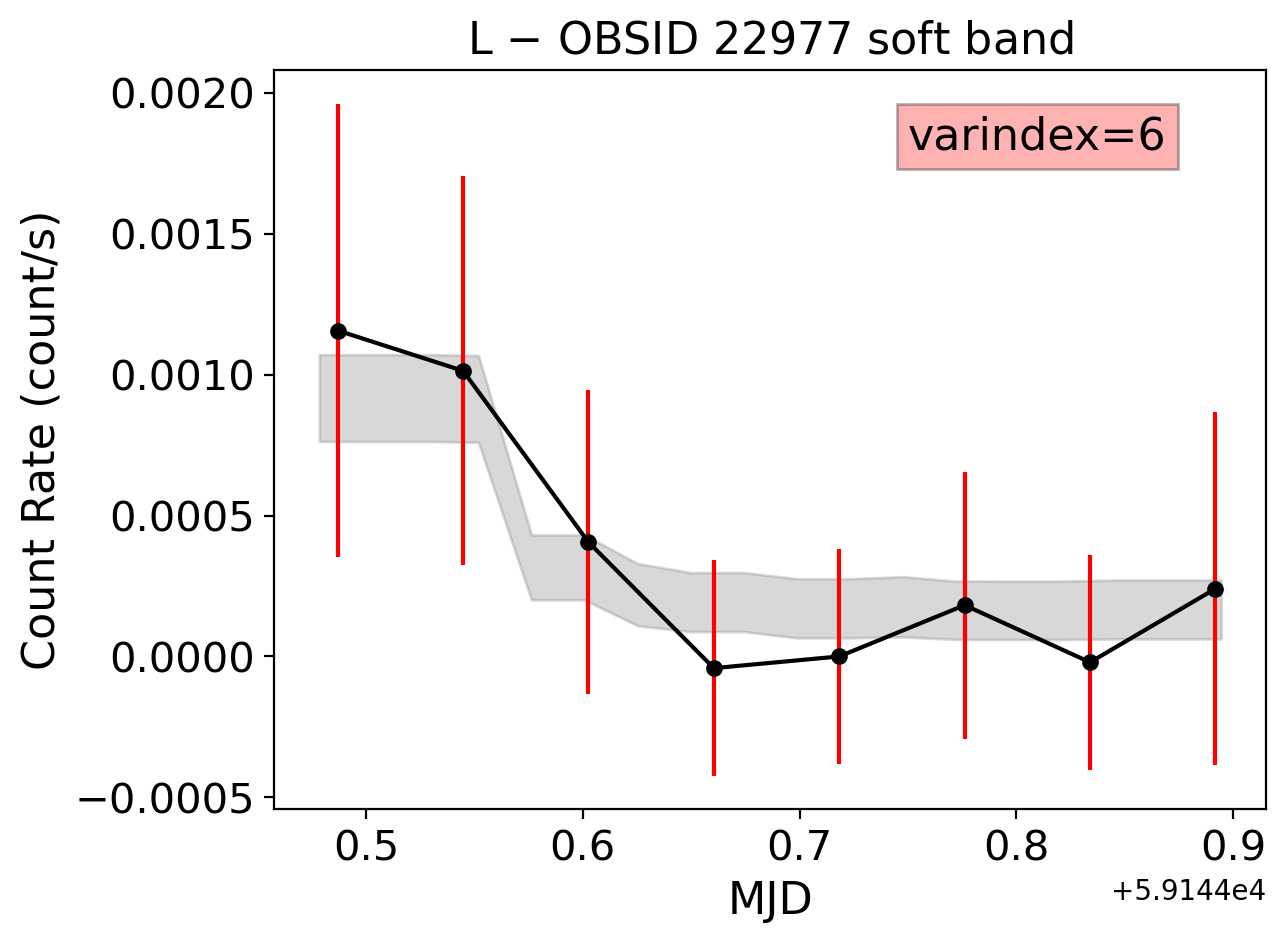}
\caption{Light curves of the single observations where short-term variability is detected. The raw light curves produced with the \textit{dmextract} tool are shown with the black line, with a time binning of 5ks. The probability-weighted light curves and their 3$\sigma$ errors obtained with the \textit{glvary} tool are shown with the shaded area.} 
 \label{fig.lc_singleobs}
 \end{figure*}
 
\FloatBarrier

\section{Long-term light curves, periodograms, and folded light curves of the Wolf-Rayet stars with defined period measurements.}

\begin{figure*}[!h]
        \centering      
        \includegraphics[width=0.40\columnwidth]{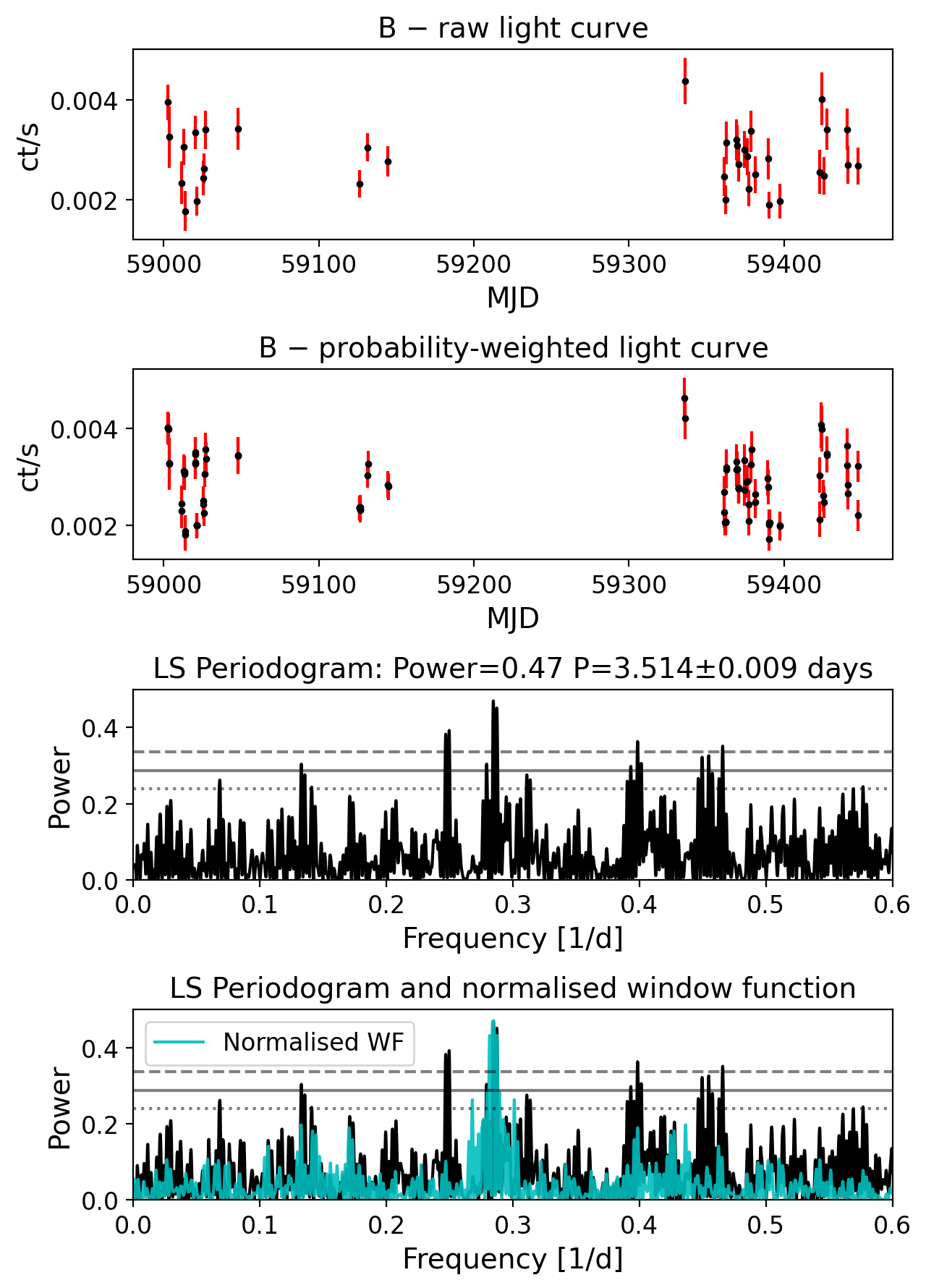}
         \includegraphics[width=0.40\columnwidth]{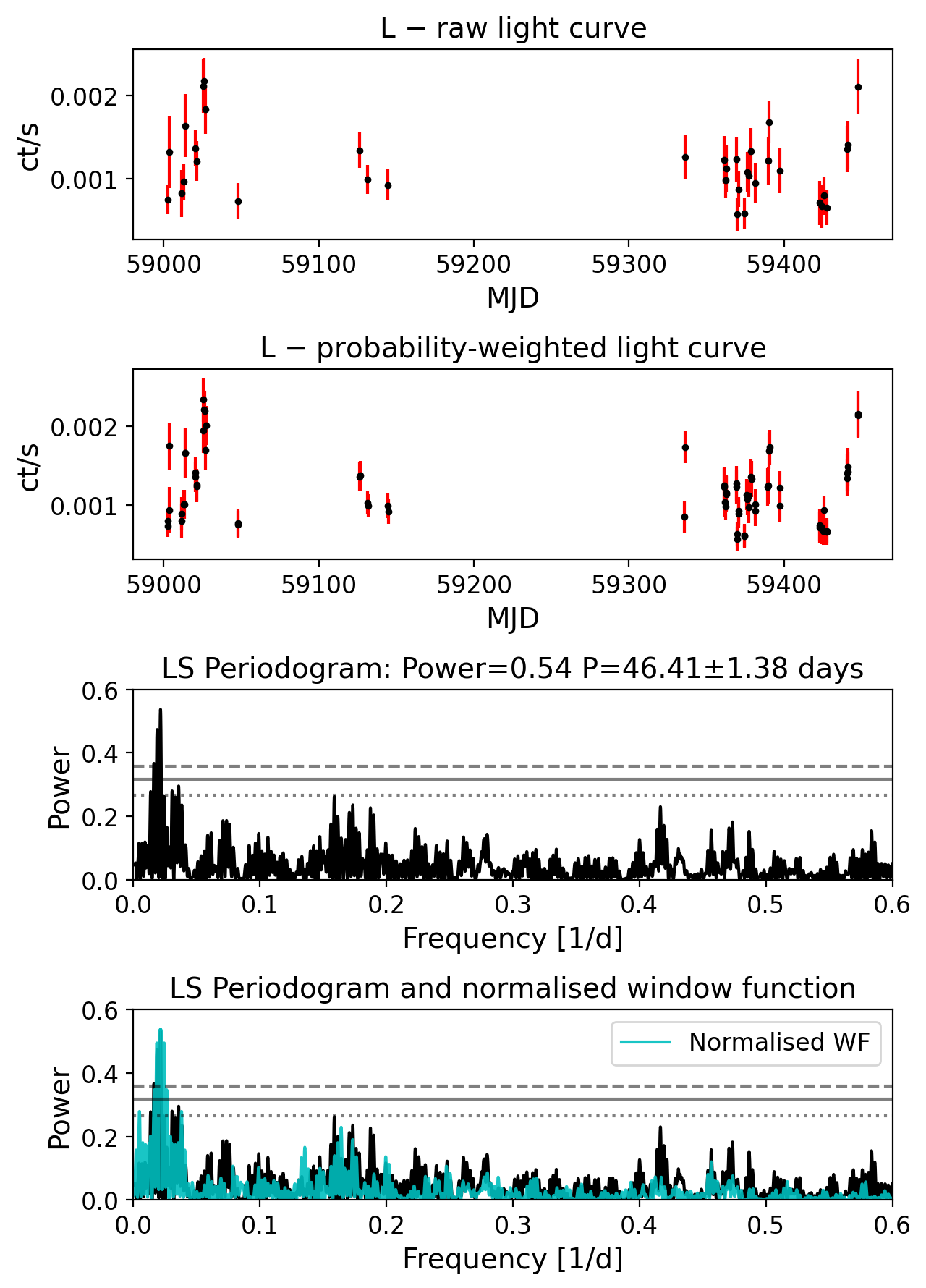}   
   \caption{Long-term light curves and periodograms of WR stars with detected periods. The FAP levels of 0.01, 0.1, and 1 percent are marked with the dashed, solid, and dotted horizontal lines.}
                \label{fig.lcbroad_all}
 \end{figure*}

        \begin{figure*}[!h]
        \ContinuedFloat
        \centering       
     \includegraphics[width=0.40\columnwidth]{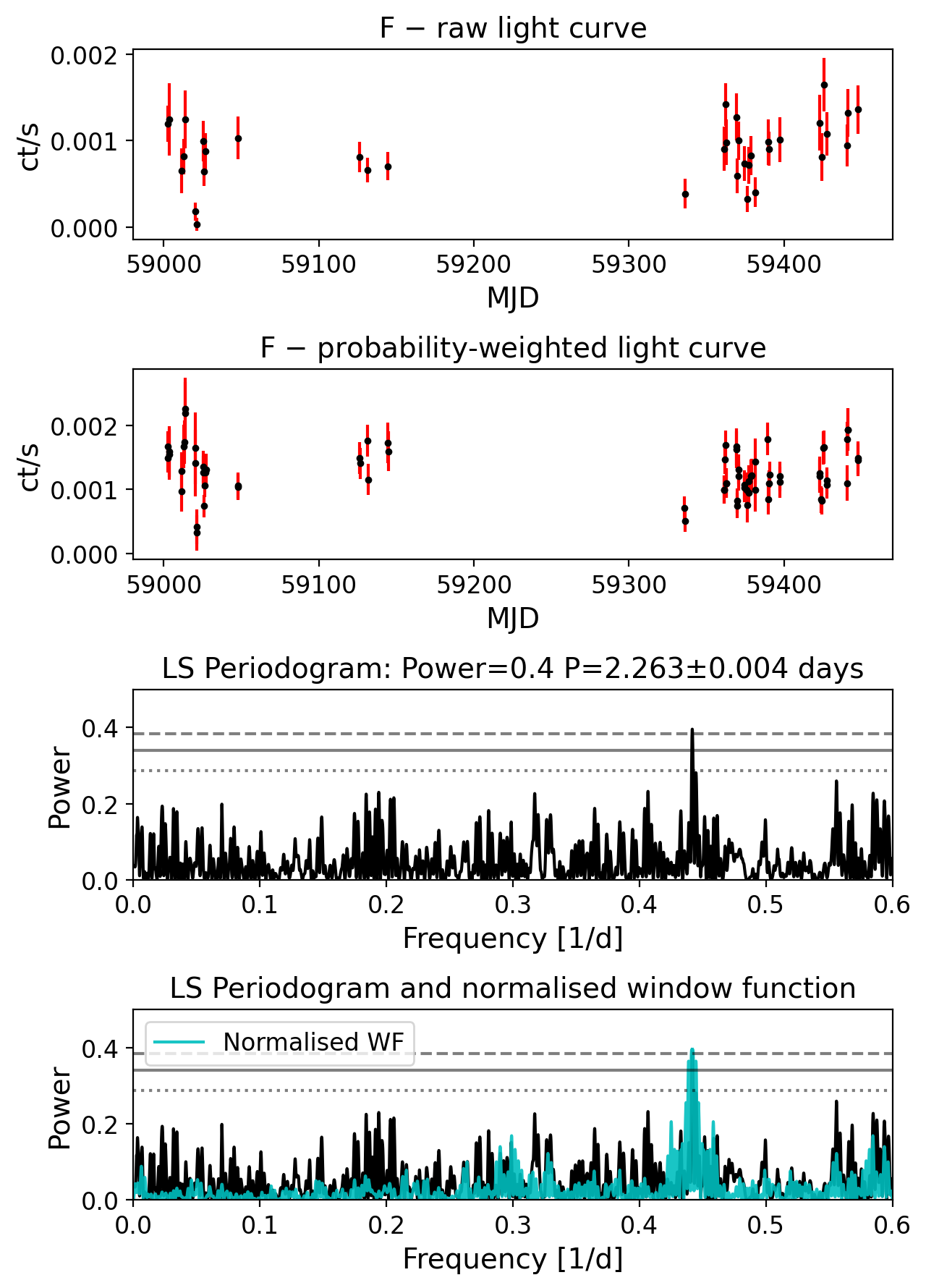}
        \includegraphics[width=0.40\columnwidth]{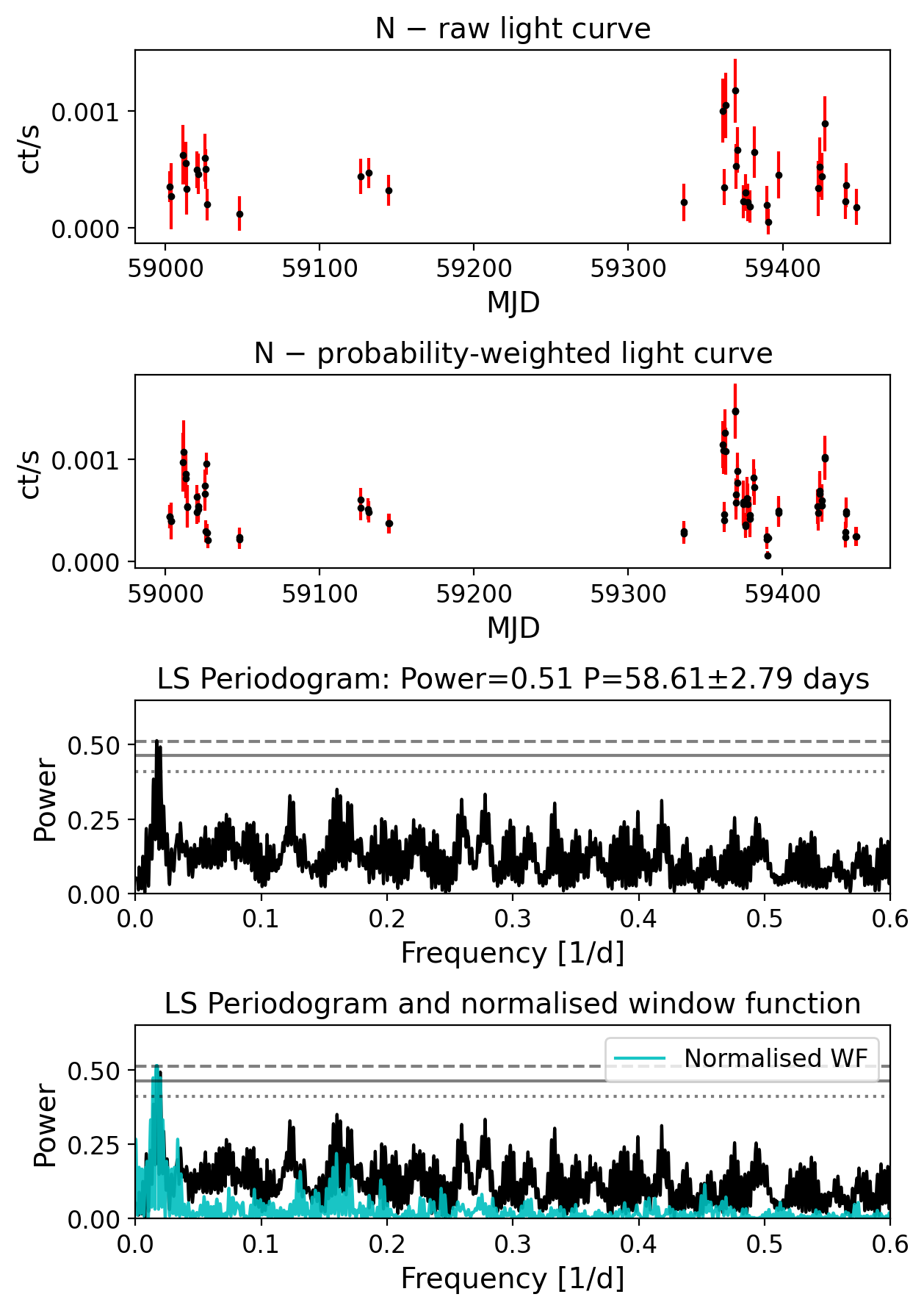}
          \includegraphics[width=0.40\columnwidth]{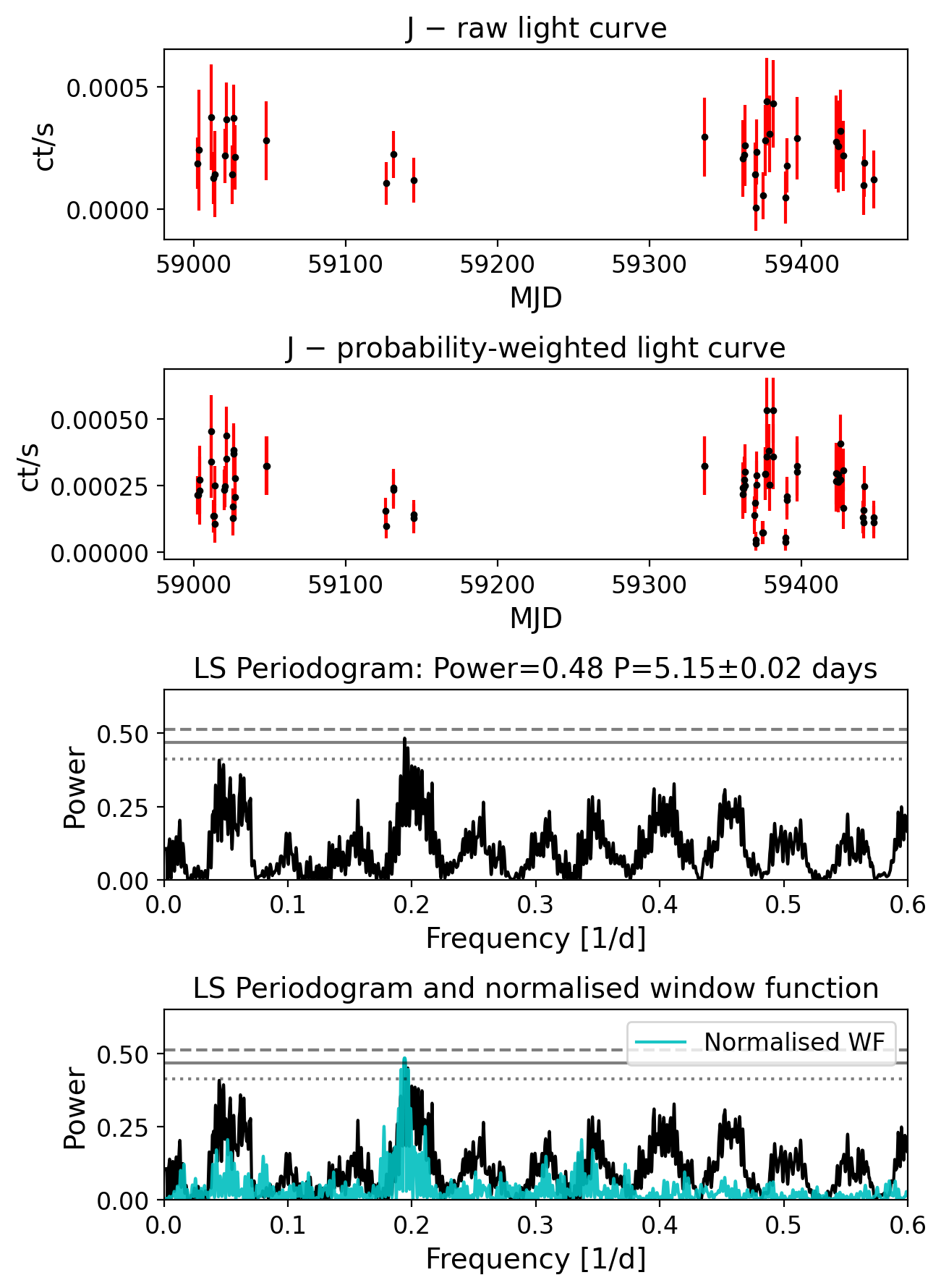}
        \includegraphics[width=0.40\columnwidth]{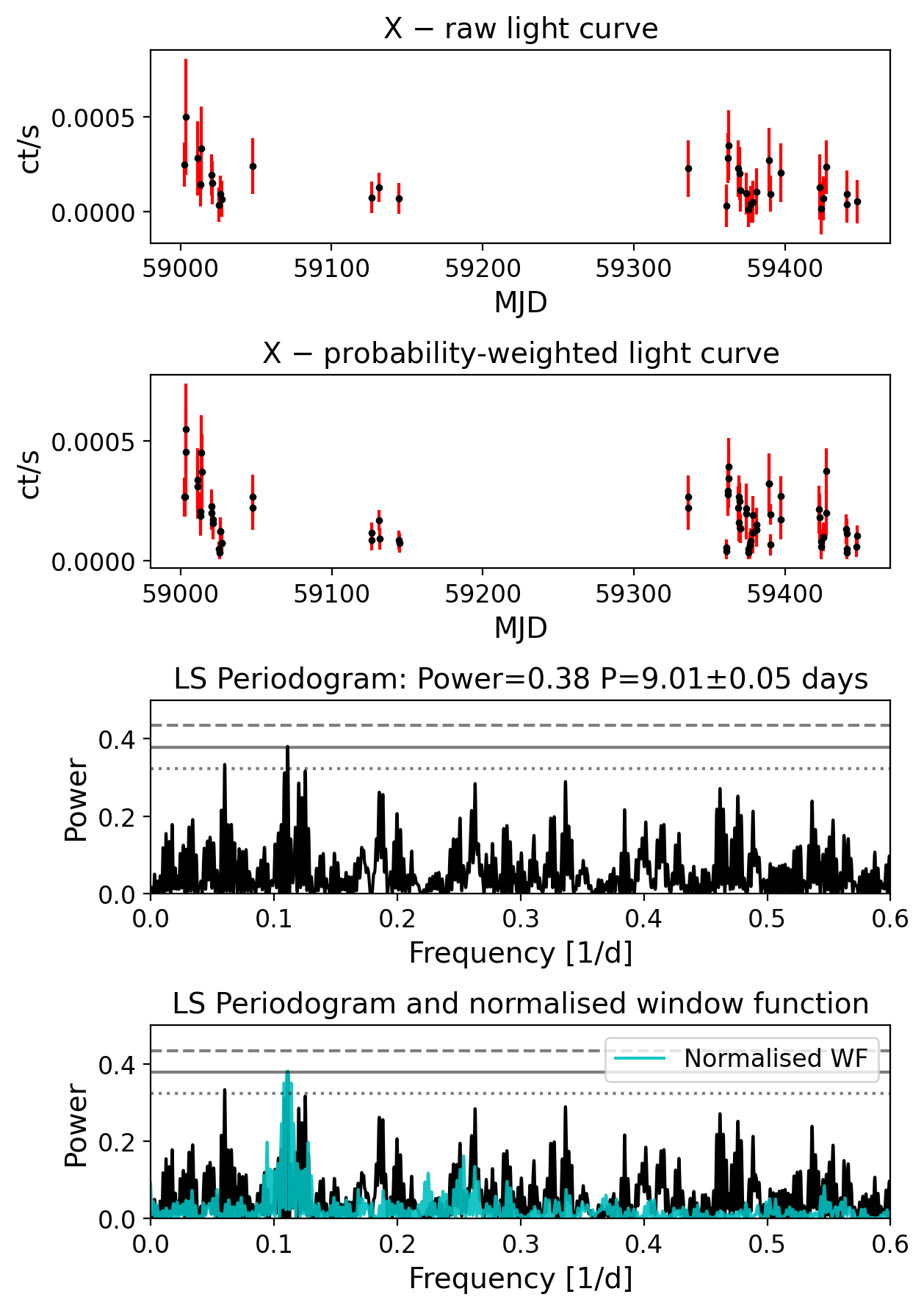}
        \caption{Continued: Long-term light curves and periodograms of WR stars with detected periods}.
                \label{fig.lcbroad_all}
 \end{figure*}

 \begin{figure*}[!h]
        \ContinuedFloat
        \centering       
   \includegraphics[width=0.40\columnwidth]{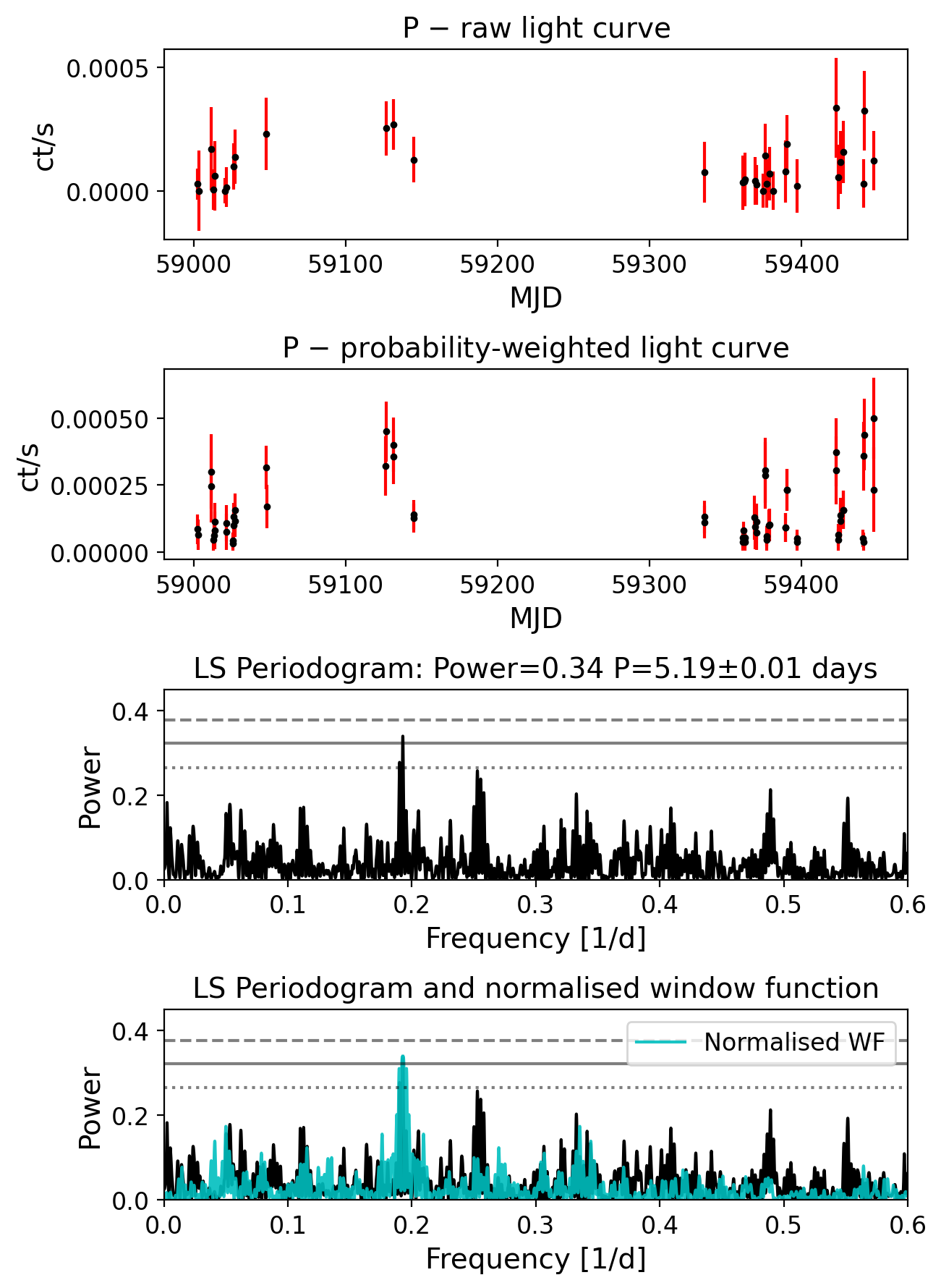}
  \includegraphics[width=0.40\columnwidth]{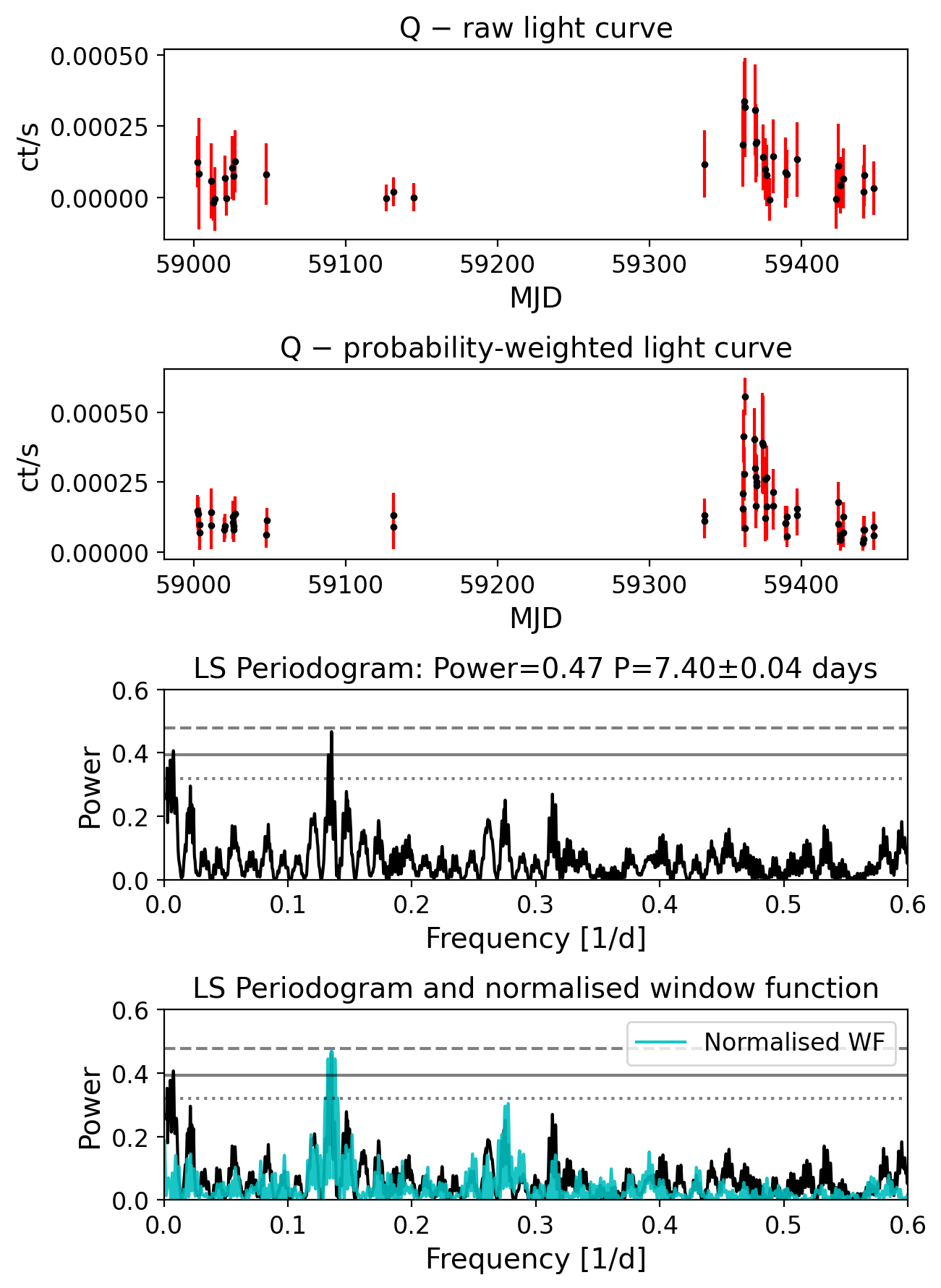}
        \caption{Continued: Long-term light curves and periodograms of WR stars with detected periods}.
                \label{fig.lcbroad_all}
 \end{figure*}

\begin{figure*}[!h]
 \centering           
         \includegraphics[width=0.27\columnwidth]{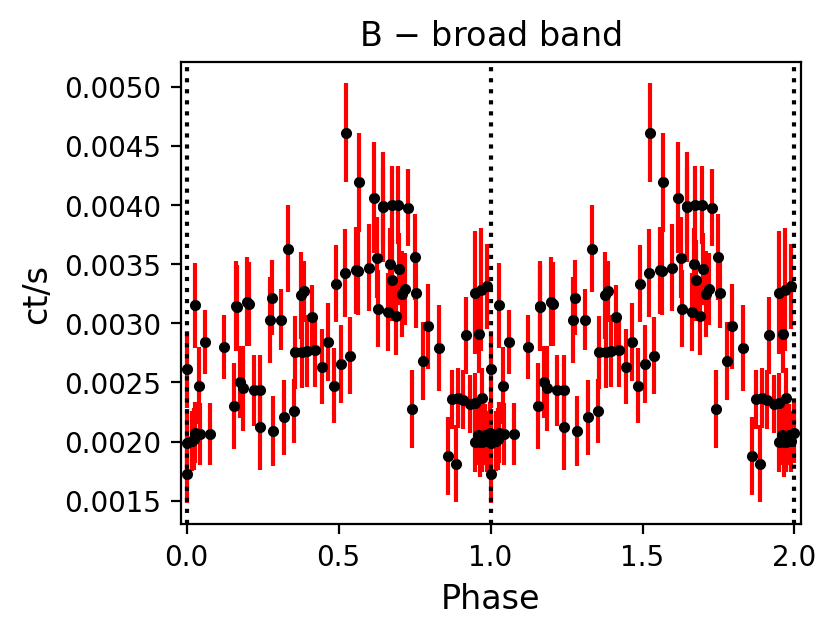}
         \includegraphics[width=0.27\columnwidth]{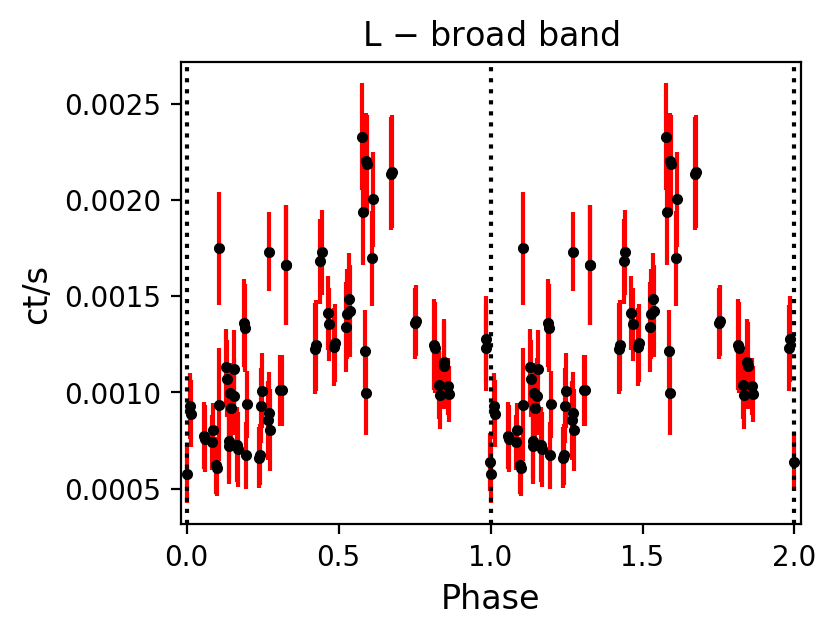}
         \includegraphics[width=0.27\columnwidth]{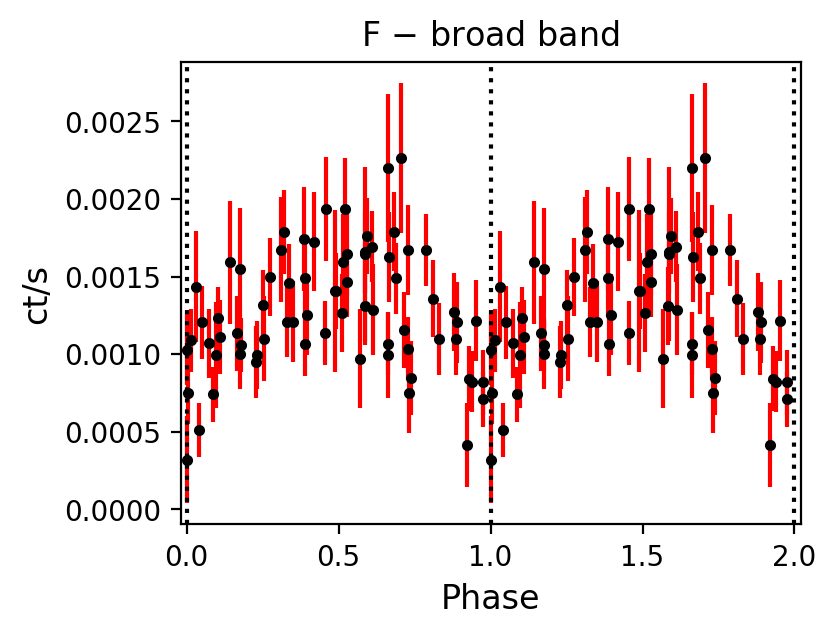}   
         \includegraphics[width=0.27\columnwidth]{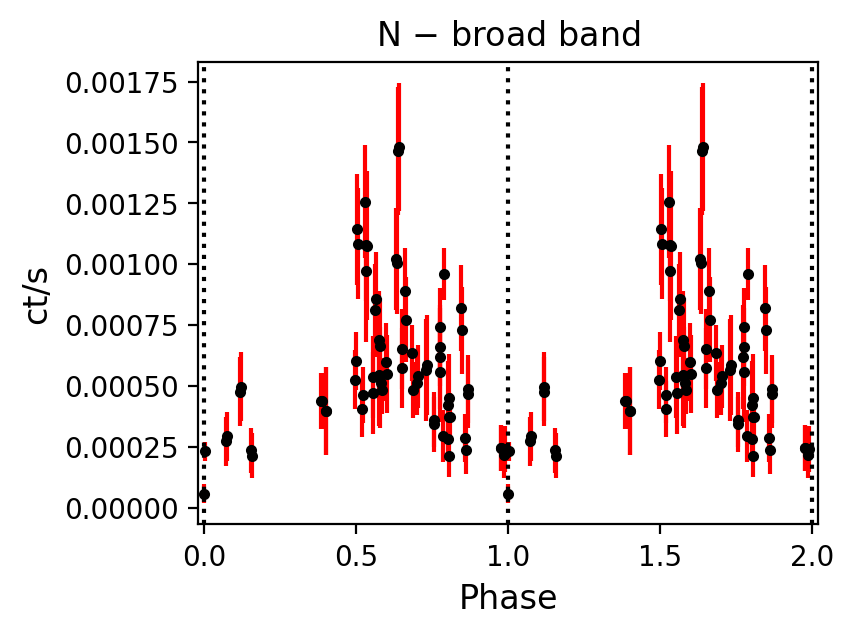}
         \includegraphics[width=0.27\columnwidth]{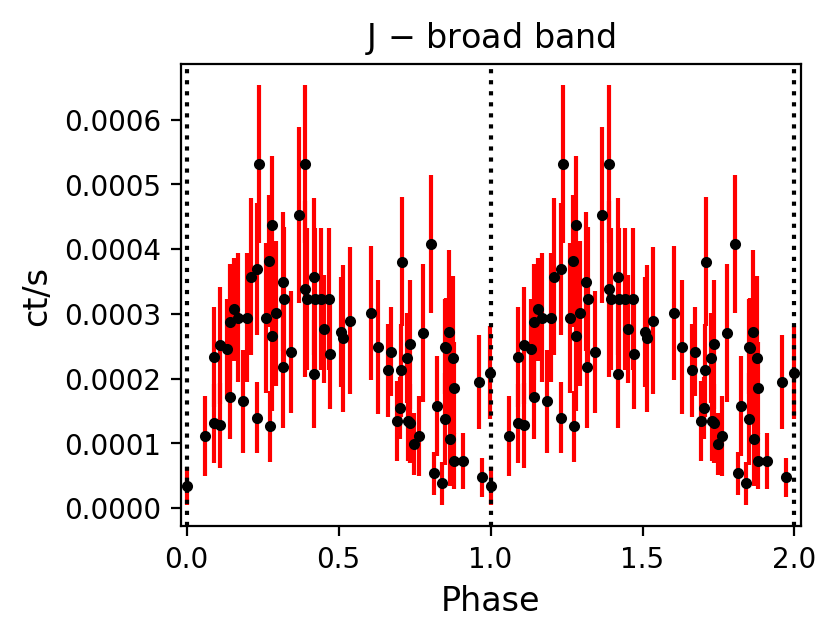}
         \includegraphics[width=0.27\columnwidth]{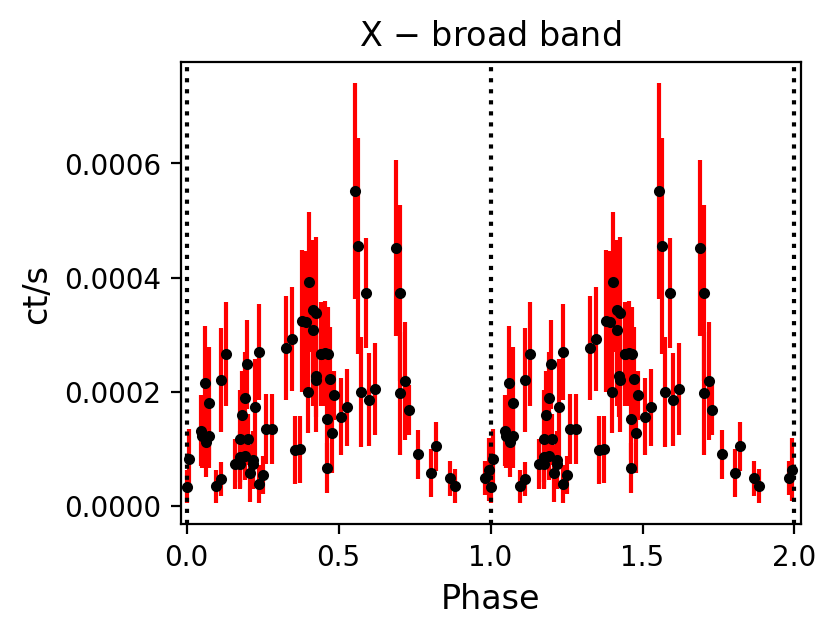}
         \includegraphics[width=0.27\columnwidth]{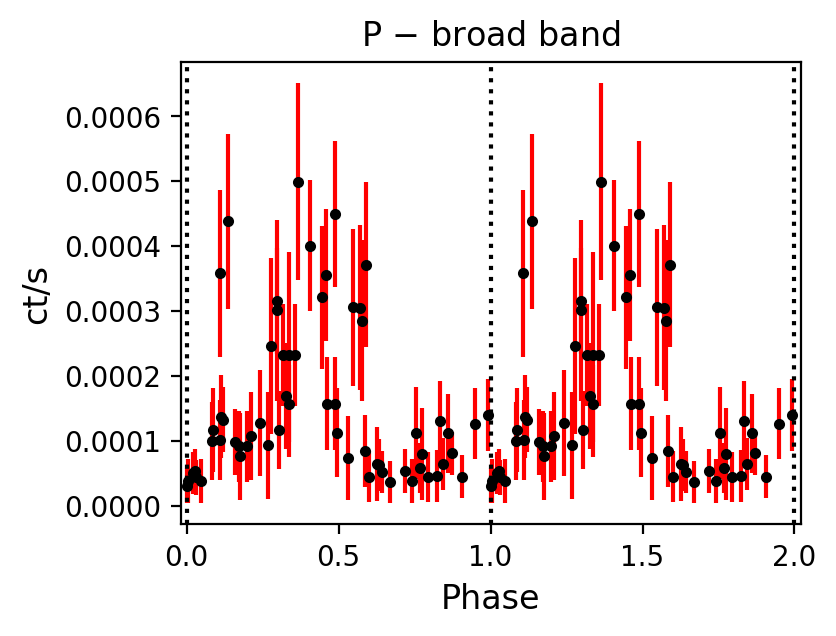}    
         \includegraphics[width=0.27\columnwidth]{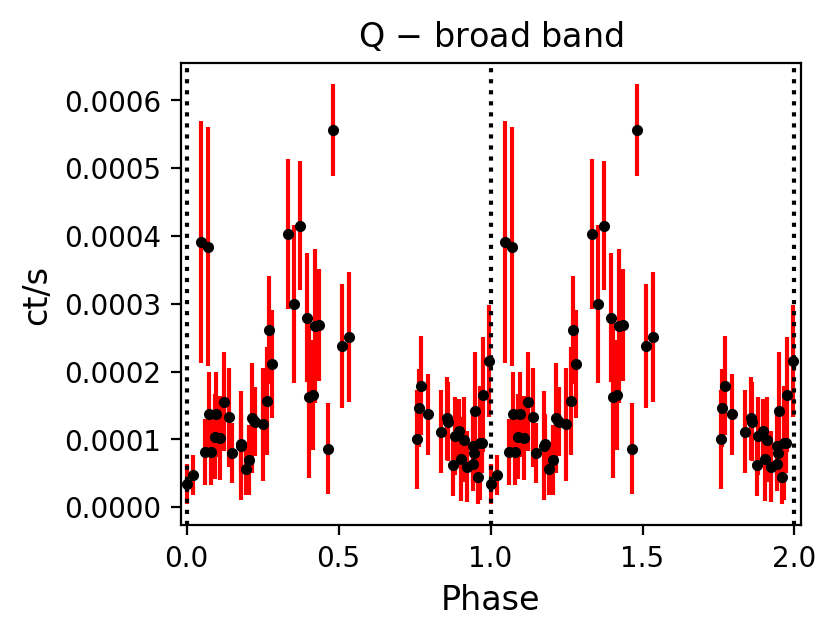} 
         \caption{Phase-folded probability-weighted light curves of WR stars in Wd1 with detected periods.}
                \label{fig.lcbroad_phase_all}
 \end{figure*}

\FloatBarrier

\section{Long-term light curves, periodograms, and folded light curves of the Wolf-Rayet stars with non-detected periods.}

%%%%%non detected
\begin{figure*}[!h]
        \centering      
\includegraphics[width=0.40\columnwidth]{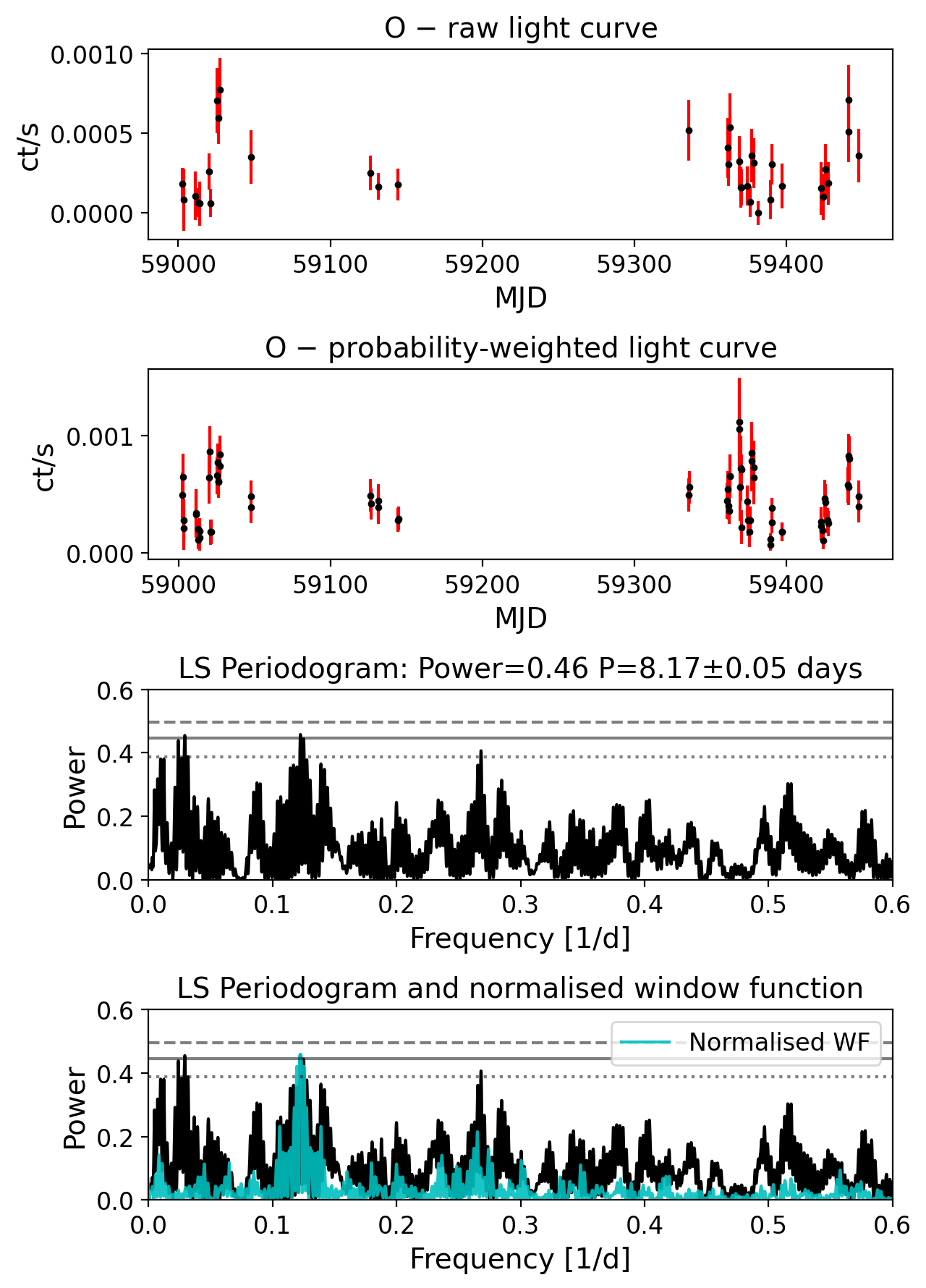}
\includegraphics[width=0.40\columnwidth]{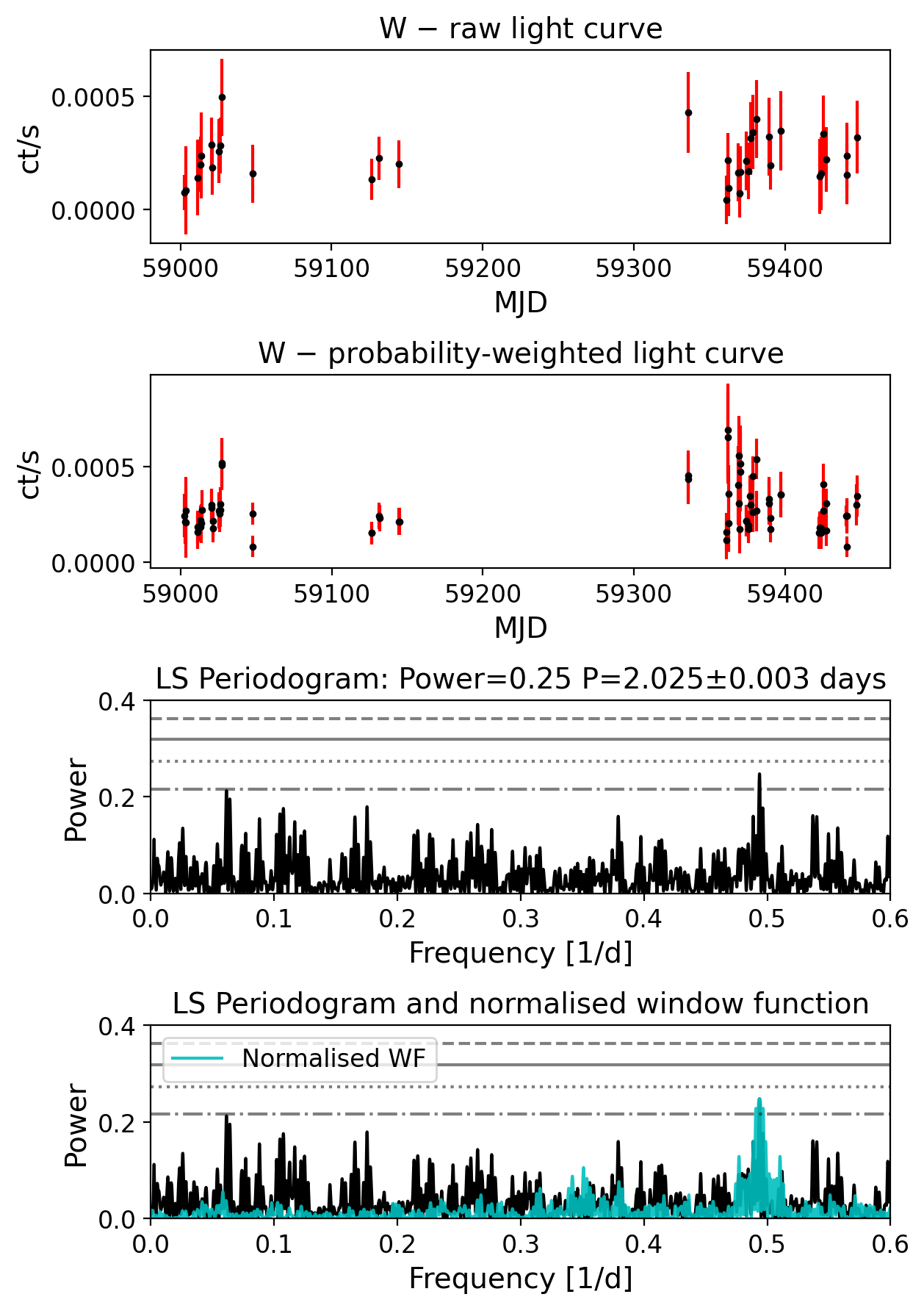}
 \includegraphics[width=0.40\columnwidth]{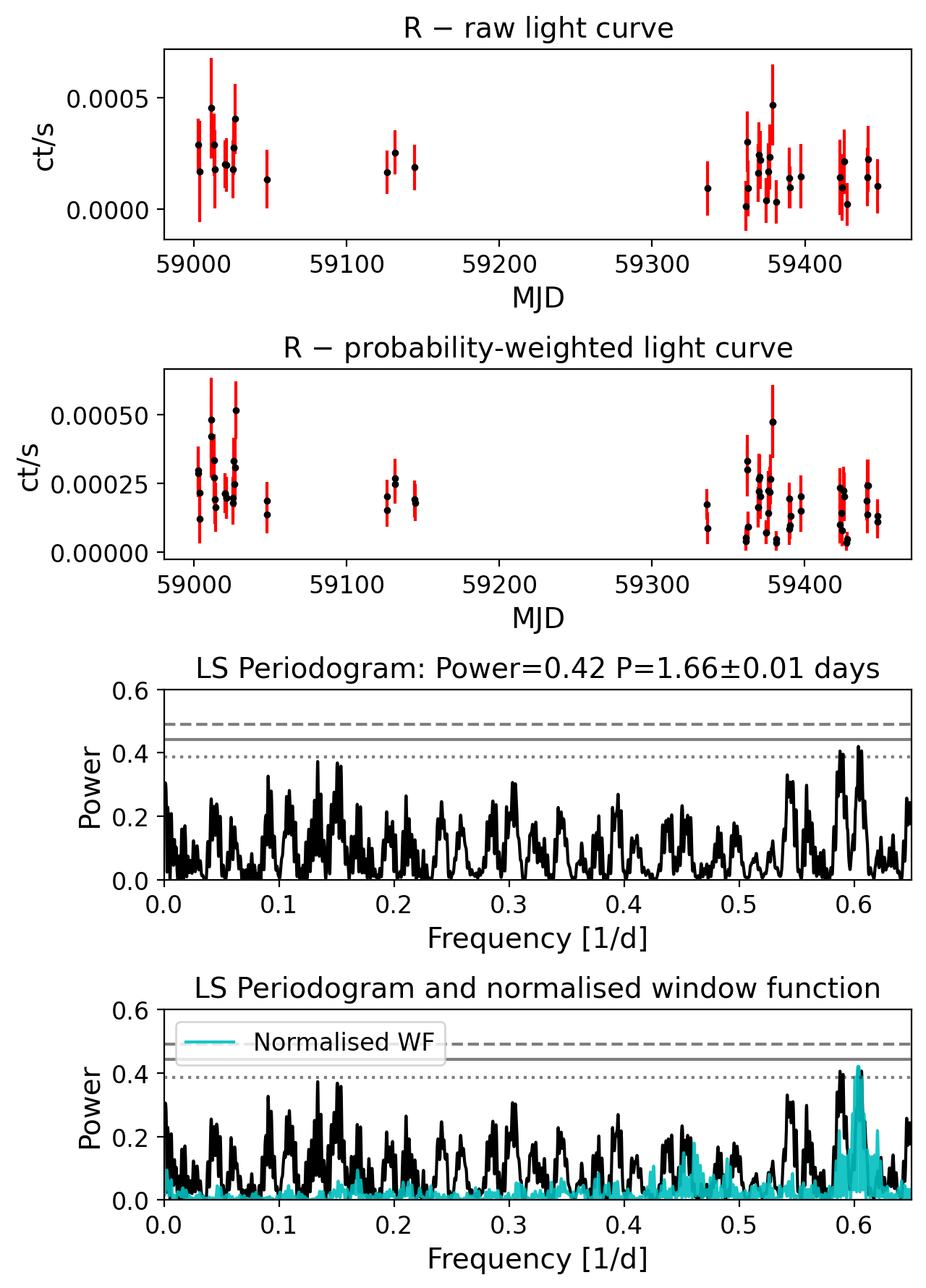}
\caption{Long-term light curves and periodograms of WR stars with non-detected periods. The FAP levels of 0.01, 0.1, and 1 percent are marked with the dashed, solid, and dotted horizontal lines. In some cases, in order to show the significance of the highest peak an additional FAP level of 10 percent is shown with a dash-dotted horizontal line.}
                \label{fig.lcbroad_all_nondetected}
 \end{figure*}

  \begin{figure*}[!h]
  \ContinuedFloat
        \centering       
   \includegraphics[width=0.40\columnwidth]{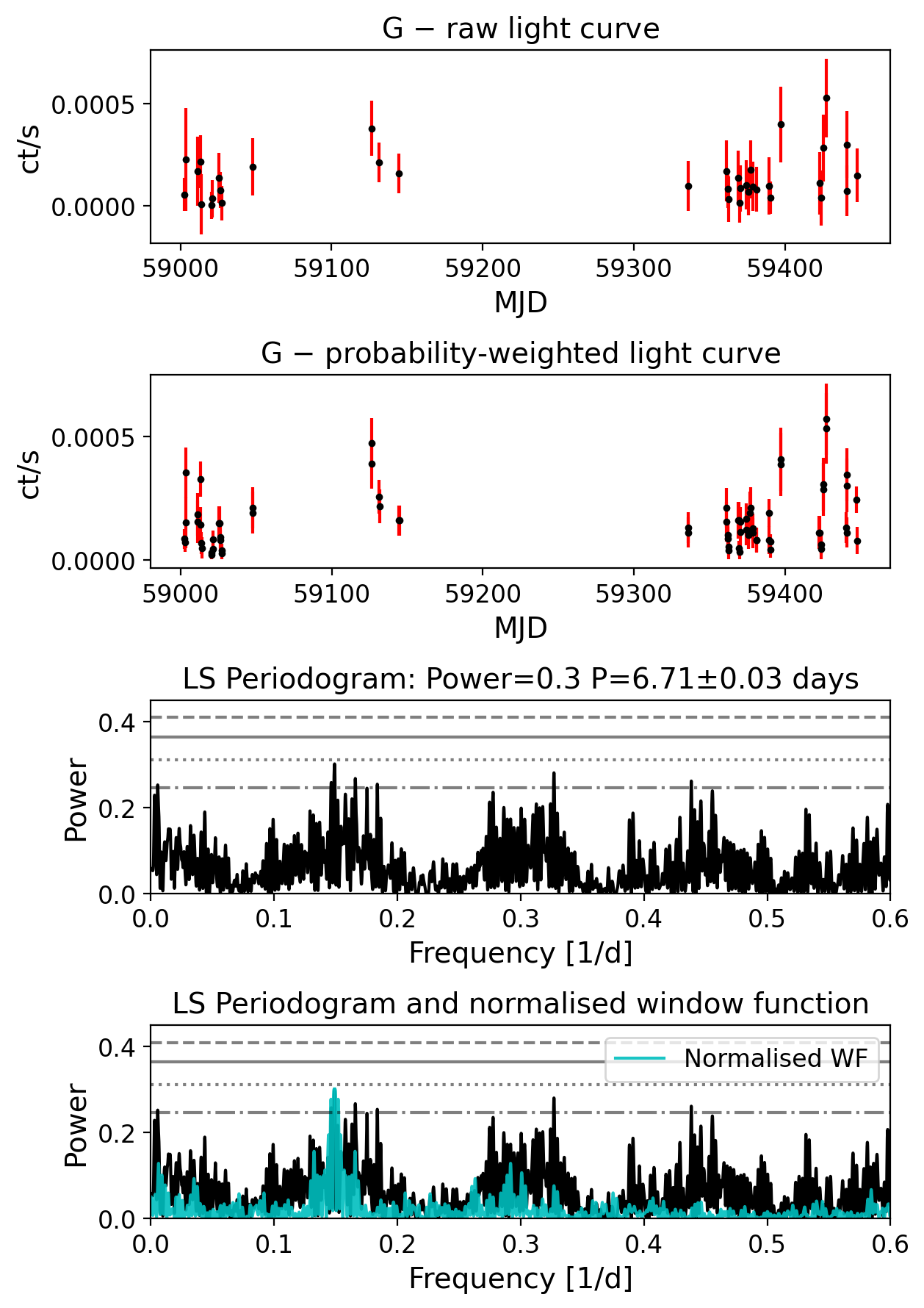}
  \includegraphics[width=0.40\columnwidth]{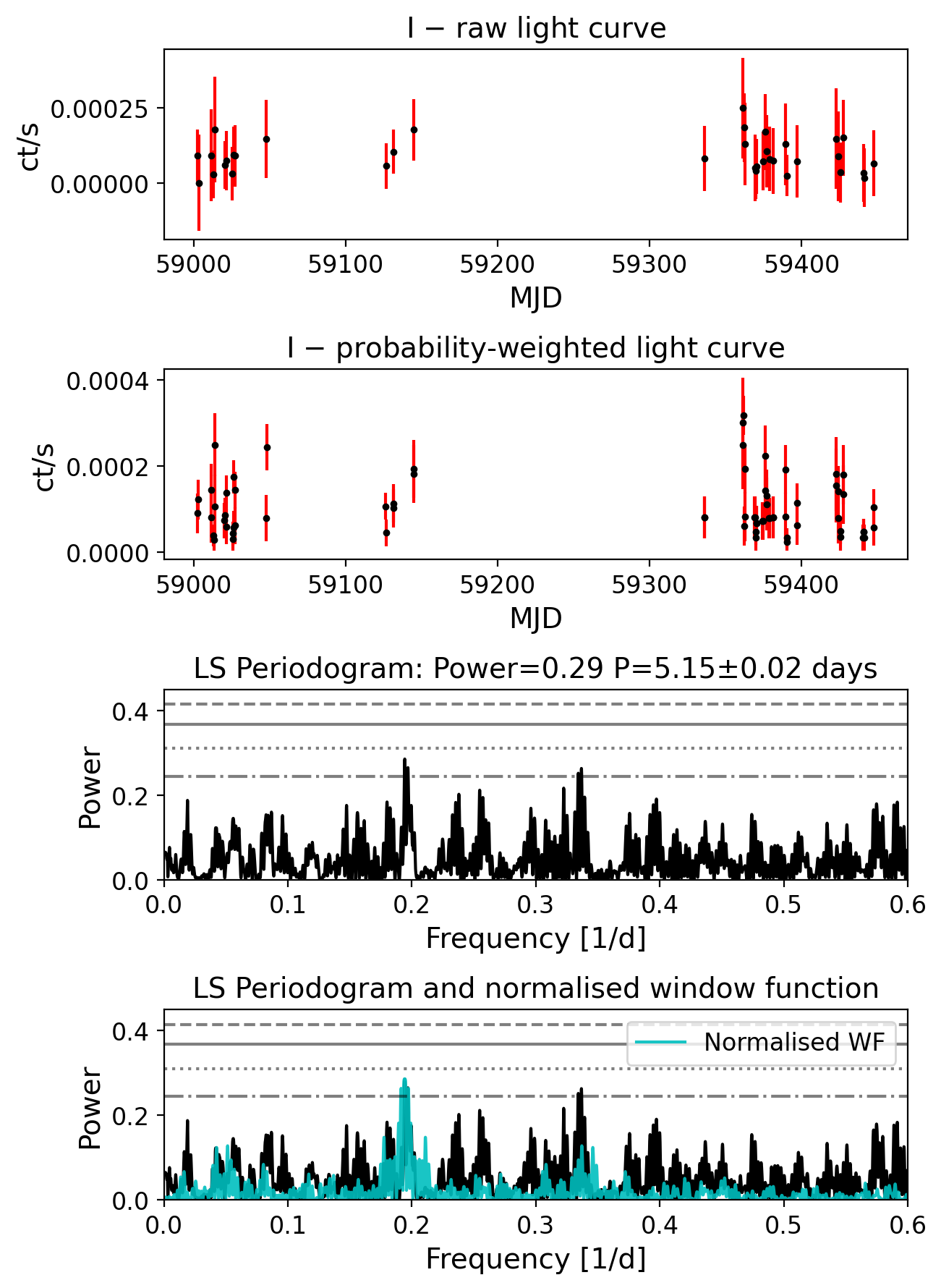}
 \caption{Continued: Long-term light curves and periodograms of WR stars with non-detected periods.}
                \label{fig.lcbroad_all_nondetected}
  \end{figure*}

\begin{figure*}[!h]
 \centering        
\includegraphics[width=0.30\columnwidth]{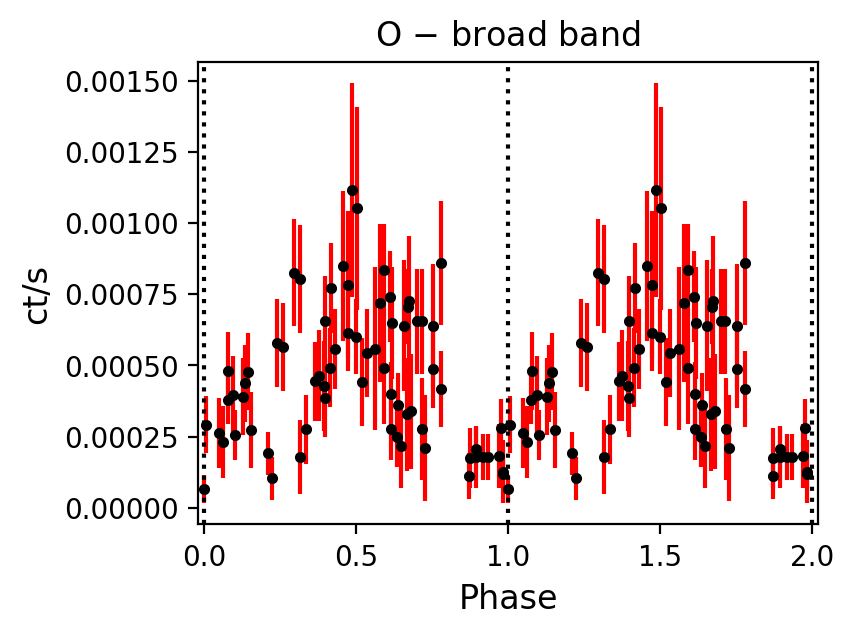}
 \includegraphics[width=0.30\columnwidth]{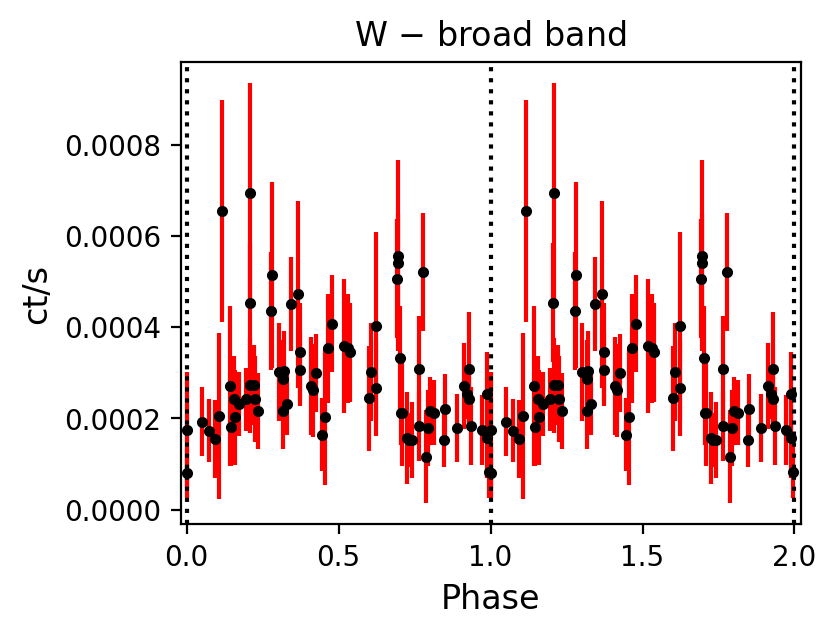}
 \includegraphics[width=0.30\columnwidth]{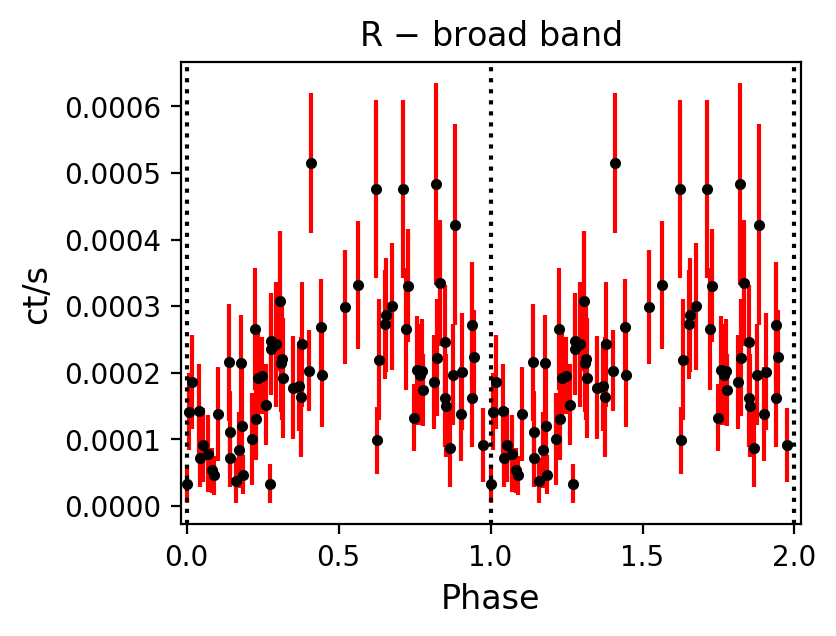}
         \includegraphics[width=0.30\columnwidth]{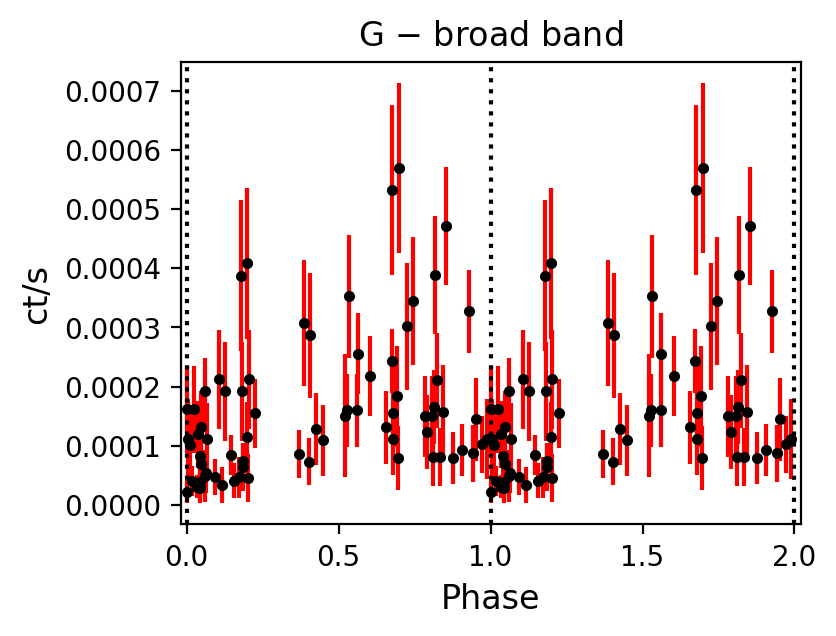}
         \includegraphics[width=0.30\columnwidth]{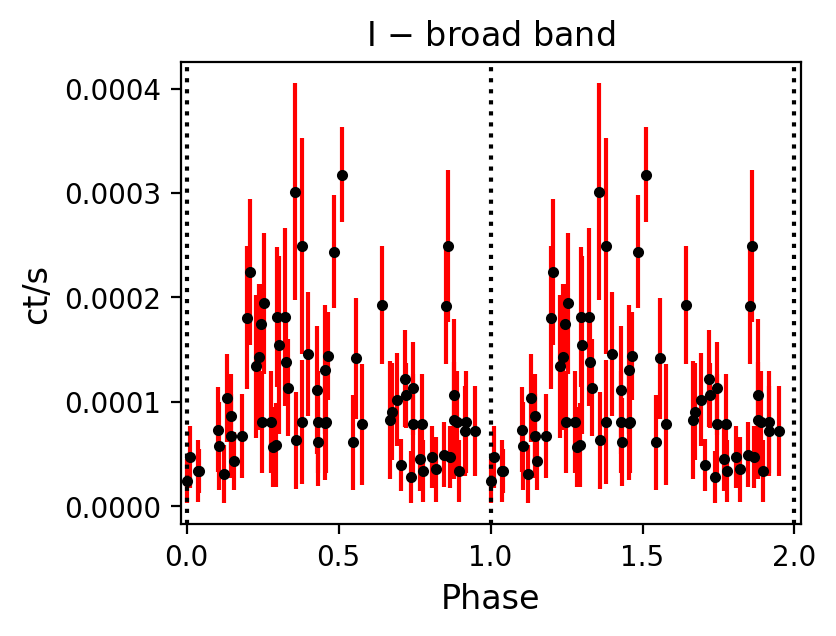}     
\caption{Phase-folded probability-weighted light curves of WR stars in Wd1 with non-detected periods.}
                \label{fig.lcbroad_phase_all_notdetected}
 \end{figure*}

\FloatBarrier

\section{Residuals of long-term light curves}

\begin{figure*}[!h]
    \centering
\includegraphics[width=0.40\linewidth]{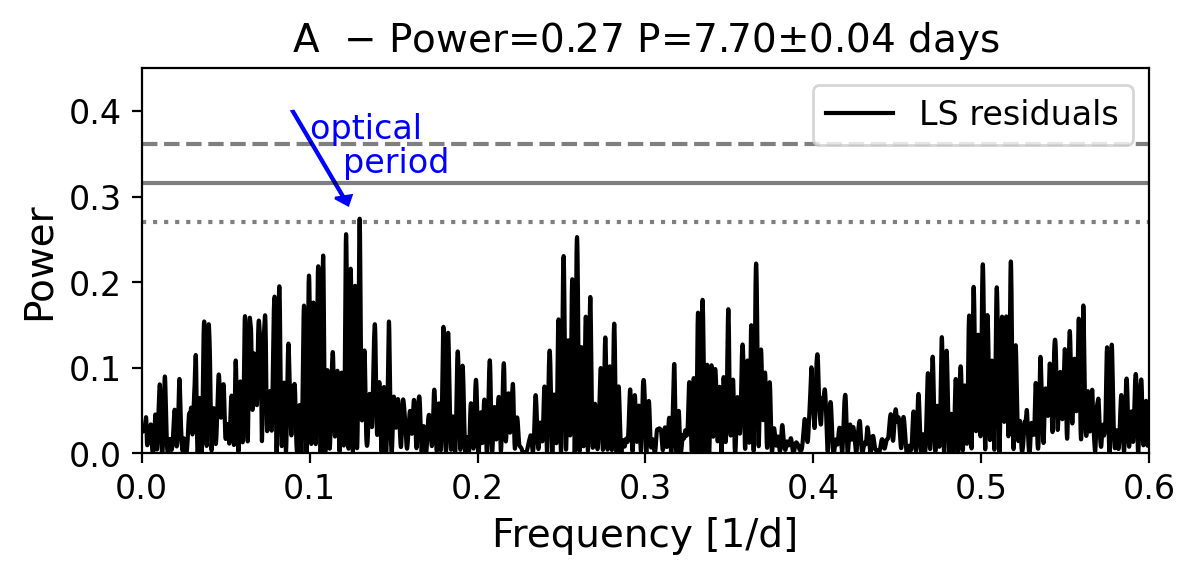}
       \includegraphics[width=0.40\linewidth]{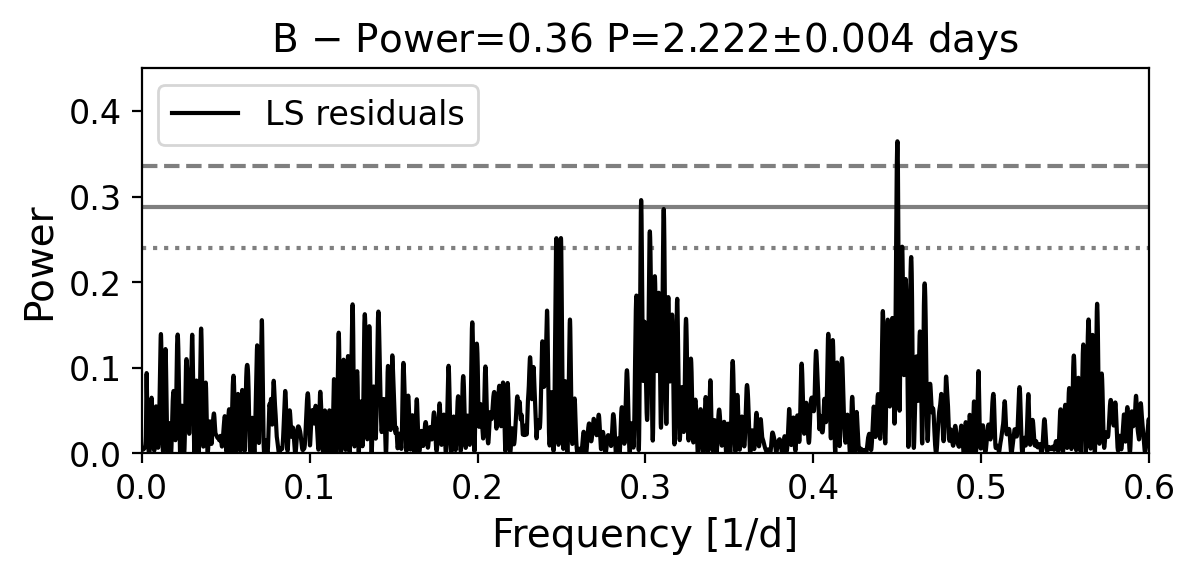} 
       \includegraphics[width=0.40\linewidth]{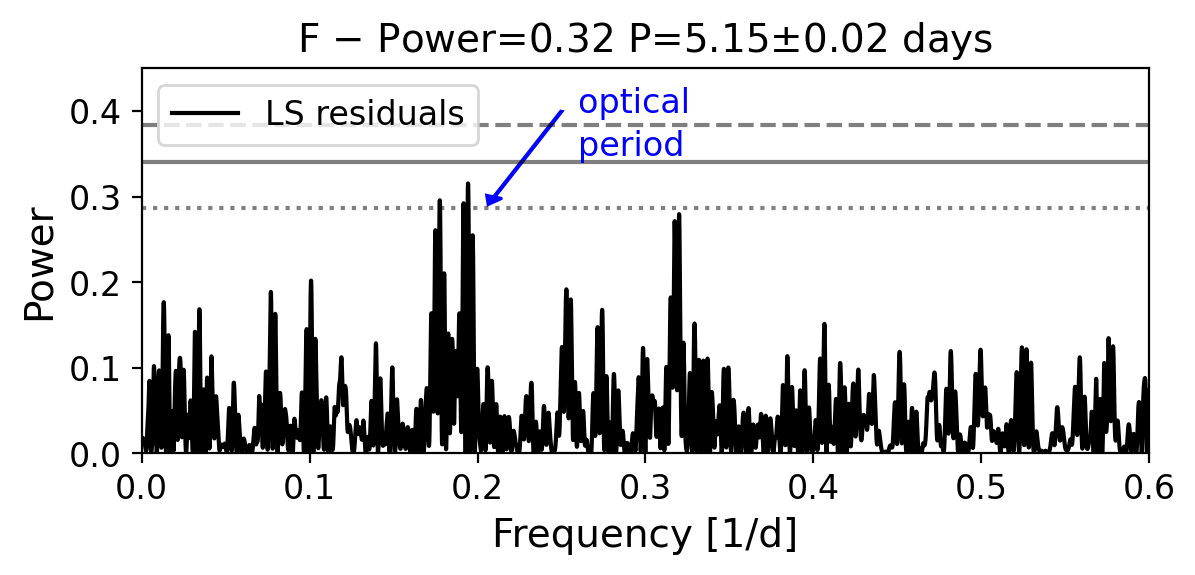} 
 \caption{LS periodograms of the residual time series of stars A, B, and F, after the removal of the main period. The FAP levels of 0.01, 0.1, and 1 percent are marked with the dashed, solid, and dotted horizontal lines.}
    \label{fig.lcsecondary}
\end{figure*}

\begin{figure*}[!h]
    \centering
     \includegraphics[width=0.35\linewidth]{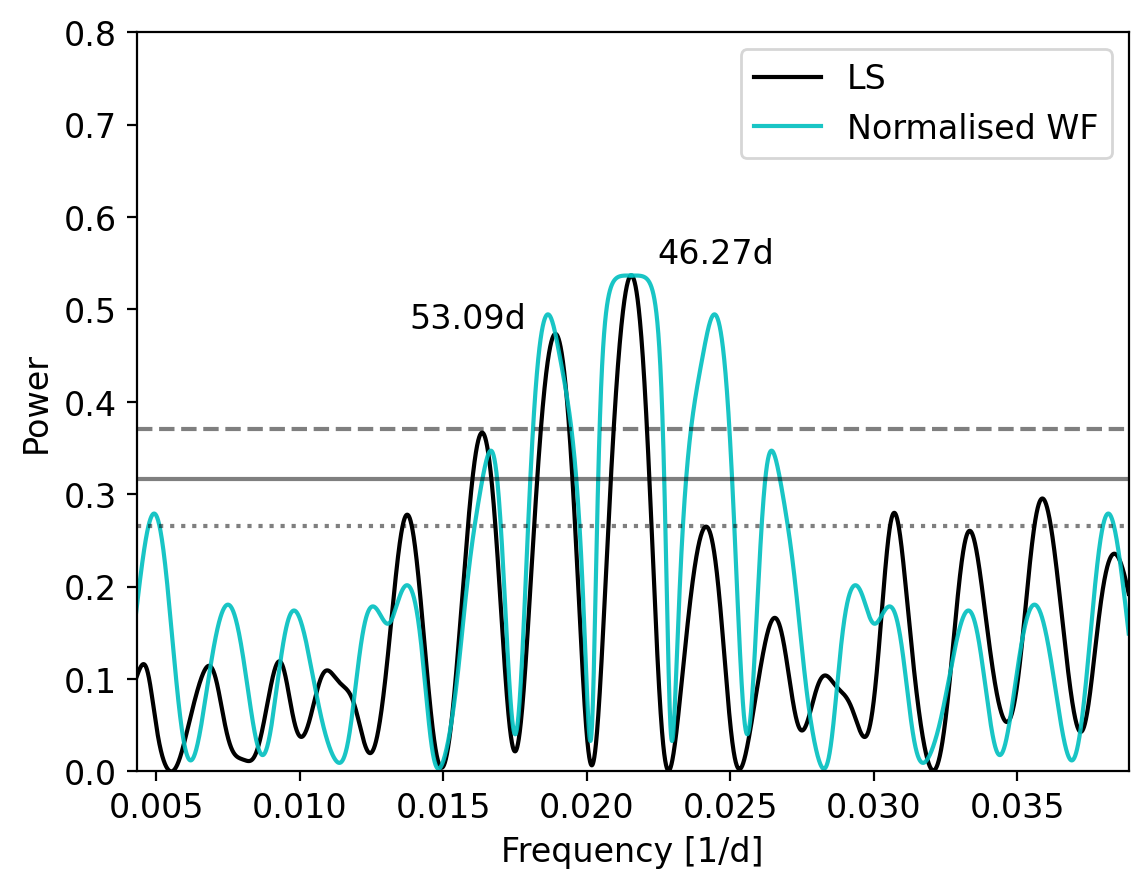}
       \includegraphics[width=0.35\linewidth]{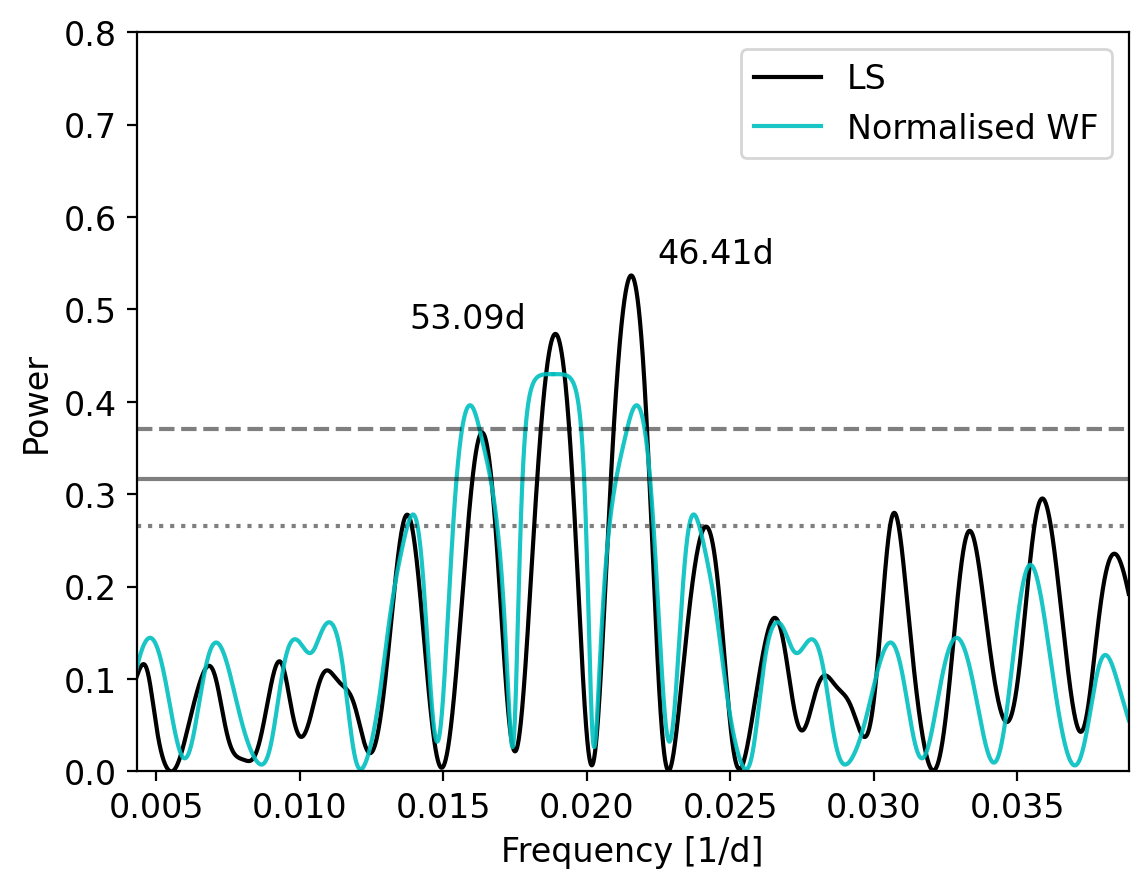}    
 \caption{Star L normalised window function (cyan line) overlaid on the periodogram (black line) to reveal the aliases using as main period the $\sim$46 (left panel) and the $\sim$53 days (right panel) respectively.  The FAP levels of 0.01, 0.1 and 1 percent are marked with the dashed, solid, and dotted horizontal lines.}
    \label{fig.lcw44}
\end{figure*}

\FloatBarrier

\section{X-ray colour variability of four of the brightest WR stars}

\begin{figure*}[!h]
        \centering
\includegraphics[width=0.47\columnwidth]{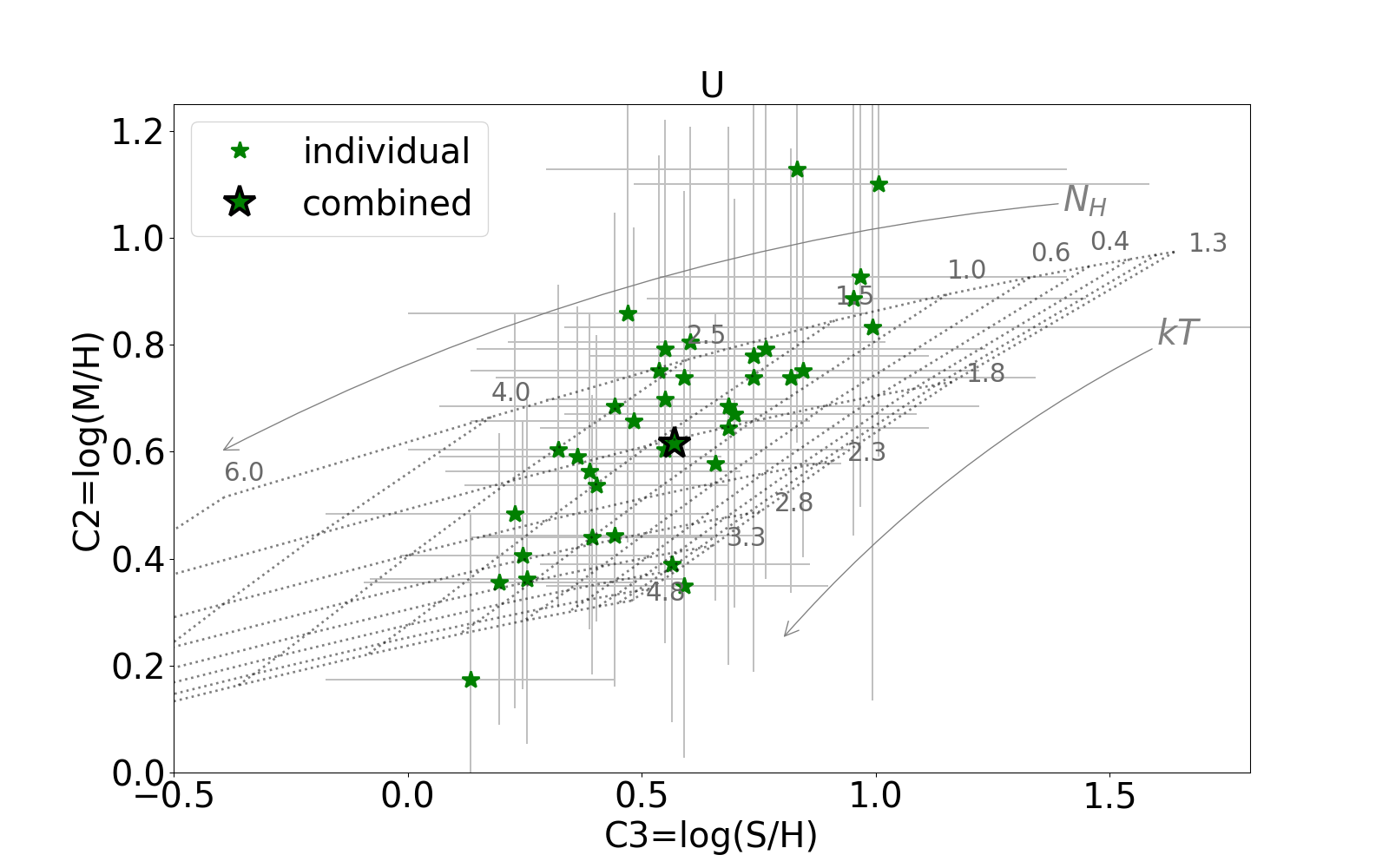}  
\includegraphics[width=0.47\columnwidth]{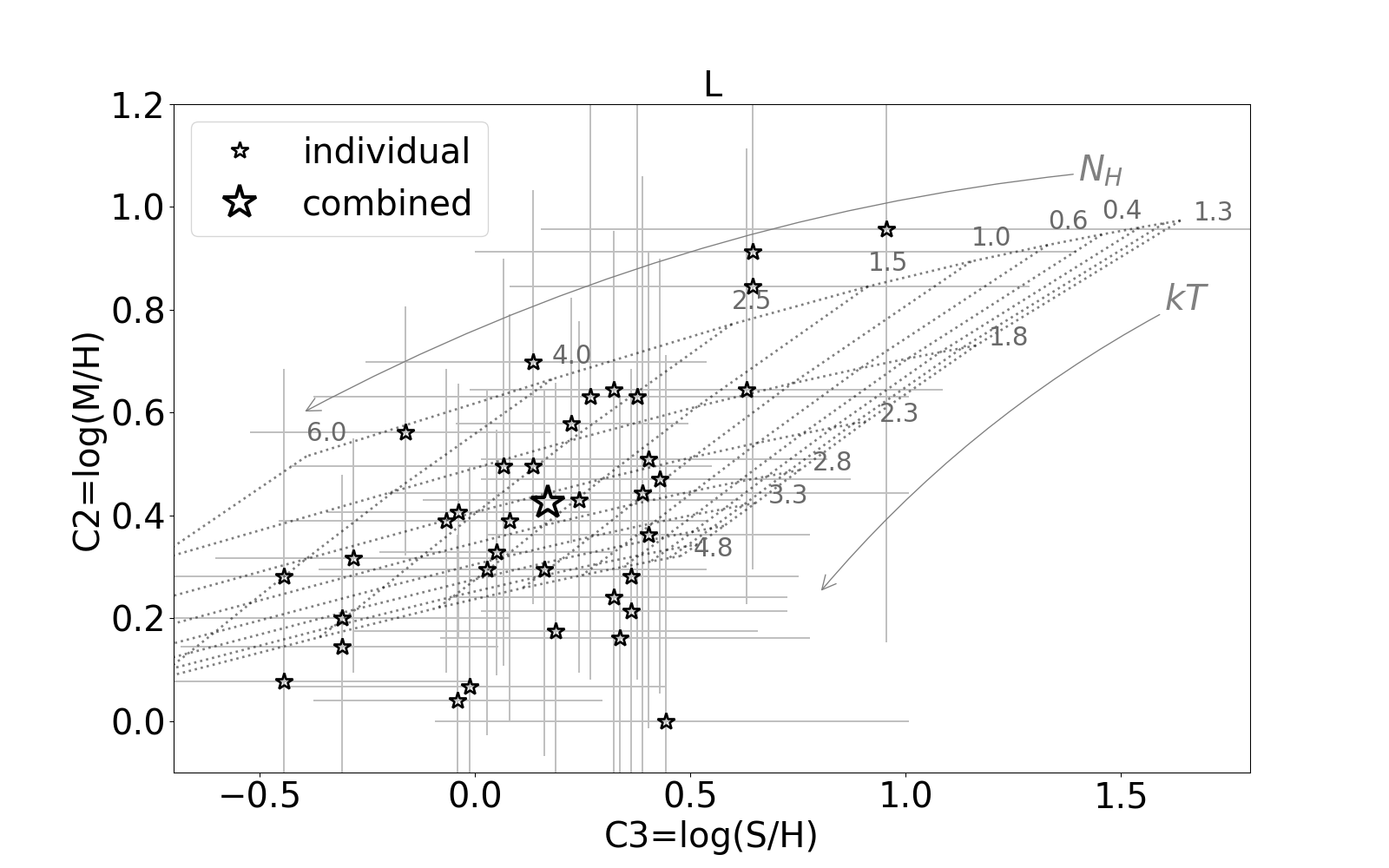}    
  \caption{Colour--colour diagrams for 4 of the brightest Wolf-Rayet stars. Large symbols correspond to the combined exposure while small symbols to the individual observations. WN type stars are marked with star symbols, WC type stars are marked with circles, while optically confirmed binaries are marked with open symbols. }
                \label{fig.color23_individual_extra}
 \end{figure*}

 \begin{figure*}[!h]
 \ContinuedFloat
        \centering
\includegraphics[width=0.47\columnwidth]{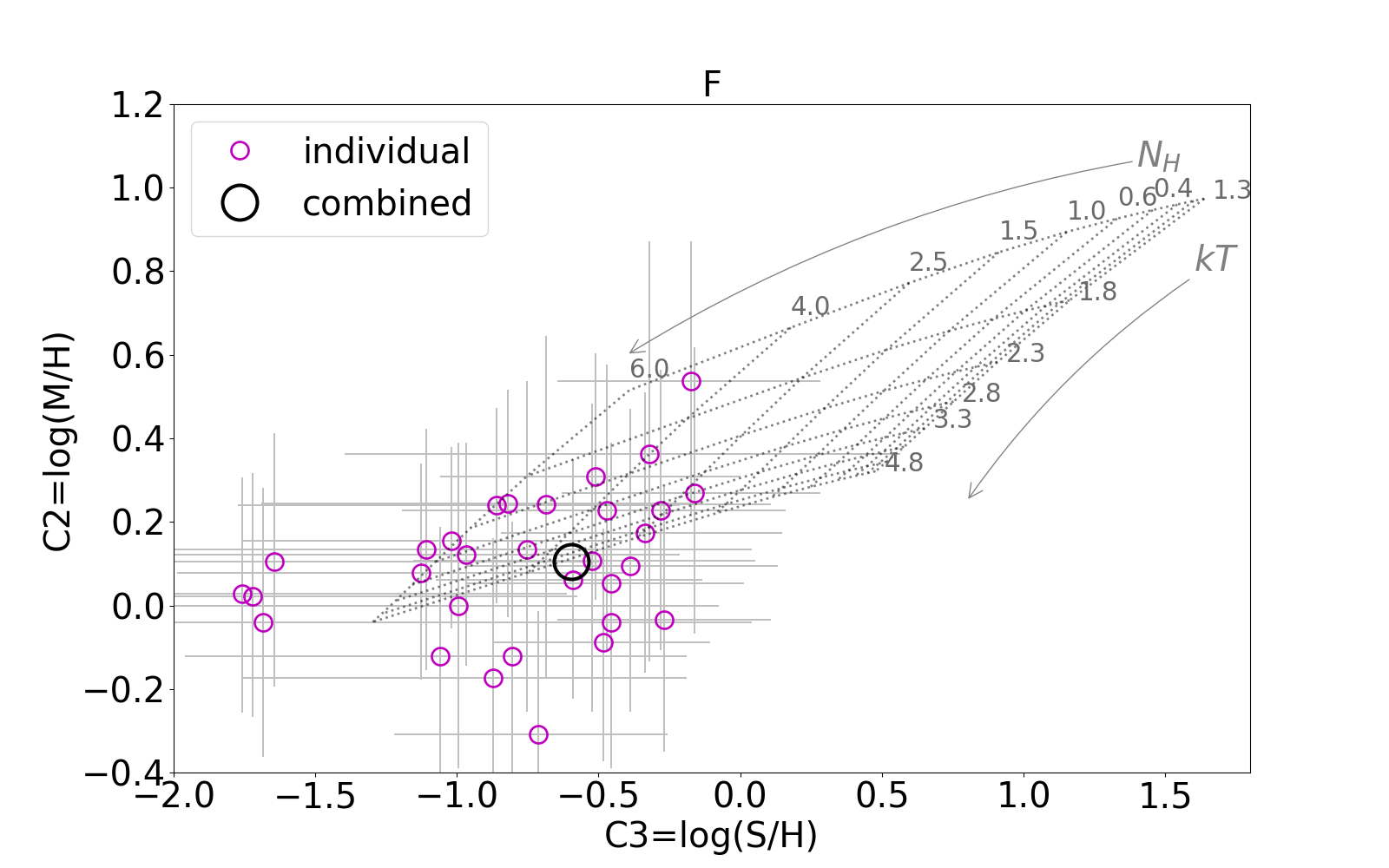}  
\includegraphics[width=0.47\columnwidth]{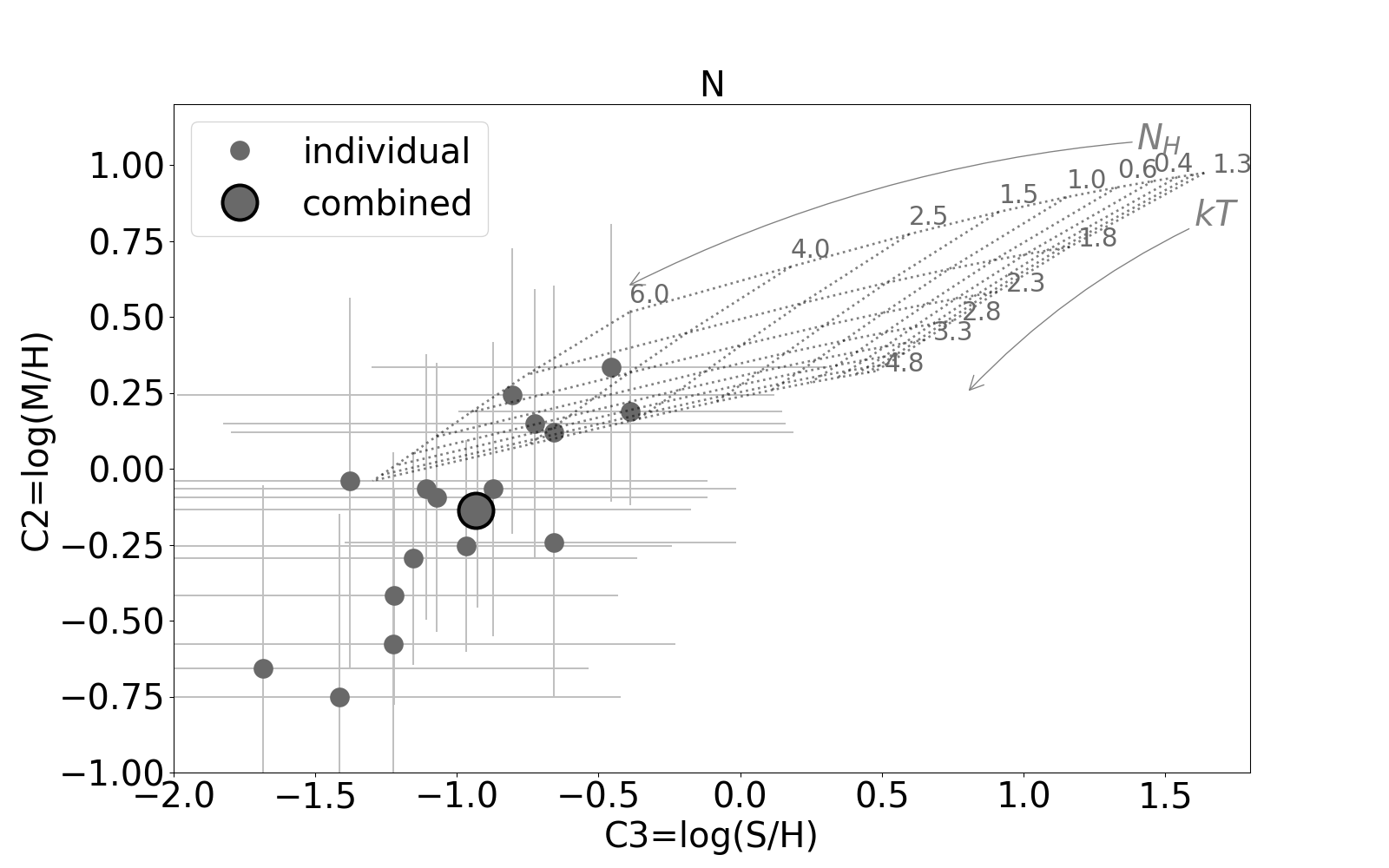}  
        \caption{Continued: Colour--colour diagrams for 4 of the brightest Wolf-Rayet stars. }
                \label{fig.color23_individual_extra}
 \end{figure*}

\FloatBarrier

\section{X-ray spectra and fitting of the WR stars in Wd1}

\begin{figure*}[htbp]
    \centering
  \renewcommand{\thesubfigure}{\arabic{subfigure}}     

    \begin{subfigure}[b]{0.47\linewidth}
        \includegraphics[width=\linewidth]{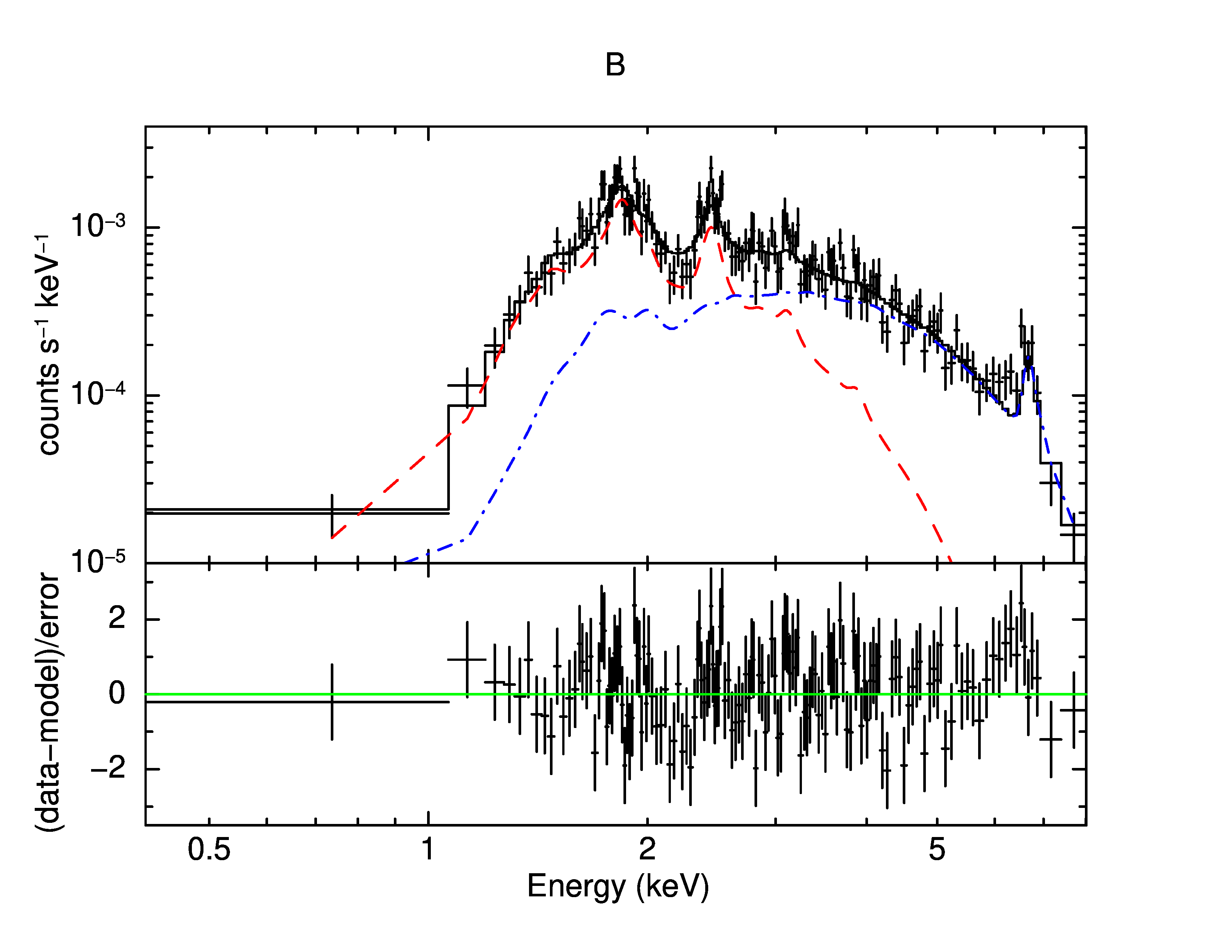} 
       \caption{}
        \label{subfig:panel3}
    \end{subfigure}    %\vspace{0.5cm}
    %\hspace{0.05\linewidth}
    \begin{subfigure}[b]{0.47\linewidth}
        \includegraphics[width=\linewidth]{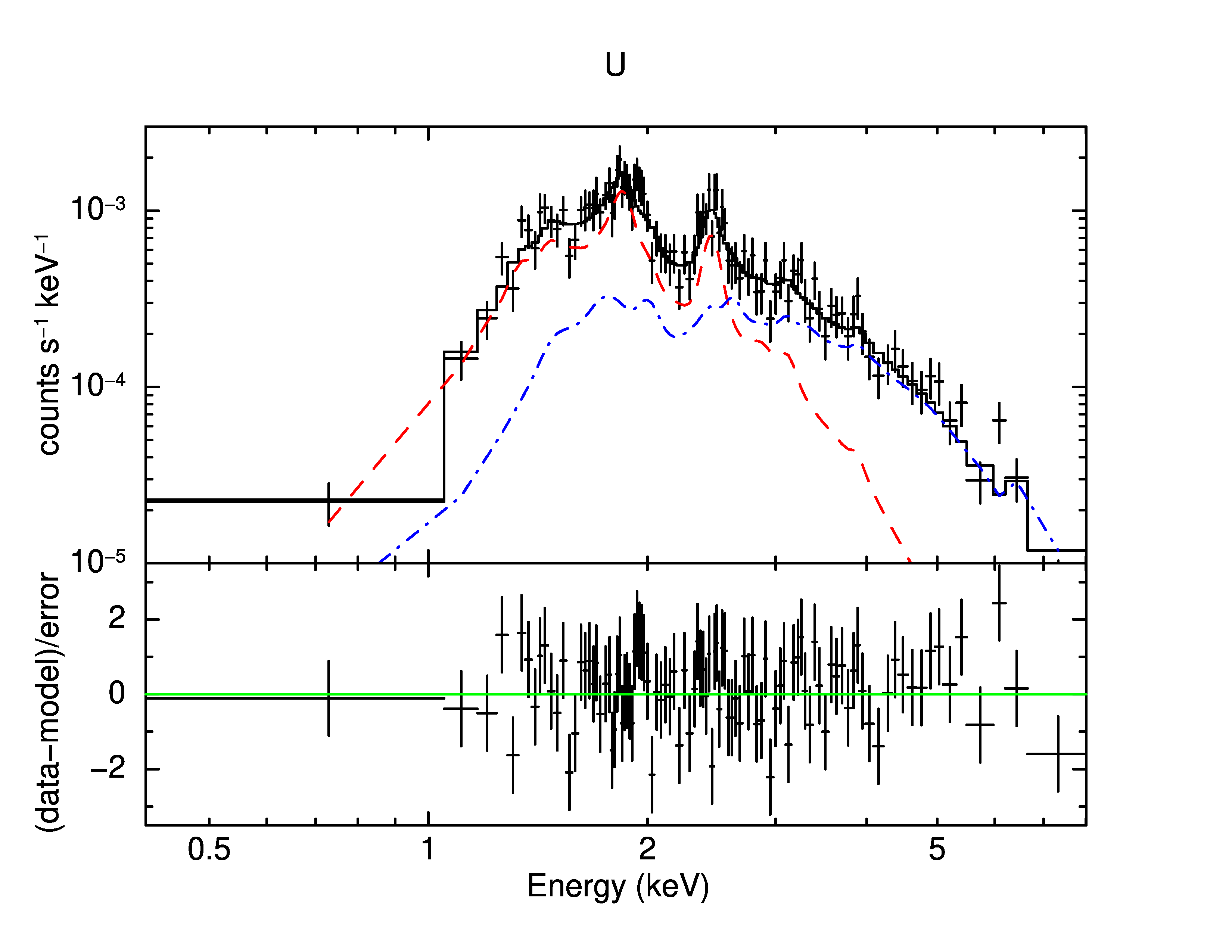} 
        \caption{}
        \label{subfig:panel4}
    \end{subfigure}
    %\hspace{0.05\linewidth}  
    \begin{subfigure}[b]{0.47\linewidth}
        \includegraphics[width=\linewidth]{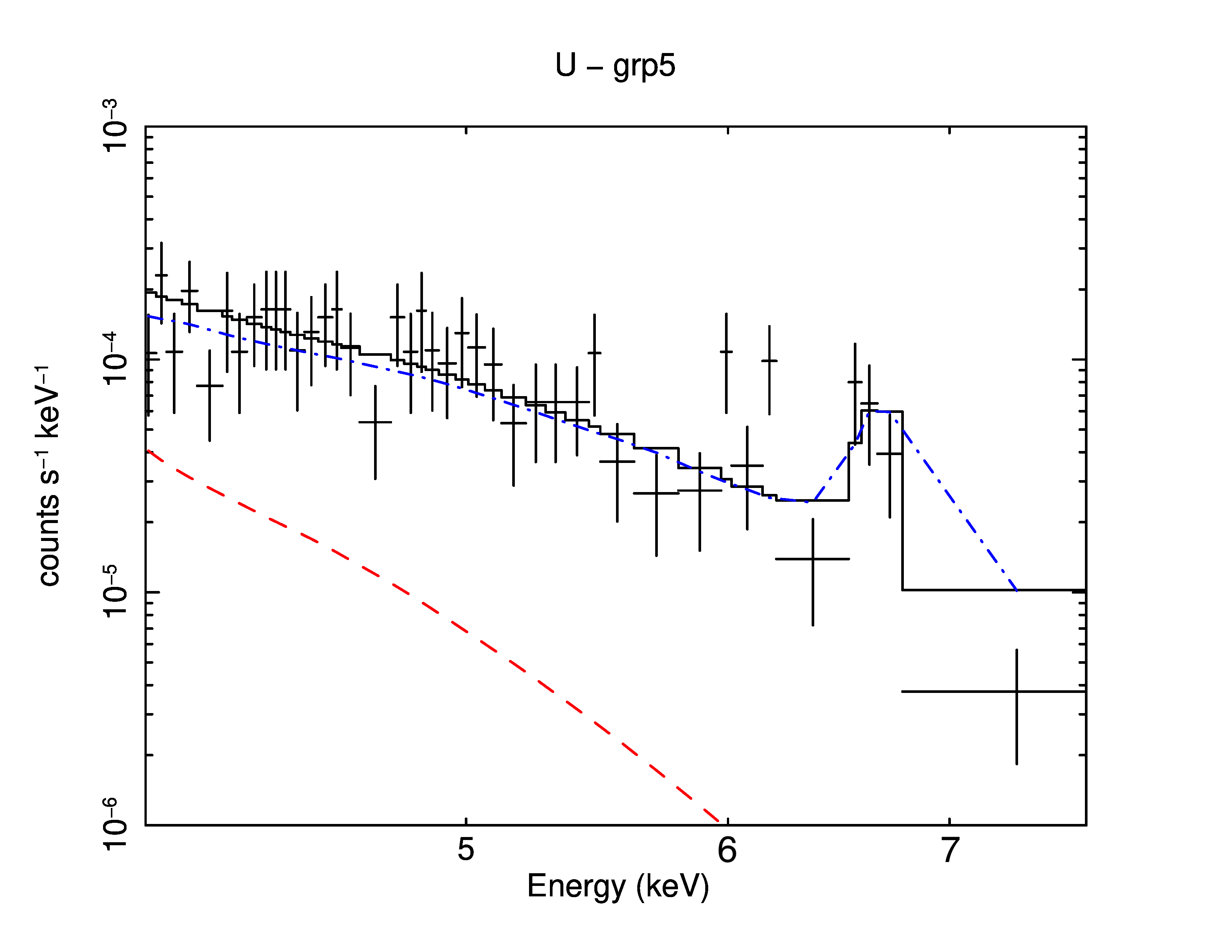} 
        \caption{}
        \label{subfig:panel5}
    \end{subfigure}   
    \begin{subfigure}[b]{0.47\linewidth}
\includegraphics[width=\linewidth]{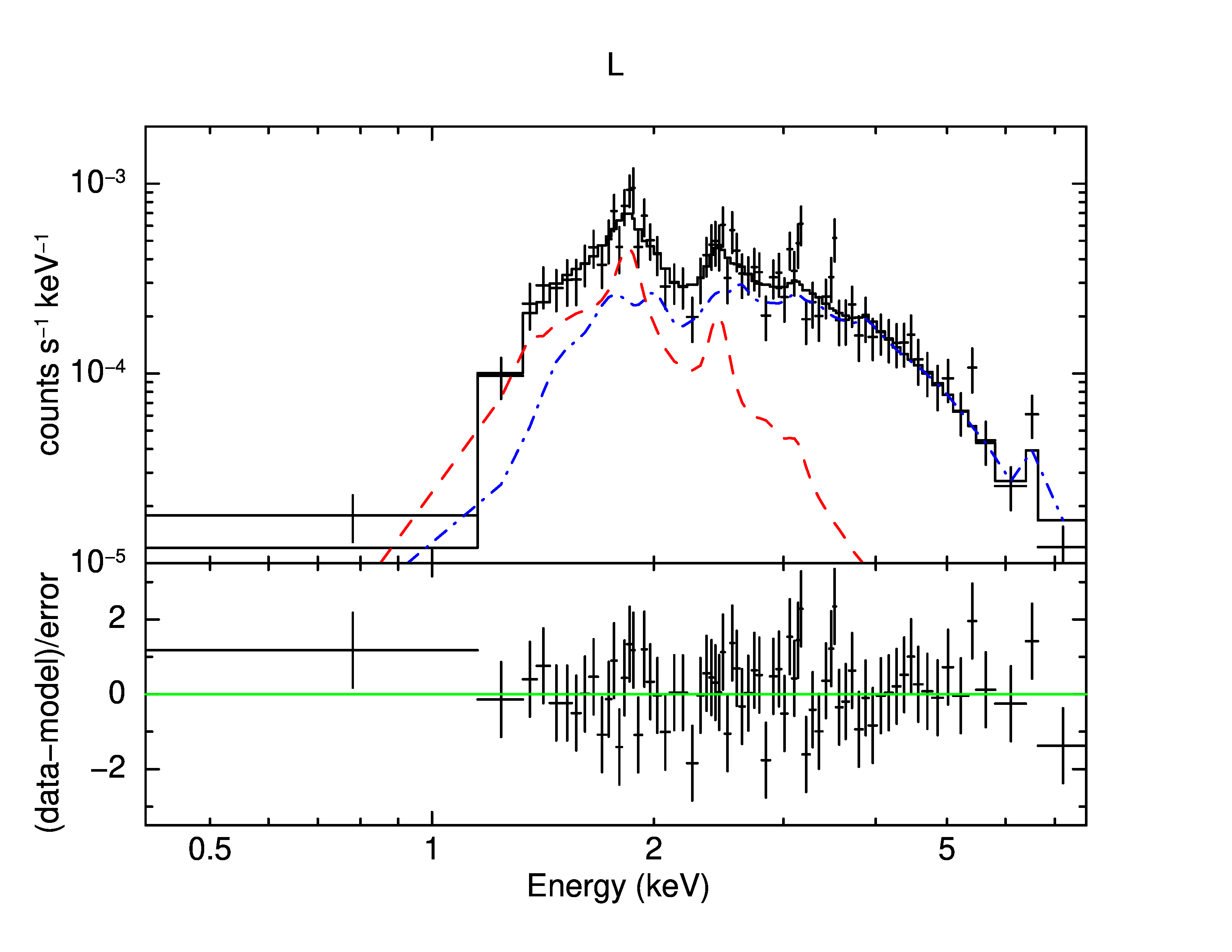}         
        \caption{}
        \label{subfig:panel6}
    \end{subfigure}
    
 \caption{Spectra of the WR stars in Wd1 along with their best-fit models. Spectra are presented from the X-ray brightest to the X-ray faintest source.}
    \label{fig.spectra}
\end{figure*}

\begin{figure*}[htbp]
    \ContinuedFloat
    \centering
  \renewcommand{\thesubfigure}{\arabic{subfigure} } 
  
    \begin{subfigure}[b]{0.47\linewidth}
        \includegraphics[width=\linewidth]{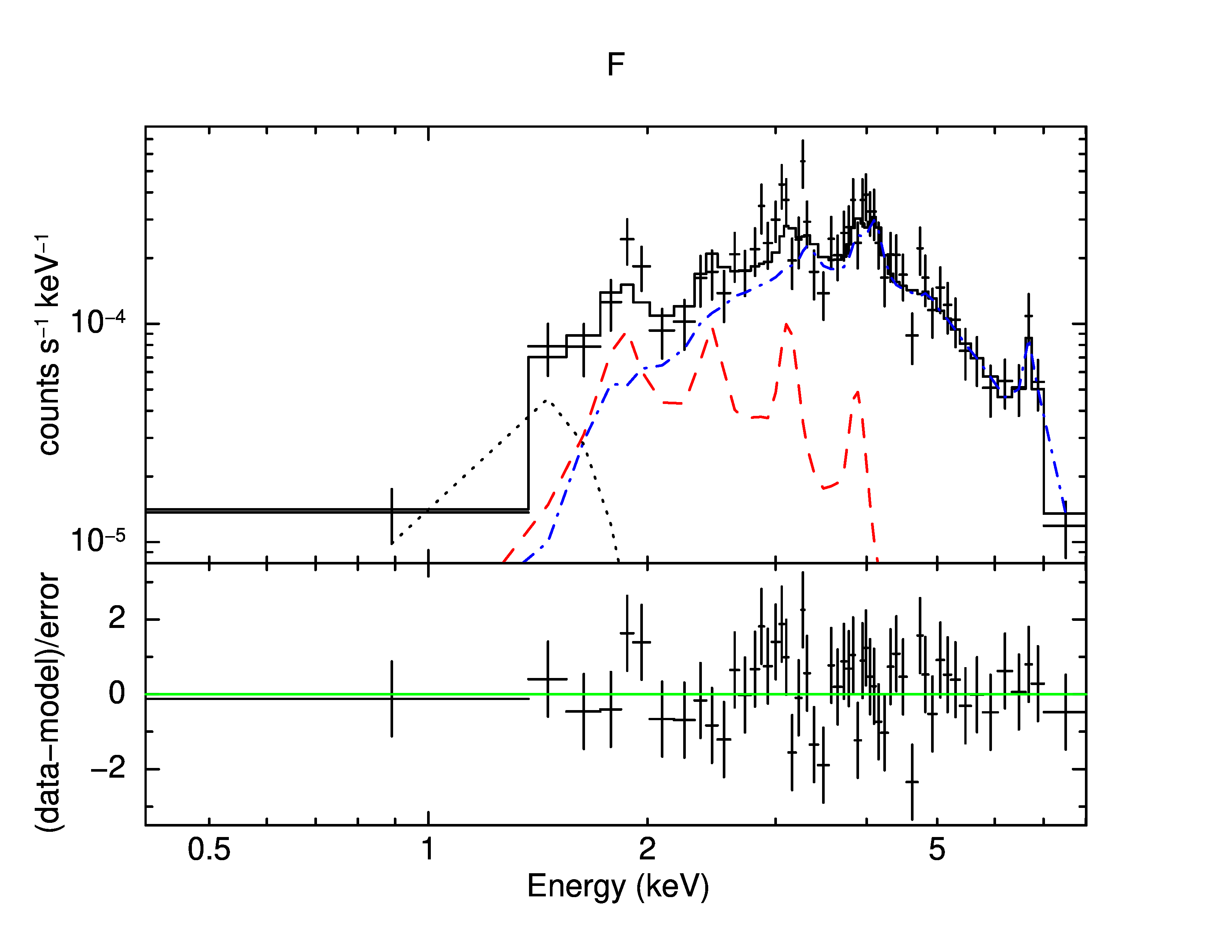} 
       \caption{}
        \label{subfig:panel8}
    \end{subfigure}    %\vspace{0.5cm}
    %\hspace{0.05\linewidth}
    \begin{subfigure}[b]{0.47\linewidth}
        \includegraphics[width=\linewidth]{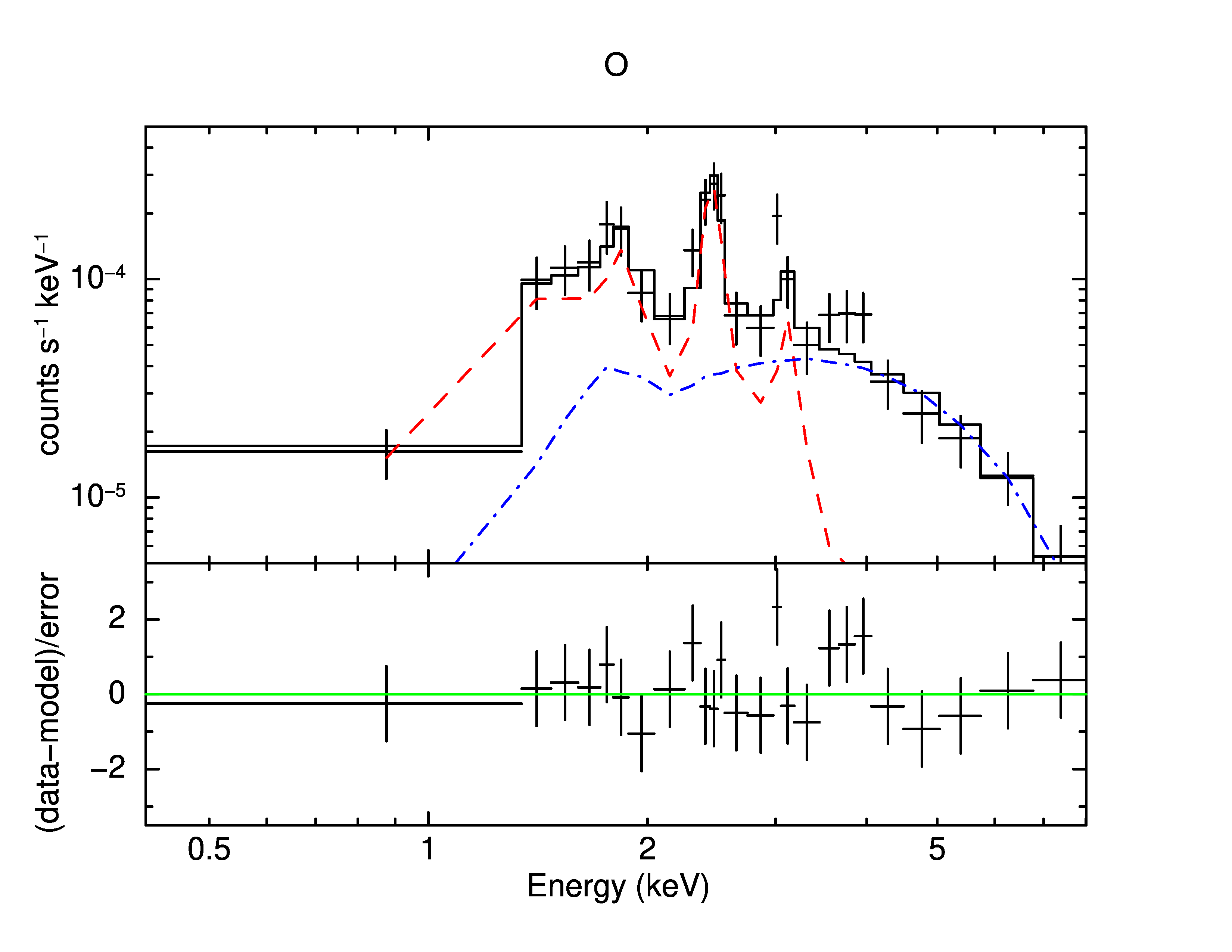} 
        \caption{}
        \label{subfig:panel9}
    \end{subfigure}
    %\hspace{0.05\linewidth}
    \begin{subfigure}[b]{0.47\linewidth}
        \includegraphics[width=\linewidth]{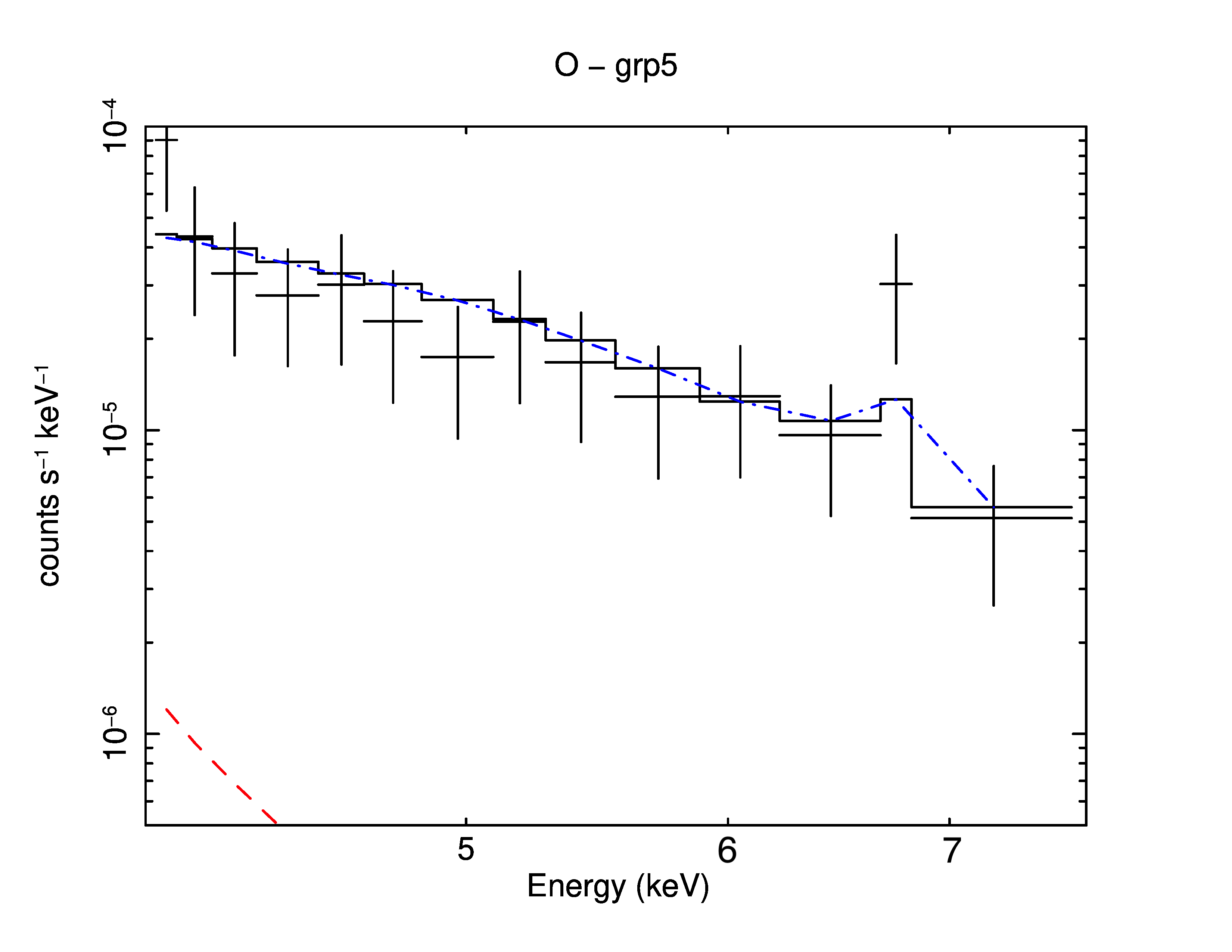} 
       \caption{}
        \label{subfig:panel11}
    \end{subfigure}    %\vspace{0.5cm}
    %\hspace{0.05\linewidth}
    \begin{subfigure}[b]{0.47\linewidth}
        \includegraphics[width=\linewidth]{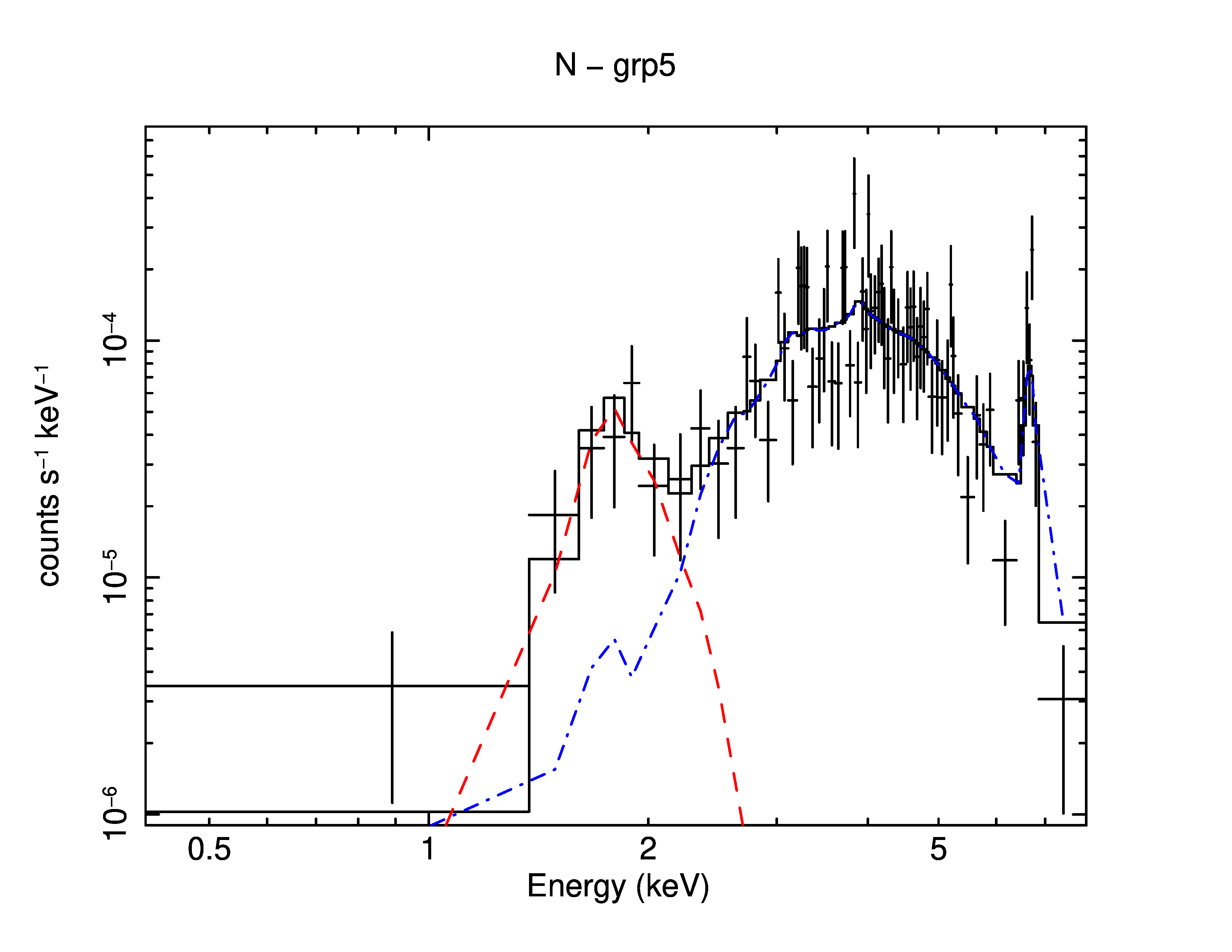} 
        \caption{}
        \label{subfig:panel13}
    \end{subfigure}
    \begin{subfigure}[b]{0.47\linewidth}
        \includegraphics[width=\linewidth]{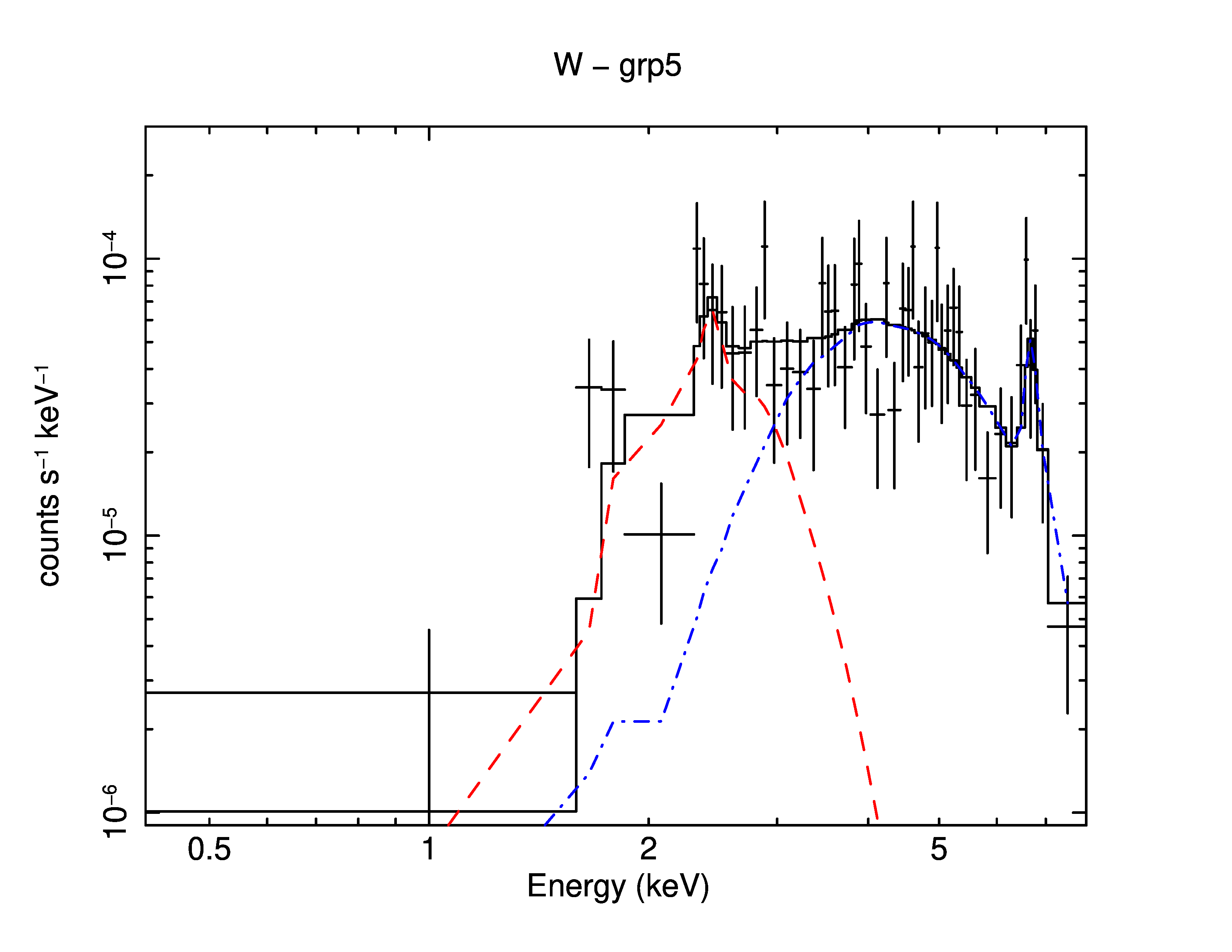} 
       \caption{}
        \label{subfig:panel14}
    \end{subfigure}    %\vspace{0.5cm}
       \begin{subfigure}[b]{0.47\linewidth}
        \includegraphics[width=\linewidth]{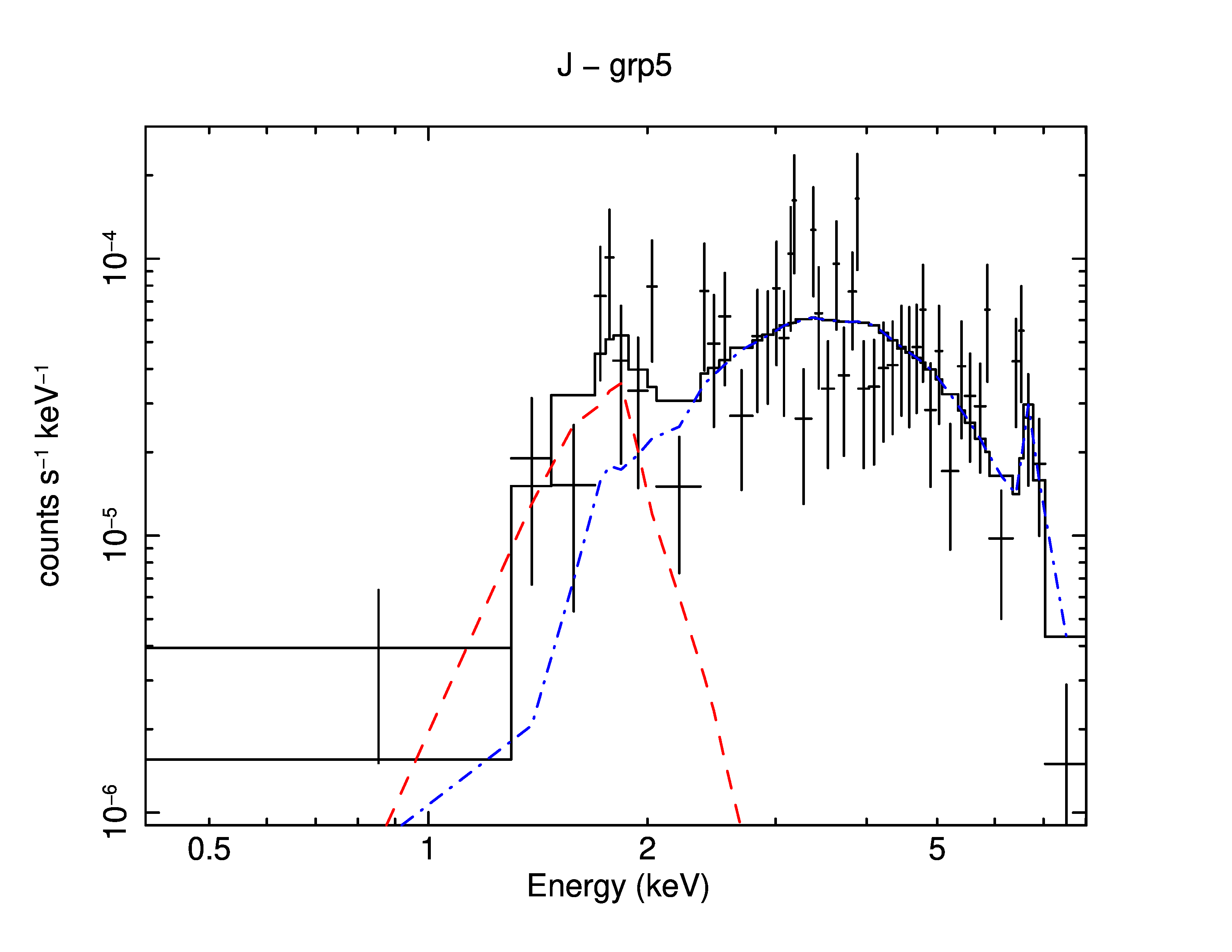} 
        \caption{}
        \label{subfig:panel15}
         \end{subfigure}    
    \caption{Continued: Spectra of the WR stars in Wd1.}
    \label{fig:spectra}
\end{figure*}

\begin{figure*}[htbp]
    \ContinuedFloat
    \centering
   \renewcommand{\thesubfigure}{\arabic{subfigure}}    
\begin{subfigure}[b]{0.47\linewidth}
        \includegraphics[width=\linewidth]{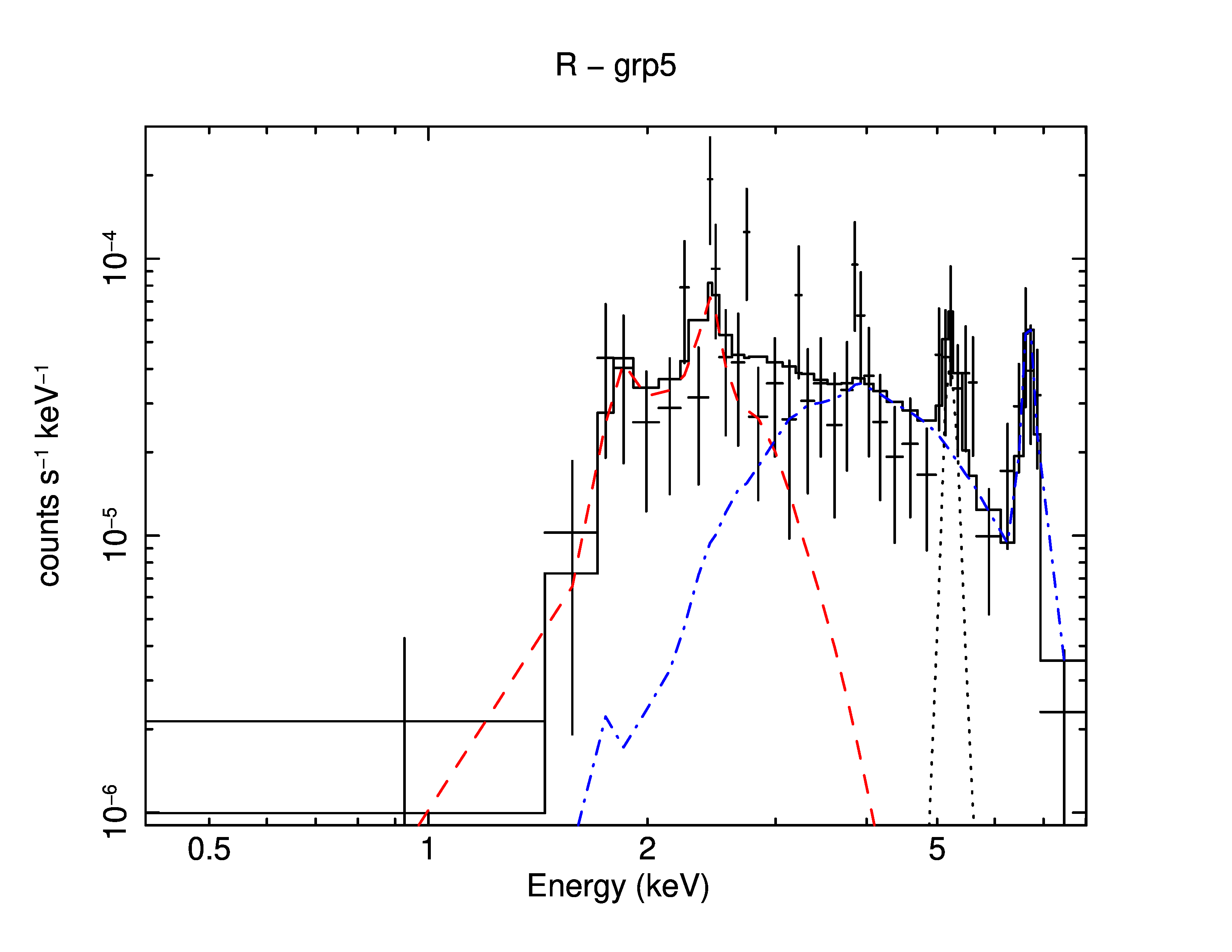} 
       \caption{}
        \label{subfig:panel18}
    \end{subfigure}    %\vspace{0.5cm}
  \begin{subfigure}[b]{0.47\linewidth}
        \includegraphics[width=\linewidth]{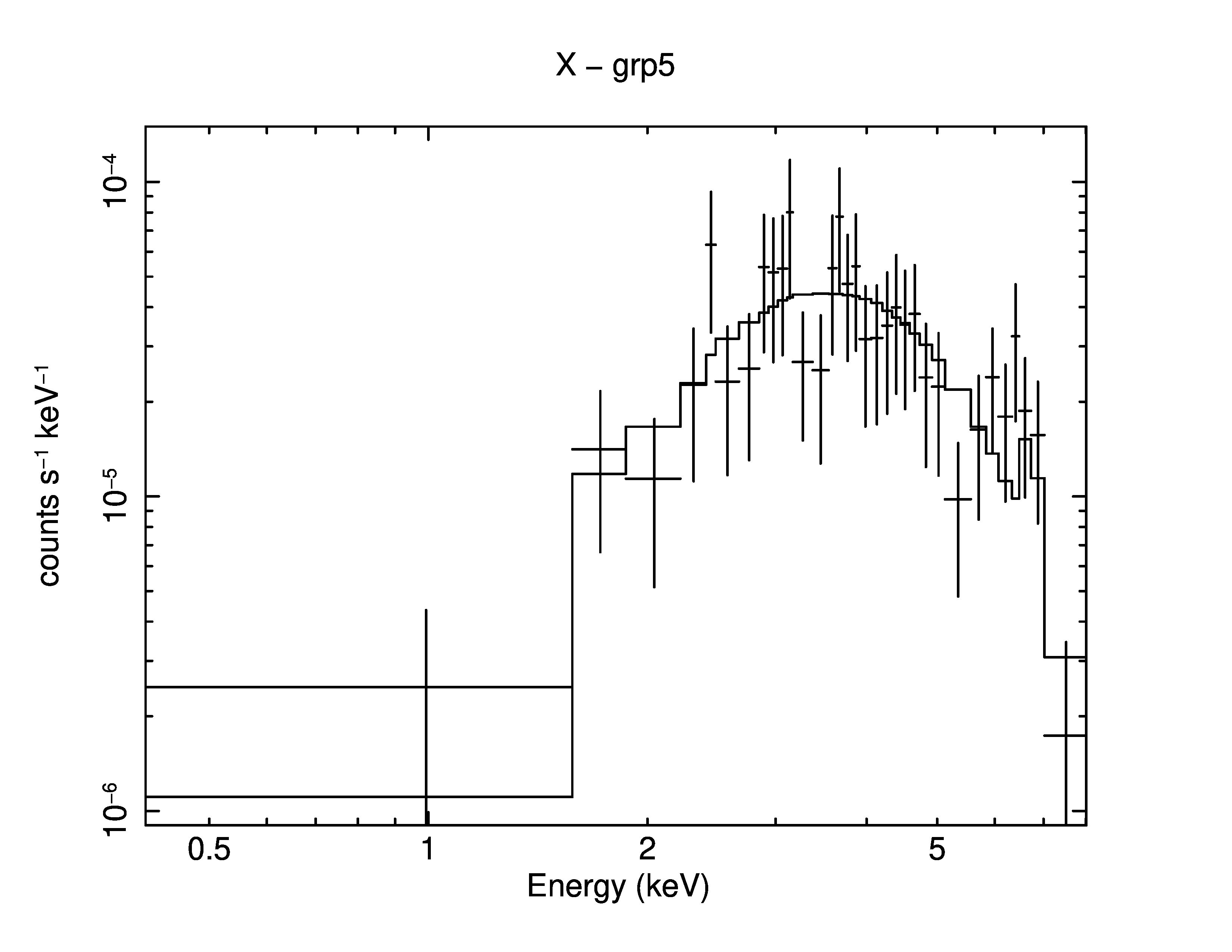} 
        \caption{}
        \label{subfig:panel20}
    \end{subfigure}
     \begin{subfigure}[b]{0.47\linewidth}
        \includegraphics[width=\linewidth]{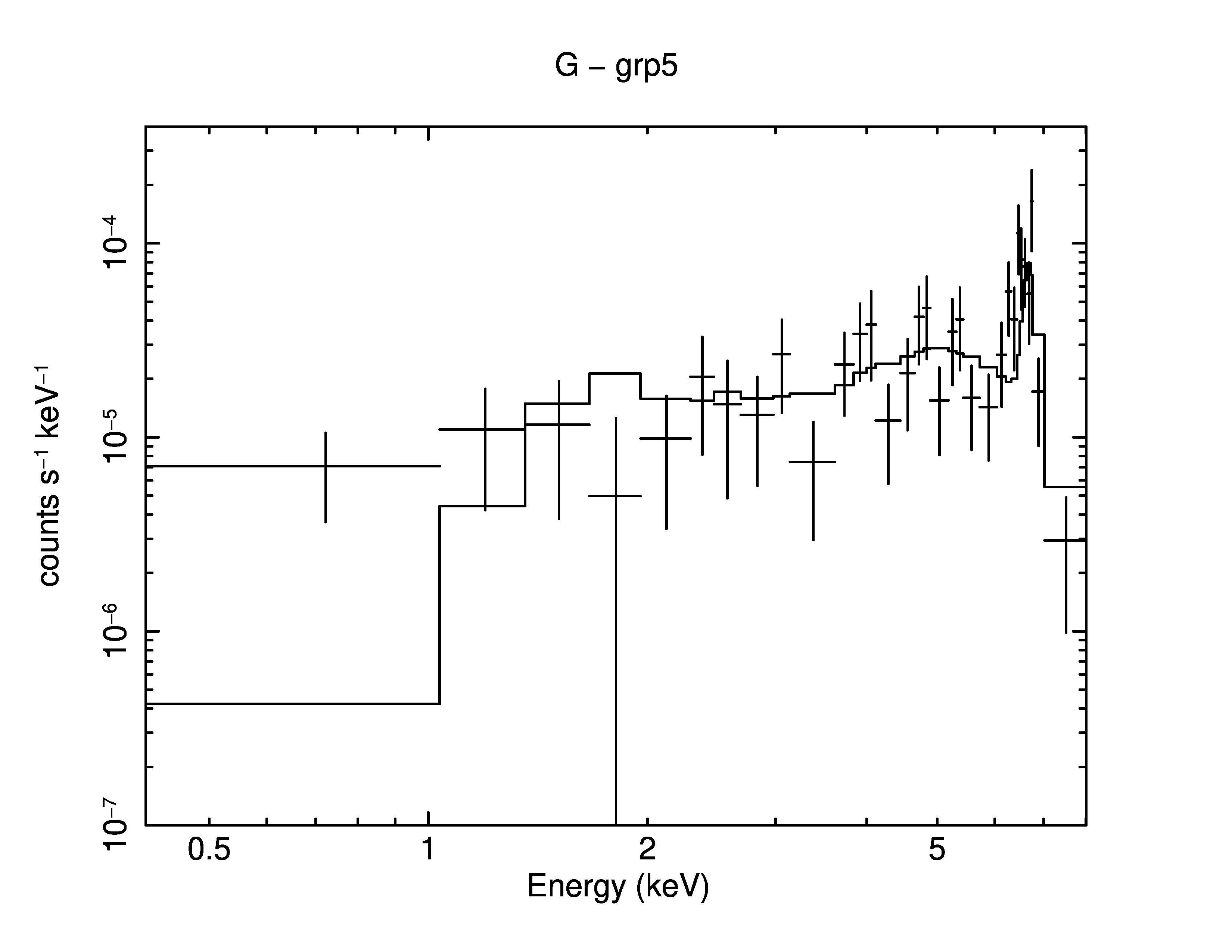} 
        \caption{}
        \label{subfig:panel22}
        \end{subfigure} 
        \begin{subfigure}[b]{0.47\linewidth}
        \includegraphics[width=\linewidth]{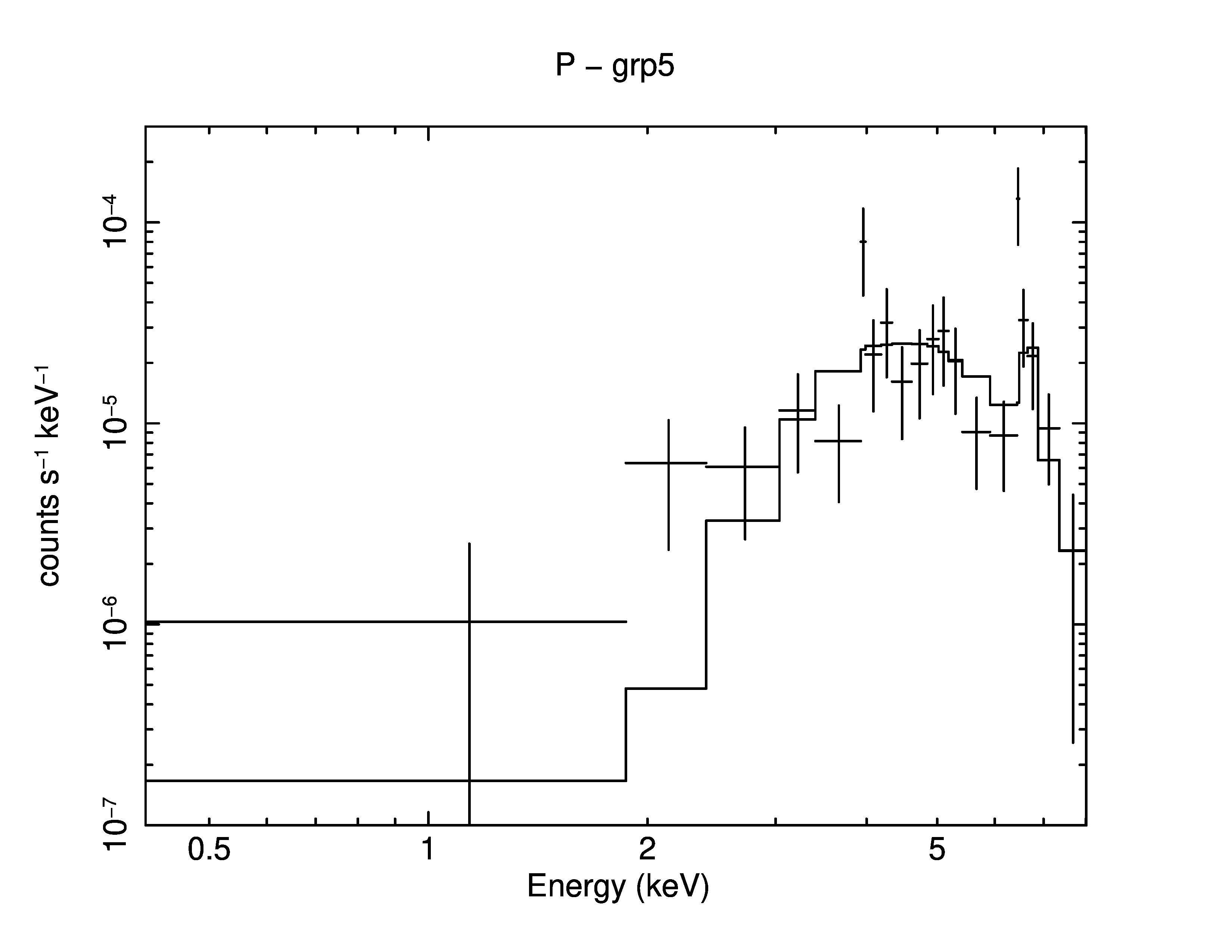} 
       \caption{}
        \label{subfig:panel24}
    \end{subfigure}    %\vspace{0.5cm}
       \begin{subfigure}[b]{0.47\linewidth}
        \includegraphics[width=\linewidth]{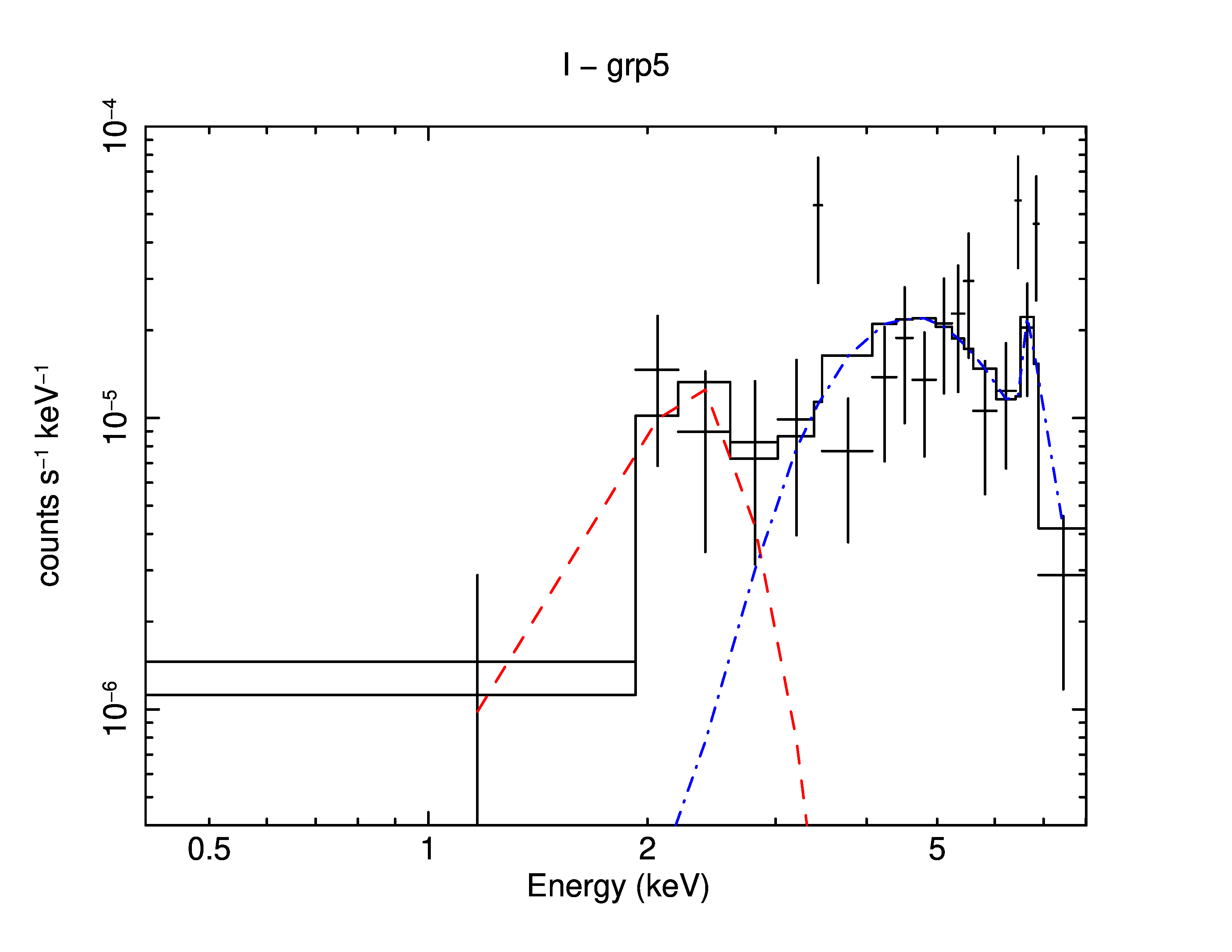} 
        \caption{}
        \label{subfig:panel26}
    \end{subfigure}
    \begin{subfigure}[b]{0.47\linewidth}
       \includegraphics[width=\linewidth]{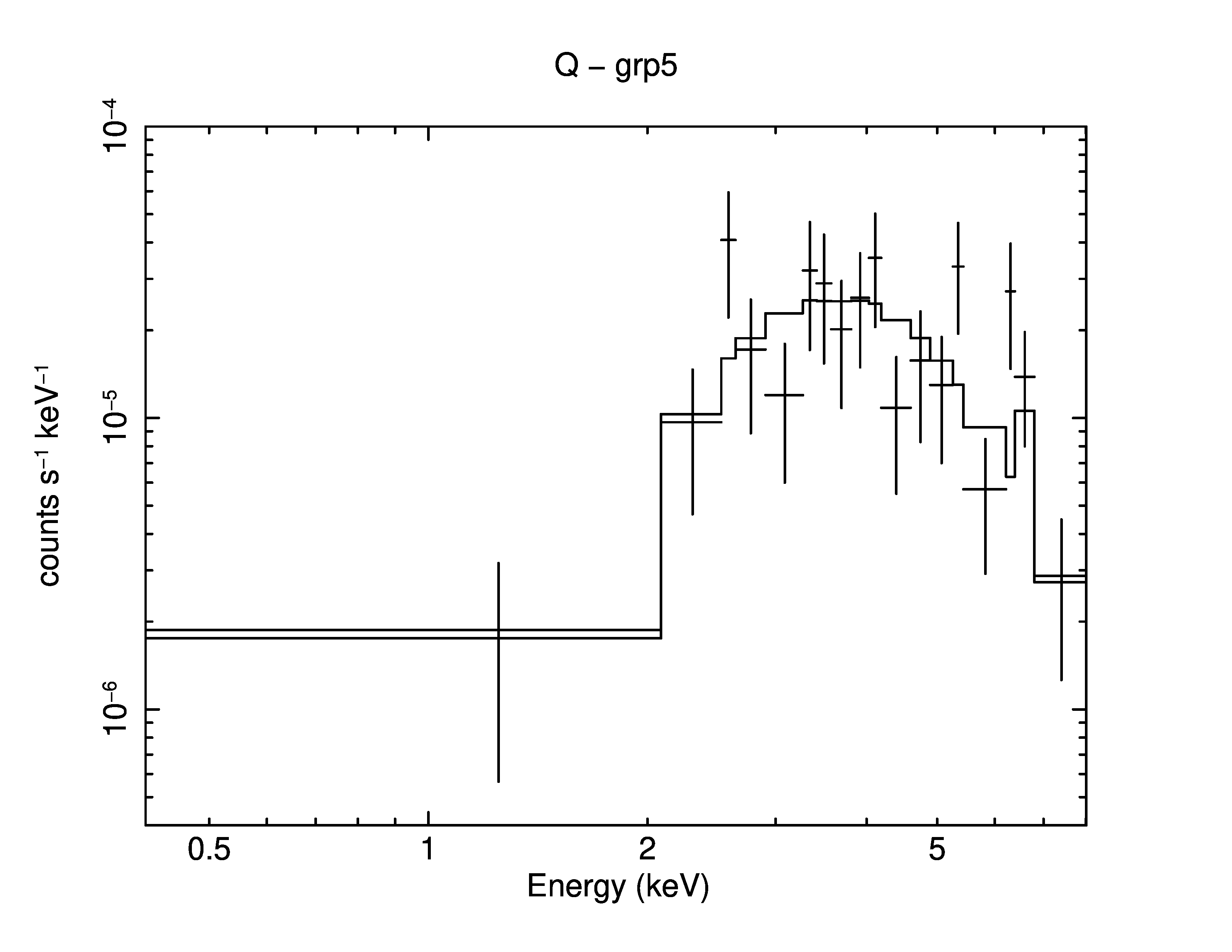} 
       \caption{}
        \label{subfig:panel30}
    \end{subfigure}    %\vspace{0.5cm}
 \caption{Continued: Spectra of the WR stars in Wd1.}
    \label{fig:spectra2}
\end{figure*}

\begin{figure*}[htbp]
    \ContinuedFloat
    \centering
     %\newcounter{subfigurecontinued}
     %\renewcommand{\thesubfigurecontinued}{\alph{subfigurecontinued}}    
      \renewcommand{\thesubfigure}{\arabic{subfigure}} 
\begin{subfigure}[b]{0.47\linewidth}
        \includegraphics[width=\linewidth]{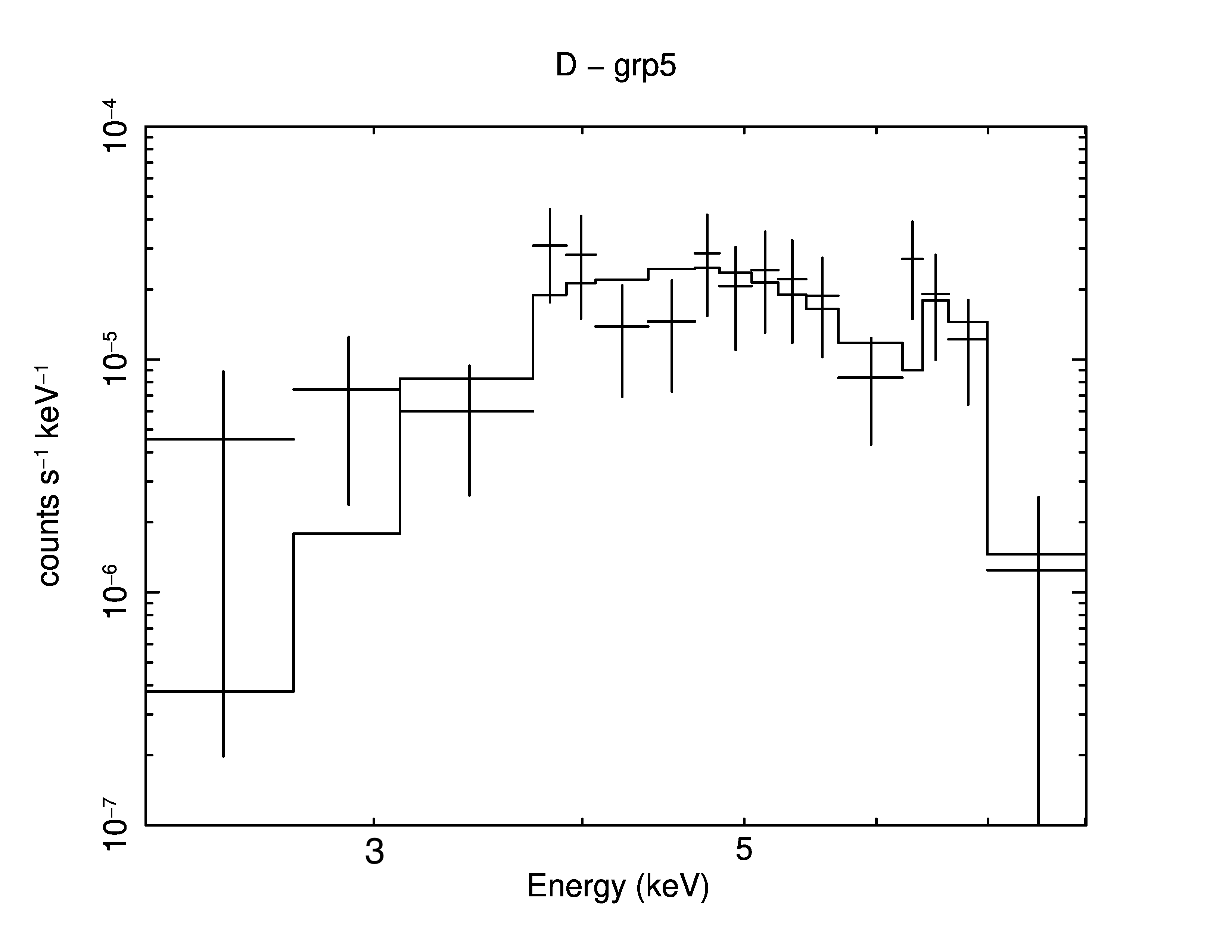} 
        \caption{}
        \label{subfig:panel32}
    \end{subfigure}
    \begin{subfigure}[b]{0.47\linewidth}
        \includegraphics[width=\linewidth]{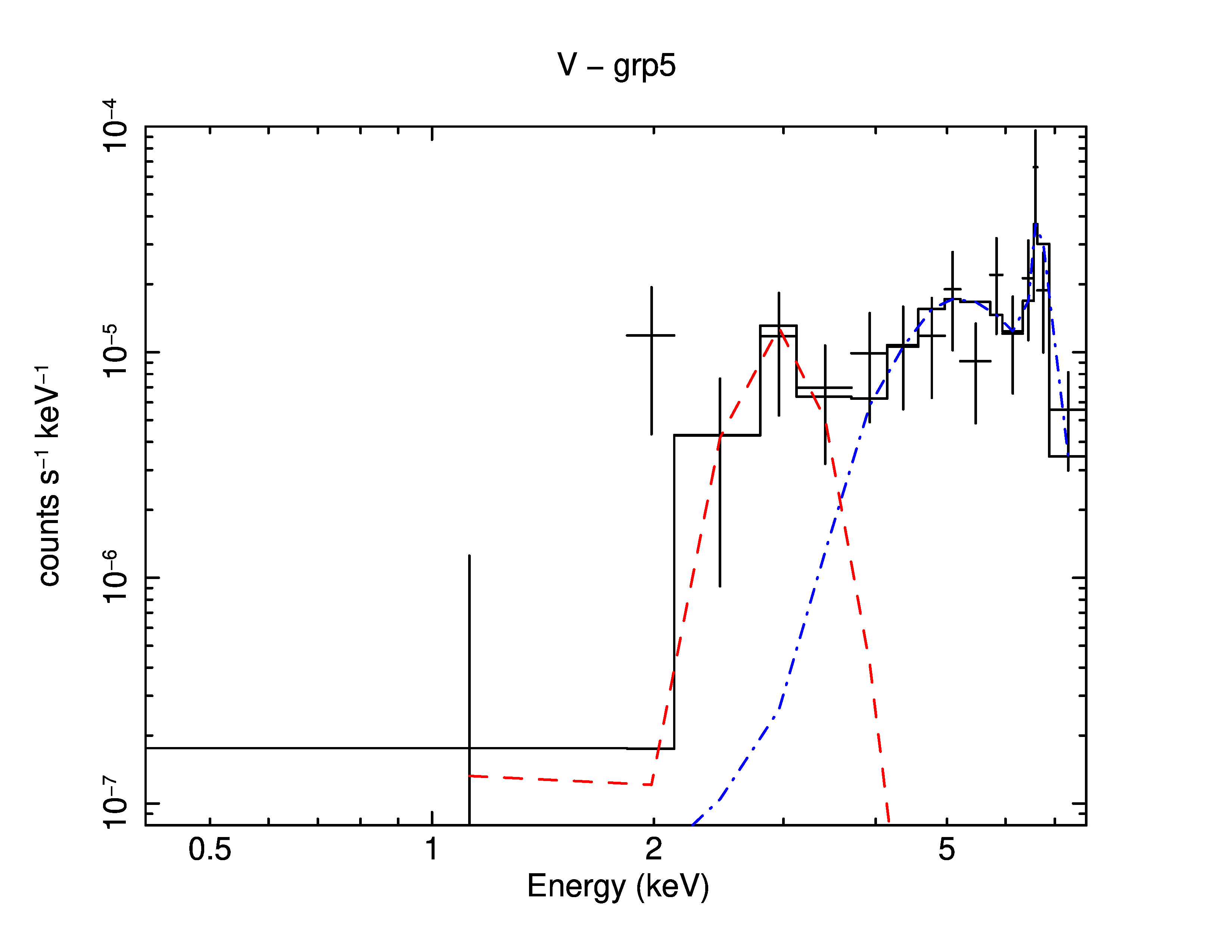} 
        \caption{}
        \label{subfig:panel34}
    \end{subfigure}    
\begin{subfigure}[b]{0.47\linewidth}
        \includegraphics[width=\linewidth]{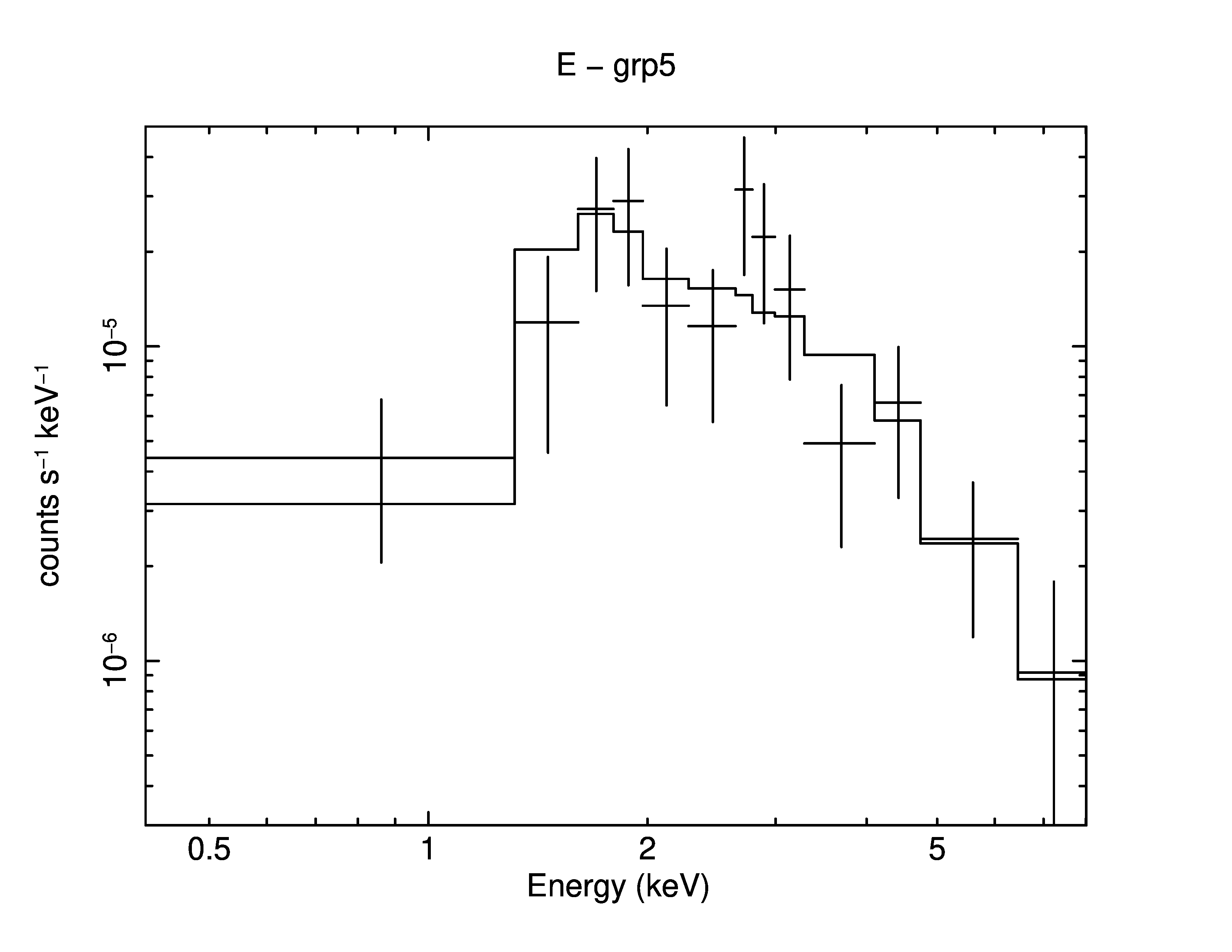} 
       \caption{}
        \label{subfig:panel36}
    \end{subfigure}    %\vspace{0.5cm}
     \begin{subfigure}[b]{0.47\linewidth}
        \includegraphics[width=\linewidth]{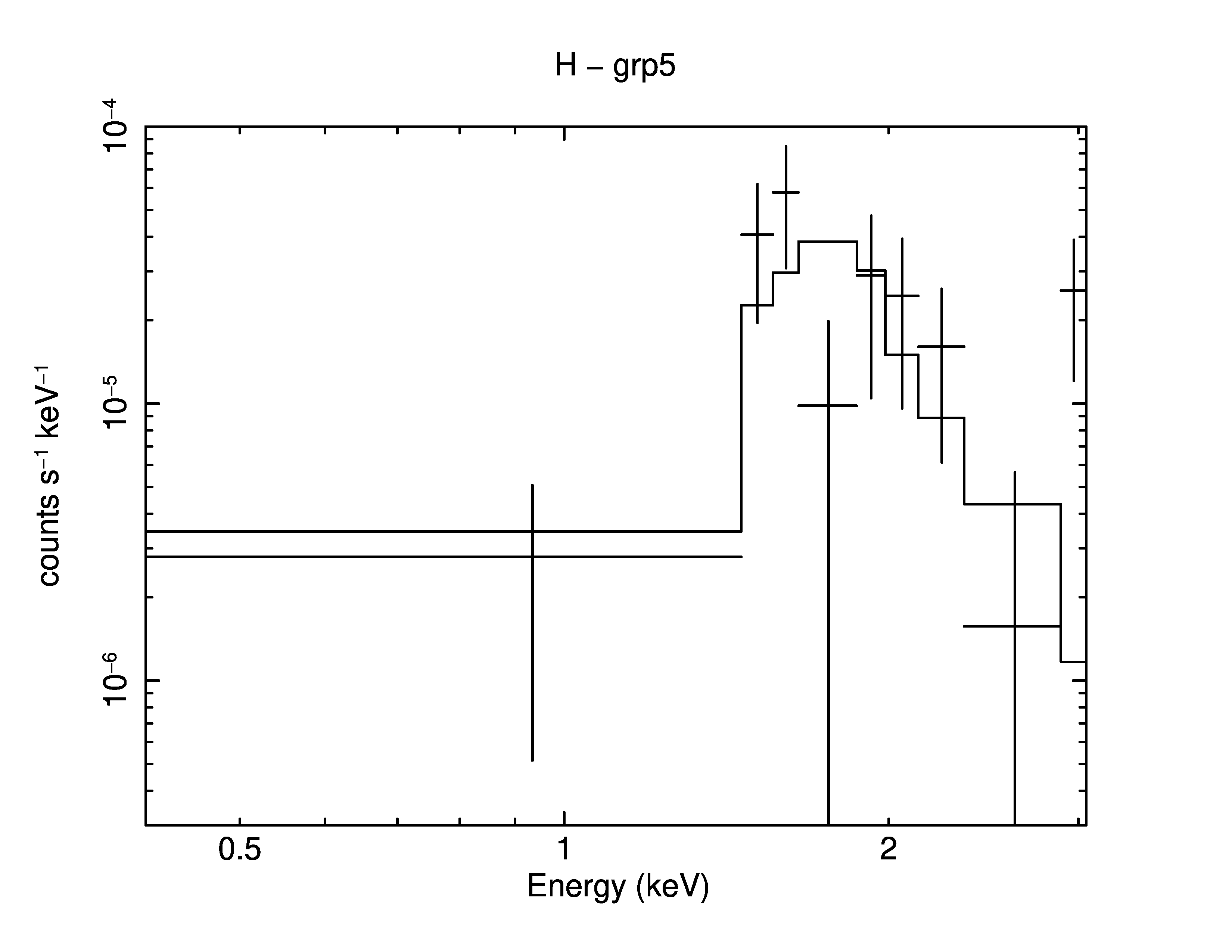} 
        \caption{}
        \label{subfig:panel37}
    \end{subfigure}    
    \begin{subfigure}[b]{0.47\linewidth}
        \includegraphics[width=\linewidth]{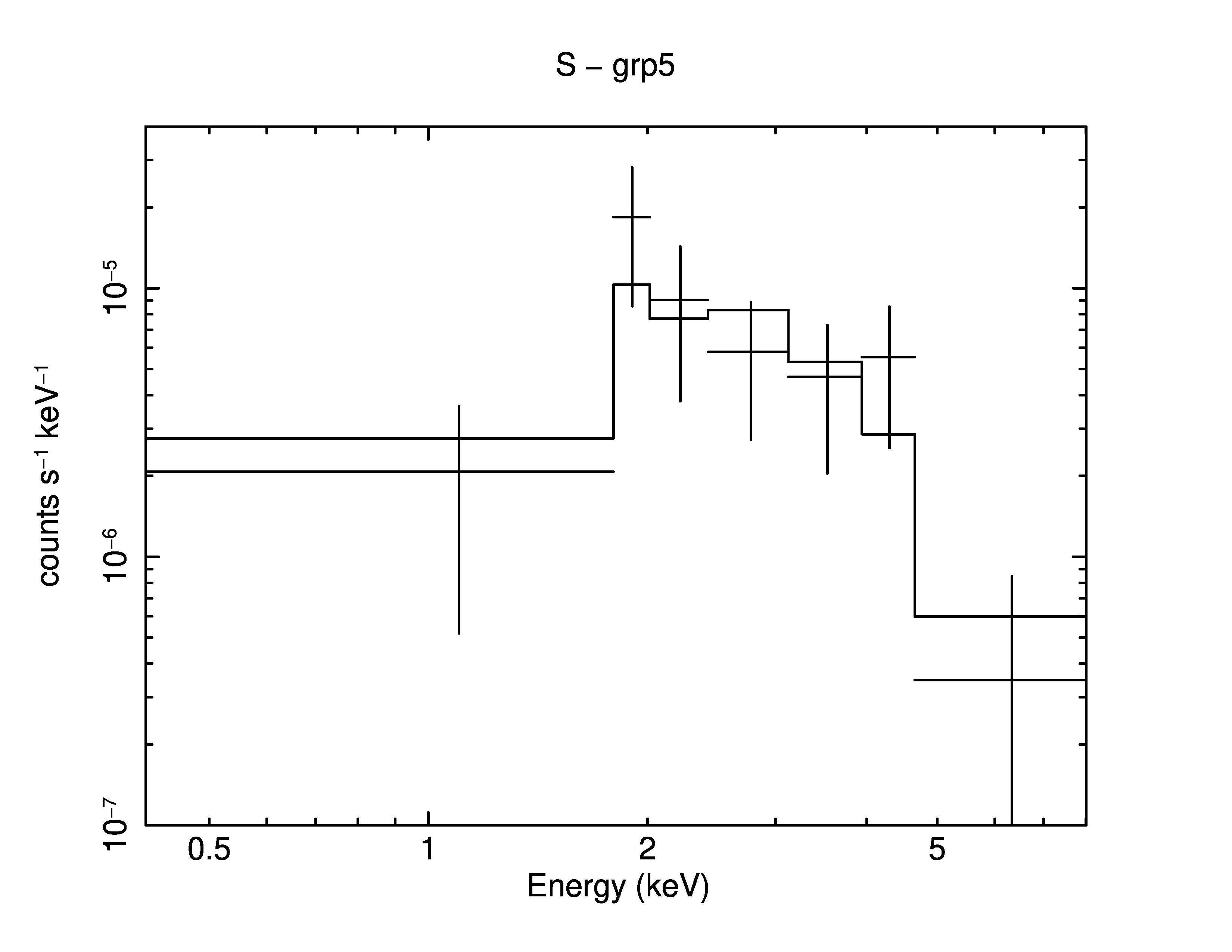} 
        \caption{}
        \label{subfig:panel38}
    \end{subfigure}    
\caption{Continued: Spectra of the WR stars in Wd1.}
    \label{fig:spectra}
\end{figure*}

\end{appendix}

% Don't change these lines
%\bsp   % typesetting comment
%\label{lastpage}
\end{document}